%% file: DGS_KF_dust.tex
\documentclass[structabstract]{aa}  

\usepackage{tabularx}
\usepackage{graphicx}
\usepackage{natbib}
\usepackage{txfonts}
\usepackage{longtable}
\usepackage{multirow}
\usepackage{lscape}
\usepackage{amssymb}

\newcommand{\hersc}{{\it Herschel}}

\newcommand{\wise}{WISE}
\newcommand{\spitz}{{\it  Spitzer}}

\newcommand{\iras}{IRAS}

\newcommand{\lsun}{$L_\odot$}
\newcommand{\msun}{$M_\odot$}

\newcommand{\zsun}{$Z_\odot$}
\newcommand{\mic}{$\mu$m}
\newcommand{\sfr}{$\rm M_\odot\, yr^{-1}$}

\newcommand{\HII}{H{\sc ii}}

\newcommand{\mgas}{{\it M$_{gas}$}}

\newcommand{\mdust}{{\it M$_{{\rm dust}}$}}
\newcommand{\ltir}{$L_{{\rm TIR}}$}
\newcommand{\lfir}{$L_{{\rm FIR}}$}
\newcommand{\uav}{$\langle U\rangle$}
\newcommand{\sigU}{$\sigma U$}
\newcommand{\DU}{$\Delta U$}
\newcommand{\umin}{$U_{{\rm min}}$}
\newcommand{\ssfr}{$sSFR$}
\newcommand{\fpah}{$f_{{\rm PAH}}$}
\newcommand{\fvsg}{$f_{{\rm vsg}}$}
\newcommand{\mstar}{$M_{{\rm star}}$}
\newcommand{\Ao}{$Z$}

\renewcommand*{\figurename}{ Fig.}
\renewcommand*{\tablename}{Table}
\newlength{\pointwidth}
\DeclareGraphicsExtensions{.pdf,.png,.jpg, .eps}

\def\rev{}

\begin{document}

  \title{Linking dust emission to fundamental properties in galaxies: \\
  	 The low-metallicity picture}
  
  \author{A. R\'emy-Ruyer\inst{1} 
  	\and S. C. Madden\inst{2}
  	\and F. Galliano\inst{2}
	\and V. Lebouteiller\inst{2}
	\and M. Baes\inst{3}
	\and G. J. Bendo\inst{4}
	\and A. Boselli\inst{5}
	\and L. Ciesla\inst{6}
	\and D. Cormier\inst{7}
	\and A. Cooray\inst{8}
	\and L. Cortese\inst{9}
	\and I. De Looze\inst{3,10}
	\and V. Doublier-Pritchard\inst{11}
	\and M. Galametz\inst{12}
	\and A. P. Jones\inst{1}	      
	\and O. \L. Karczewski\inst{13} 
	\and N. Lu\inst{14}
	\and L. Spinoglio\inst{15}
	}

	      \institute{Institut d'Astrophysique Spatiale, CNRS, UMR8617, 91405, Orsay, France \\
	      \email{aurelie.remyruyer@ias.u-psud.fr} 
	      \and Laboratoire AIM, CEA/IRFU/Service d'Astrophysique, Universit\'{e} Paris Diderot, Bat. 709, 91191 Gif-sur-Yvette, France 
	      \and Sterrenkundig Observatorium, Universiteit Gent, Krijgslaan 281 S9, B-9000 Gent, Belgium 
	      \and UK ALMA Regional Centre Node, Jodrell Bank Centre for Astrophysics, School of Physics \& Astronomy, University of Manchester, Oxford Road, Manchester M13 9PL, UK 
	      \and Laboratoire dÕAstrophysique de Marseille - LAM, Universit\'e d'Aix-Marseille \& CNRS, UMR7326, 38 rue F. Joliot-Curie, 13388 Marseille Cedex 13, France 
	      \and Department of Physics, University of Crete, GR-71003, Heraklion, Greece 
	      \and Zentrum f\"ur Astronomie der Universit\"at Heidelberg, Institut f\"ur Theoretische Astrophysik, Albert-Ueberle-Str. 2, 69120 Heidelberg, Germany 
	      \and Center for Cosmology, Department of Physics and Astronomy, University of California, Irvine, CA 92697, USA
	      \and Centre for Astrophysics \& Supercomputing, Swinburne University of Technology, Mail H30, PO Box 218, Hawthorn, VIC 3122, Australia 
	      \and Institute of Astronomy, University of Cambridge, Madingley Road, Cambridge, CB3 0HA, UK 
	      \and Max Planck f\"ur Extraterrestrische Physik, Giessenbachstr. 1, 85748 Garching-bei-M\"unchen, Germany 
	      \and European Southern Observatory, Karl-Schwarzschild-Str. 2, D-85748 Garching-bei-M\"unchen, Germany 
	      \and Department of Physics and Astronomy, University of Sussex, Brighton, BN1 9QH, UK 
	      \and NASA Herschel Science Center, MS 100-22, California Institute of Technology, Pasadena, CA 91125, USA 
	      \and Instituto di Astrofisica e Planetologia Spaziali, INAF-IAPS, Via Fosso del Cavaliere 100, I-00133 Roma, Italy 
	      }

\date{Received date/Accepted date}


 \abstract
{}
{In this work, we aim at providing a consistent analysis of the dust properties from metal-poor to metal-rich environments by linking them to fundamental galactic parameters.
}
{We consider two samples of galaxies: the Dwarf Galaxy Survey (DGS) and the Key Insights on Nearby Galaxies: a Far-Infrared Survey with \hersc\ (KINGFISH), totalling 109 galaxies, spanning almost 2 dex in metallicity. We collect infrared (IR) to submillimetre (submm) data for both samples and present the complete data set for the DGS sample. 
We model the observed spectral energy distributions (SED) with a physically-motivated dust model to access the dust properties: dust mass, total-IR luminosity, polycyclic aromatic hydrocarbon (PAH) mass fraction, dust temperature distribution, and dust-to-stellar mass ratio. 
}
{Using a different SED model (modified blackbody), different dust composition (amorphous carbon in lieu of graphite), or a different wavelength coverage at submm wavelengths results in differences in the dust mass estimate of a factor two to three, showing that this parameter is subject to non-negligible systematic modelling uncertainties.
We find half as much dust with the amorphous carbon dust composition. For eight galaxies in our sample, we find a rather small excess at 500 \micÊ\ ($\leq$ 1.5 $\sigma$). 
We find that the dust SED of low-metallicity galaxies is broader and peaks at shorter wavelengths compared to more metal-rich systems, a sign of a clumpier medium in dwarf galaxies. The PAH mass fraction and the dust temperature distribution are found to be driven mostly by the specific star-formation rate, \ssfr, with secondary effects from metallicity.
The correlations between metallicity and dust mass or total-IR luminosity are direct consequences of the stellar mass-metallicity relation. The dust-to-stellar mass ratios of metal-rich sources follow the well-studied trend of decreasing ratio for decreasing \ssfr. The relation is more complex for highly star-forming low-metallicity galaxies and depends on the chemical evolutionary stage of the source (i.e., gas-to-dust mass ratio). Dust growth processes in the ISM play a key role in the dust mass build-up with respect to the stellar content at high \ssfr\ and low metallicity.
}
{We conclude that the evolution of the dust properties from metal-poor to metal-rich galaxies derives from a complex interplay between star-formation activity, stellar mass and metallicity.}
  

     \authorrunning{A. R\'emy-Ruyer et al.}
     \titlerunning{Linking dust emission to fundamental properties in galaxies}

 \maketitle


\section{Introduction}\label{sec:intro}

The processes by which galaxies evolve from primordial environments to present-day galaxies are still widely debated, but the seeds of this evolution lie in the star-formation histories of the galaxies, and in their interaction with their environment through gas infall, outflows or mergers. The interstellar medium (ISM) plays a key role in this evolution, being the site of stellar birth and the repository of stellar ejecta. 

Although interstellar dust represents only $\sim$ 1\%  of the total mass of the ISM it is an important agent in star formation. Dust absorbs the stellar radiation that would otherwise dissociate molecules and thus participates actively in the cooling of the ISM. It is also a catalyser for molecular gas formation by providing a surface where atoms can react \citep{Hasegawa1993, Vidali2004, LeBourlot2012, Bron2014}. The presence of dust can increase the H$_2$ formation rate by about two orders of magnitudes compared to H$_2$ formation without dust \citep{Tielens2005}, thus facilitating star formation.

Dust forms from the available heavy elements in the explosively ejected material from core-collapse supernov\ae\ (SN) and in the quiescent outflows from low-mass stars {\rev \citep{Todini2001, Gomez2012, Gomez2012b, Indebetouw2014, Rowlands2014a, Matsuura2015}}. The refractory dust grains may, after their injection into the ISM, grow by accretion or coagulation in dense molecular clouds \citep{Bazell1990, Stepnik2001, Stepnik2003, Kohler2012, Kohler2015}, locking even more heavy elements in the solid phase of the ISM \citep{SavageSembach1996, Whittet2003}.
Through destructive processes \citep[such as erosion or sputtering, see][]{Jones1994, Jones1996, SerraDiaz-Cano2008, Bocchio2012, Bocchio2014}, elements are released again into the gas phase. Metallicity, defined as the mass fraction of heavy elements, or ``metals'', in the ISM, is thus a key parameter in studying the evolution of galaxies.

Understanding how dust properties evolve as a function of metal enrichment can provide important constraints for galaxy evolution studies. Dwarf galaxies in the local Universe are ideal targets for such a study as many of them have low metallicity and high star-formation activity. As such, they present star-formation properties and ISM conditions that are the closest analogues to those thought to be present in the primordial environments of the early Universe \citep{Madau1998}.

Previous studies already demonstrated that the dust properties in low-metallicity galaxies were notably different from that of metal-rich sources. Low-metallicity galaxies harbour warmer dust \citep[e.g.,][]{Helou1986, MelisseIsrael1994, Galliano2005, Rosenberg2006, Cannon2006}, lower polycyclic aromatic hydrocarbon (PAH) abundances \citep[e.g.,][]{Madden2000, Boselli2004, Engelbracht2005, Wu2006, OHalloran2006, Draine2007, Galliano2008, Gordon2008, Wu2011, Sandstrom2012}, higher gas-to-dust mass ratios \citep[e.g.,][]{Issa1990, LisenfeldFerrara1998, Hirashita2002, James2002, Draine2007, Engelbracht2008, Galliano2008, MunozMateos2009, Bendo2010b, Galametz2011, Magrini2011}. Submillimetre (submm) excess emission, presently not accounted for by dust models, is observed in numerous dwarf galaxies or low-mass spirals \citep[e.g.,][]{Galliano2003, Galliano2005, Dumke2004, Bendo2006, Galametz2009, Galametz2012, Zhu2009, Bot2010, Grossi2010, Dale2012, Ciesla2014, Grossi2015, Galametz2014, Gordon2014}. Low-mass galaxies also show a broadening of the infrared (IR) peak of the SED \citep[e.g.,][]{Boselli2012, SmithD2012, Ciesla2014}, and a flattening of the far-infrared (FIR) slope \citep[e.g.,][]{Boselli2010b, Cortese2014a}.
However, all of these studies do not extend to very low-metallicities and/or have a limited number of sources below $\sim$1/5~\zsun\footnote{Throughout the paper, we assume (O/H)$_\odot$= 4.90$\times$10$^4$, i.e., 12+log(O/H)$_\odot$=~8.69 \citep{Asplund2009}} (12+log(O/H) = 8.0).

In \cite{RemyRuyer2013}, we derived the dust properties in a systematic way for a large number of galaxies (109), among which more than half are dwarf galaxies, and $\sim$ 35 \% have Z $\leq$ 1/5 \zsun. We confirmed on this significant sample of low-metallicity sources that the dust is warmer at low metallicities and identified several galaxies with submm excess. This study was made using only \hersc\ data and a modified blackbody to model the dust emission. In \cite{RemyRuyer2014}, we confirmed the higher gas-to-dust mass ratios (G/D) at low metallicities using the sample of \cite{RemyRuyer2013}, and a semi-empirical spectral energy distribution (SED) model over the whole IR range. We showed that the G/D is actually higher at low metallicities than that expected from a simple description of the dust evolution in the ISM. The large scatter in the observed G/D is intrinsic to the galaxies and reflects a non-universal dust-to-metal mass ratio \citep[see also][]{Dwek1998, DeCia2013, Zafar2013, Mattsson2014a}. We showed that the metallicity was the main driver of the G/D but that the scatter was controlled by the different star-formation histories of the sources. Thus metallicity is not the only parameter shaping the dust properties.

As a follow-up of \cite{RemyRuyer2013, RemyRuyer2014}, we aim at providing a consistent picture of the evolution of the dust properties from metal-poor to metal-rich galaxies. 
{\rev On the observational side, we present the complete catalog of IR-to-submm flux densities for the Dwarf Galaxy Survey (DGS).}
We use a semi-empirical dust SED model that accounts for starlight intensity mixing in the ISM, to interpret the whole IR-to-submm observed SEDs.
In this work, we extend the range of dust properties and look at the dust mass, the total infrared (TIR) luminosity, the PAH mass fraction and the dust temperature distribution. The dust temperature distribution is directly linked to the SED shape and will provide valuable insight on the average dust temperature, but also on the broadness of the FIR peak of the SED. 
We also discuss the evolution of dust-to-stellar mass ratios and link it to the evolution of the G/D observed in \cite{RemyRuyer2014}. For the analysis we do not consider only metallicity as our main parameter, but also include the specific star-formation rate and the stellar mass. 

The paper is organised as follows: in Section \ref{sec:data}, we present the sample used for the study and the whole IR-to-submm dataset. Section \ref{sec:model} presents the dust model, together with the free parameters (Sects. \ref{ssec:descmodel} and \ref{ssec:modelparam}), and errors on the best-fit parameters (Sect. \ref{ssec:errors}). In this section we also address how the use of a realistic dust model (Sect. \ref{ssec:compBB}), {\rev or a different interstellar radiation field (ISRF, Sect. \ref{ssec:isrf})}, or a different dust composition (Sect. \ref{ssec:amcar}), or a different wavelength coverage at 500 \mic\ (Sect. \ref{ssec:submm}) impacts the estimated dust properties (especially the dust mass). Section \ref{sec:dustprop} is dedicated to the derived dust properties and their variation with metallicity and star-formation activity. In Section \ref{sec:compview}, we discuss our results on the dust-to-stellar mass ratio in the context of galaxy evolution.

The paper also contains several appendices\footnote{{\rev online version only}} in order to facilitate the reading of the main body of the paper. Interested readers can find more details there. Appendix \ref{app:updateHersc} presents how we updated \hersc\ photometry since the studies of \cite{Dale2012} and \cite{RemyRuyer2013}. Observing logs are given in Appendix \ref{ap:ObsLogs} for the DGS IRAC and IRS \spitz\ data. Appendix \ref{app:compMIR} compares IRAC and WISE photometry for the DGS sources to other measurements in the literature. Appendix \ref{ap:IRSdata} explains in detail the data reduction and preparation of the IRS spectra. {\rev Special cases of SED modelling are discussed in Appendix \ref{ap:default}}. 

\section{Samples and Observations}\label{sec:data}

In this section we present the sample of galaxies and the set of IR-to-submm data used to build their observed dust SEDs. 
We also derive star-formation rates (SFR) and specific star-formation rates (\ssfr) to quantify the star-formation activity.

\subsection{Samples}\label{ssec:samples}

\begin{figure}[h!tbp]
\begin{center}
\includegraphics[width=8.8cm]{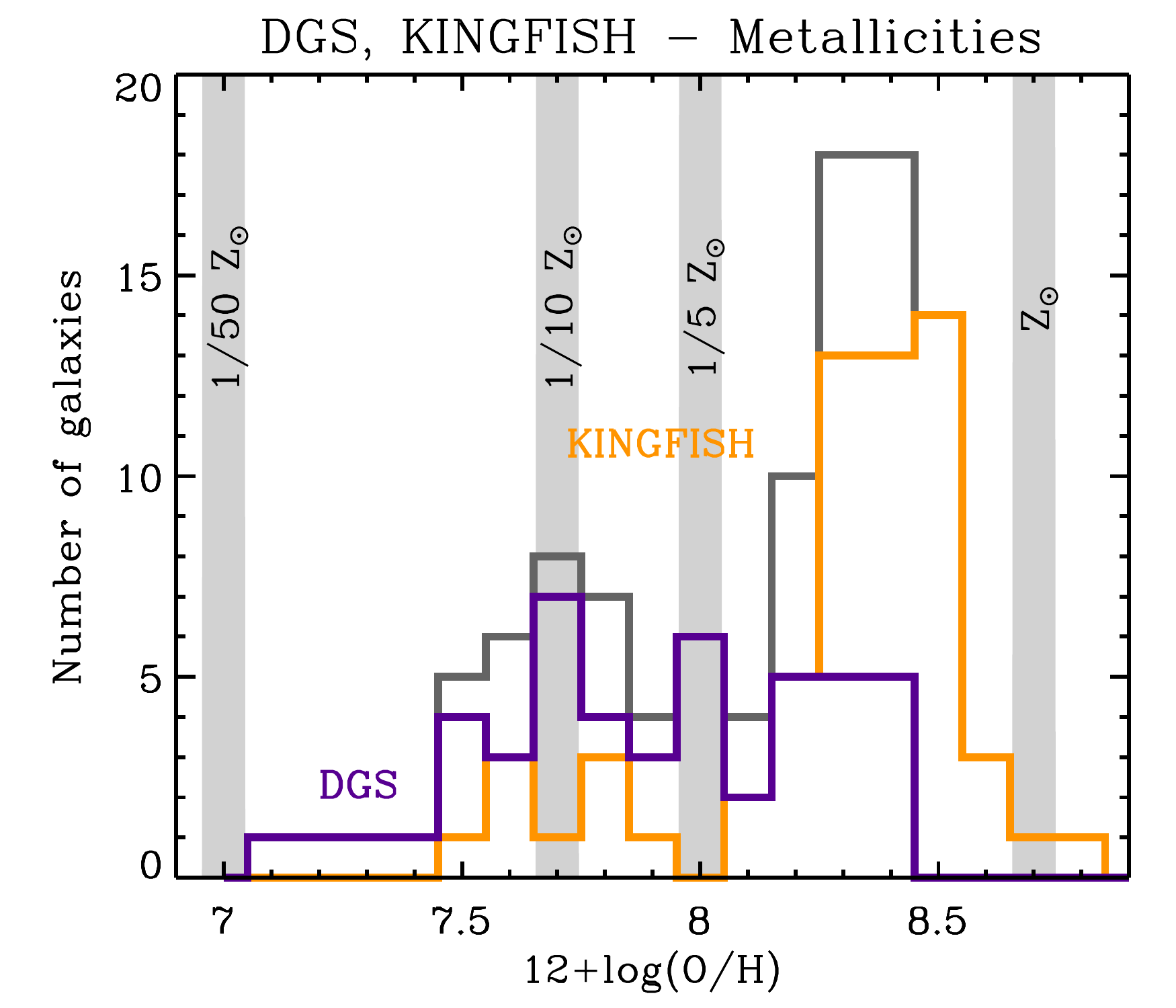}
\includegraphics[width=8.8cm]{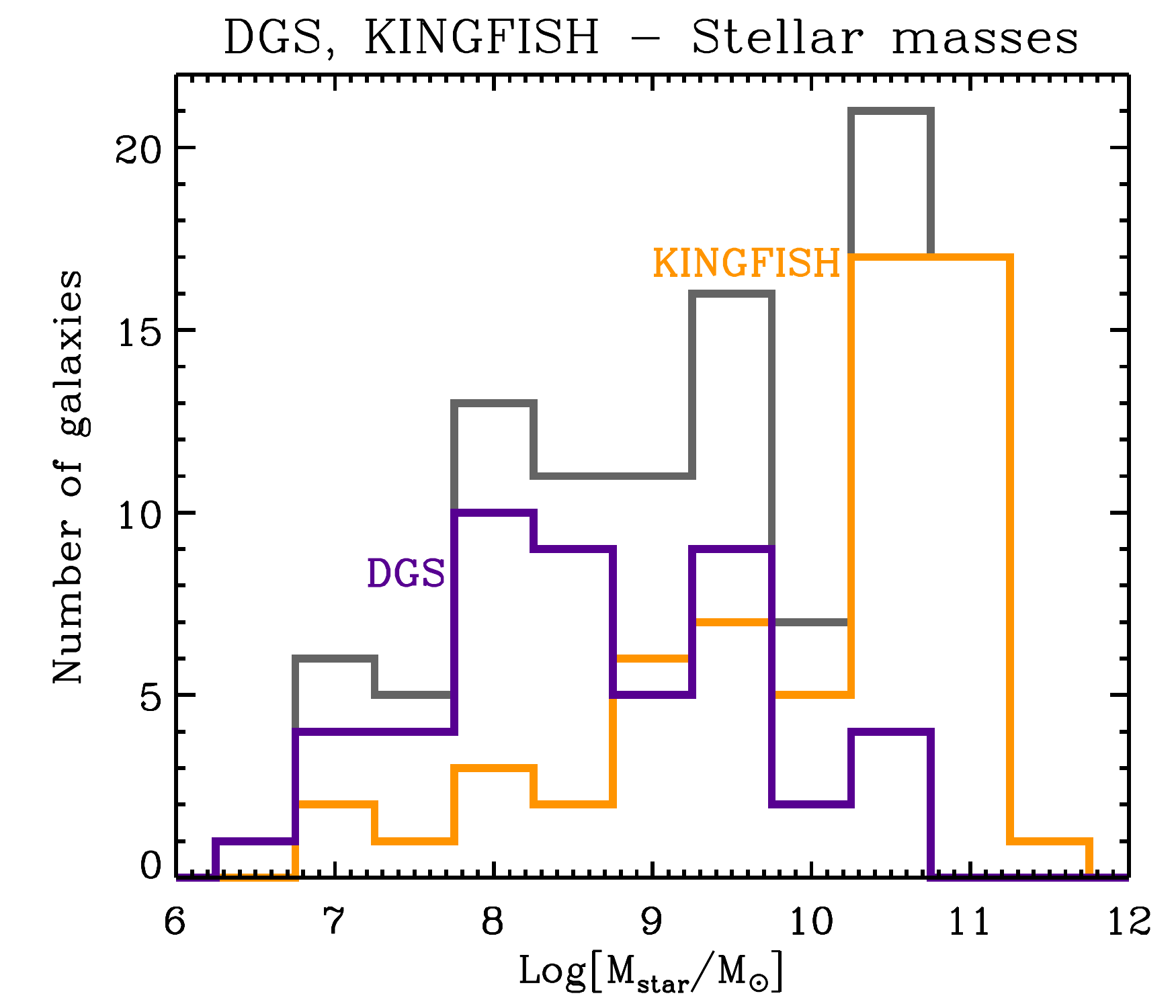}
\caption{Metallicity {\it (top)}, and stellar masses {\it (bottom)} distributions of the DGS (purple), KINGFISH (orange) samples. The total distribution is indicated in grey. In the top panel, solar metallicity is indicated here to guide the eye, as well as the 1/50, 1/10 and 1/5 \zsun values.}
\label{f:histos}
\end{center}
\end{figure}

For this study, we combine two samples of local galaxies observed with \hersc: the Dwarf Galaxy Survey \citep[DGS,][]{Madden2013} and the Key Insights on Nearby Galaxies: a Far-Infrared Survey with \hersc\ \citep[KINGFISH,][]{Kennicutt2011}. 
The DGS is a sample of 48 star-forming low-metallicity local dwarf galaxies dedicated to the study of the dust and the gas properties in low-metallicity environments. 
The DGS sample includes objects among the most metal-poor galaxies in the local Universe, with metallicities ranging from Z $\sim$ 0.03 \zsun\ to 0.55 \zsun\ (Fig. \ref{f:histos}). 
Stellar masses range over 4 orders of magnitudes, from 3$\times$10$^6$ \msun\ to 3$\times$10$^{10}$~\msun\ \citep{Madden2014}. 

The KINGFISH sample has been built from the \spitz\ Infrared Nearby Galaxies Survey \citep[SINGS,][]{Kennicutt2003}, and enables us to span a wider metallicity range by including more metal-rich galaxies. KINGFISH contains 61 galaxies, including mostly spiral galaxies together with several early-type and dwarf galaxies. The metallicities in the KINGFISH sample range from Z $\sim$ 0.07 \zsun\ to 1.20 \zsun. The two samples complement each other very well in terms of metallicities and stellar masses (Fig. \ref{f:histos}).  

The KINGFISH sources have been classified in terms of ``star-forming" or ``Active Galactic Nuclei" (AGN) type by \cite{Moustakas2010} using an optical emission-line diagnostic diagram \citep{Baldwin1981}. According to \cite{Kennicutt2011}, the detected AGNs are low-luminosity AGNs and none of the galaxies have a dominant AGN, except for NGC~1316. In terms of mid-infrared (MIR) emission line diagnostics, NGC~4725 has a [O~{\sc iv}]/[Ne~{\sc ii}] ratio of $\sim$ 1 \citep{Dale2006} and can be considered to be an AGN-dominated source. NGC~4736 has [O~{\sc iv}]/[Ne~{\sc ii}]~$\sim$~0.3 and is an ``intermediate" object with a AGN contribution of $\sim$~50\%. The other KINGFISH AGN sources have [O~{\sc iv}]/[Ne~{\sc ii}]~$\lesssim$~0.1 or no detection of the [O~{\sc iv}] 25.9 \micÊ\ line, equivalent to an AGN fraction $\lesssim$ 10 \% \citep{Genzel1998, Tommasin2010}. Except for these three galaxies, we thus do not expect the AGN to significantly impact the IR emission in the KINGFISH galaxies.

The metallicities in both samples have been determined {\rev in \cite{Madden2013} and \cite{Kennicutt2011}} using the ``strong-line'' calibration from \cite{PilyuginThuan2005}, linking a specific oxygen optical lines ratio\footnote{The R$_{23}$ ratio, R$_{23}$=([OII]$\lambda$3727+[OIII]$\lambda\lambda$4959,5007)/H$\beta$.} to the oxygen abundance in terms of 12+log(O/H). {\rev We choose this specific metallicity calibration because it is available and homogeneous for both samples, in order to avoid any biases by using different calibrations.
The stellar masses were derived by \cite{Madden2014} for the DGS using the prescription of \cite{Eskew2012} and the IRAC 3.6 \mic\ and 4.5 \mic\ flux densities. We use this prescription for the KINGFISH sample for consistency purposes and to reduce the scatter in the dust-to-stellar mass ratios \citep[as seen in][]{RemyRuyer2014b}.}
When no IRAC observed flux densities are available we use synthetic flux densities computed from the best-fit SED model (see Section \ref{sec:model}).
Coordinates, distances, and metallicities for the DGS can be found in \cite{Madden2013}, and in \cite{Kennicutt2003, Kennicutt2011} for the  KINGFISH sample. {\rev Stellar masses are tabulated in Table \ref{t:sfr}.} We have a total of 109 galaxies for this study.

\subsection{\hersc\ data}\label{ssec:hersc}
The 109 DGS and KINGFISH galaxies have been observed with \hersc\ in 6 photometric bands at 70, 100 and 160 \mic\ with the Photodetector Array Camera and Spectrometer \citep[PACS,][]{Poglitsch2010}, and at 250, 350 and 500 \mic\ with the Spectral and Photometric Imaging REceiver \citep[SPIRE,][]{Griffin2010}. 

The data reduction and photometry for the DGS and KINGFISH galaxies are presented in \cite{RemyRuyer2013} and \cite{Dale2012}, respectively. However, the data reduction techniques and the calibration of the instruments have been updated since the publication of \cite{RemyRuyer2013} and \cite{Dale2012}. In this work we therefore present the latest updated data. For the DGS data, the {\sc Scanamorphos} PACS maps are reprocessed with version 23 of {\sc Scanamorphos} \citep{Roussel2013}, and the SPIRE maps are reprocessed with version 12 of the \hersc\ Processing Interactive Environment \citep[HIPE -][]{Ott2010}. 
We find an average difference of 5\%, 8\% and 2 \% at 70, 100 and 160 \mic\ between the updated flux densities and the flux densities of \cite{RemyRuyer2013}, and in some extreme cases the difference is as high as 30\%. 
For the SPIRE flux densities, the average difference is 3\%, 5\% and 3\% at 250, 350, and 500 \mic\ respectively. The updated PACS and SPIRE flux densities for the DGS galaxies are presented in Table \ref{t:herschel}. 

For the KINGFISH data, we update the flux densities presented in \cite{Dale2012}, with an average difference with the \cite{Dale2012} flux densities of 7\% at 160 \mic, and $\sim$ 2\% for SPIRE wavelengths. Detailed information regarding the update of the DGS and KINGFISH \hersc\ data can be found in Appendix \ref{app:updateHersc}.
Colour corrections are applied to the SPIRE flux densities in both samples to account for the dependence of the SPIRE beam areas on the FIR spectral shape of the sources (see Appendix \ref{app:updateHersc}).

\subsection{\spitz\ data} 

We collect \spitz\ observations for our sample in order to constrain the warm dust and the PAH emission in MIR wavelengths, with photometric data from the InfraRed Array Camera (IRAC), and from the Multiband Imaging Photometer for \spitz\ (MIPS), and spectroscopic data from the InfraRed Spectrograph (IRS).
\spitz\ observations of several DGS sources were already available in the \spitz\ database\footnote{The query form is available at: http://sha.ipac.caltech.edu.}. {\rev New} complementary \spitz\ observations {\rev for 19 DGS galaxies} have also been obtained during the cycle 5 program to complete the set of existing \spitz\ data (Dust Evolution in Low-Metallicity Environments: P.I. F. Galliano; ID: 50550), {\rev and for which only the MIPS data at 24, 70 and 160 \mic\ have been published so far \citep[in][]{Bendo2012}}. Thus we only {\rev present} the IRAC and IRS observations for the DGS sample (Observing Logs in Appendix \ref{ap:ObsLogs}). 

\spitz\ IRAC and MIPS photometry for the KINGFISH galaxies is taken from \cite{Dale2007}. 
We do not attempt to use the IRS data existing for the KINGFISH galaxies, as the galaxies are very extended and a complete coverage of the sources with IRS is not available \citep{Dale2009b}.

\subsubsection{IRAC} 

IRAC data at 3.4, 4.5, 5.8 and 8.0 \mic\ is available for the DGS galaxies, and the IRAC maps are retrieved from the NASA/IPAC ISA database for \spitz\ data. For three galaxies, HS~1442+4250, SBS~1415+437 and UM~311, only two wavelengths are available (4.5 and 8.0 \mic).
After subtracting the background and possible contamination from background sources or foreground stars, flux densities are extracted from the maps using aperture photometry. 
In most cases, we use the same apertures as those used for \hersc\ photometry. The final IRAC flux densities and the apertures used are given in Table \ref{t:irac}.

The IRAC calibration is based on point-source photometry for a 12\arcsec\ radius aperture. An additional aperture correction is needed to account for the emission from the wings of the PSF and the scattering of diffuse emission across the IRAC focal plane. This correction is given in the IRAC Instrument Handbook (version 2.0.1 Section 4.11.1\footnote{This document is available at http://irsa.ipac.caltech.edu/data/\\ SPITZER/docs/irac/iracinstrumenthandbook.}) and depends on the source aperture radius. Following the recipe given in the IRAC Instrument Handbook, we do not apply this correction to small and compact sources (see Table \ref{t:irac}).

The uncertainty on the flux density is computed by summing in quadrature the calibration error, the error from the background determination and the error from the source flux determination. We adopt a calibration error for the four IRAC bands of 10\% as recommended in the IRAC Instrument Handbook\footnote{This can be found in Section 4.3 of the IRAC Instrument Handbook.}.
We consider that galaxies are not detected when the computed flux density is lower than three times its corresponding uncertainty in a given band. 
The final 3$\sigma$ upper limit is reported in Table \ref{t:irac}.

IRAC photometry is available in the literature for 29 DGS sources: \cite{Hunt2006, Dale2007, Dale2009, Engelbracht2008, Galametz2009}. If we compare our measurements to those from the literature we get a fairly good agreement between them, with some outliers. The details of the comparison with the literature measurements and possible explanations for the outliers are given in Appendix \ref{app:compMIR}.

\subsubsection{IRS}\label{sssec:irs}

The DGS galaxies have a very rapidly rising continuum in the MIR, and the IRS spectra gives important constraints on the continuum from 20 to 40 \mic. Moreover, the spectra can also be used to put constraints on the PAH emission (see Section \ref{ssec:modelparam}). Line intensities have been extracted by \cite{Cormier2015} for the [S{\sc IV}] 10.5 \mic, [Ne{\sc II}] 12.8 \mic, [Ne{\sc III}] 15.6 \mic, [S{\sc III}] 18.7 \mic\ and 33.5 \mic\ spectral lines, in the compact sources of the DGS.

The IRS spectra for the DGS galaxies have been extracted from the Cornell AtlaS of \spitz\ Infrared spectrograph Sources\footnote{The database is available at http://cassis.astro.cornell.edu/atlas/.} \citep[CASSIS v5,][]{Lebouteiller2011}. 
Most of the DGS galaxies were observed using the staring mode, except for a few extended sources observed with the mapping mode, for which the spectra were reduced manually with {\sc CUBISM} \citep{Smith2007b}. We were able to obtain Short-Low (i.e., short wavelength data at low spectral resolution, SL) and Long-Low (i.e., long wavelength data at low spectral resolution, LL) data for 43 galaxies in total. More details about the extraction and the data reduction of the IRS spectra can be found in Appendix \ref{ap:IRSdata}. 

It is then necessary to rescale the SL and LL spectra in order to match the photometry. We derive synthetic IRS photometry to correct the two modules and use all of the constraints we have in this wavelength range: the IRAC 5.8 and 8.0 \mic, \wise\ 12 and 22 \mic\ (see Section \ref{ssec:wise}) and MIPS 24 \mic\ bands. IRAC 5.8, 8.0 and \wise\ 12 \mic\ are used simultaneously to derive a correction factor for the SL module that depends on the wavelength.  
However, for the LL module, the two constraints do not sample well the LL spectrum, 
and the LL correction factor is thus a constant. We use MIPS 24 \mic\ in most cases (or \wise\ 22 \mic\ when not available).
SL and LL can be treated separately as they are two independent observations. More details about the rescaling of the IRS spectrum can be found in Appendix \ref{ap:IRSdata}. The median correction applied at 5.8 \mic, 8.0 \mic, 12 \mic, and 24 \mic\ are 1.7, 2.0, 1.0 and 1.2 respectively. 
Note that this correction assumes that the spectral shape of the area observed by the IRS slits likewise describes the expected spectral shape of the full galaxy. This is true for compact sources but can be erroneous for more extended galaxies, except if the region falling within the IRS slits dominates the total emission of the galaxy in the MIR. 

To improve the quality of the noisiest spectra (e.g. HS~1304+3529, Fig \ref{f:IRSex}, top panel), we smooth the spectra until we reach a signal-to-noise ratio (S/N) of 3 at every wavelength. 
This smoothing step is applied for 22 DGS galaxies. However, for three galaxies, HS~1236+3937, HS~1442+4250, Tol~0618-402, even with the smoothing step, we cannot reach a S/N of 3 for any point in the spectrum. Thus we do not consider these spectra in the SED modelling. In two galaxies, HS~2352+2733 and UGCA~20, the IRS slits are not centred on the source position, thus we do not present these spectra either. The remaining 38 IRS spectra are shown in Appendix \ref{ap:IRSdata}, Fig. \ref{f:IRSall}.

\subsection{WISE data}\label{ssec:wise}

The Wide-field Infrared Survey Explorer \citep[\wise,][]{Wright2010} observed the whole sky at 3.4, 4.6, 12 and 22 \mic, and gives an additional constraint at 12 \mic, very valuable for probing the MIR range of the observed SED. The InfraRed Astronomical Satellite \citep[\iras,][]{Neugebauer1984} also provides a constraint at 12 \mic\ (see Section \ref{ssec:allsky}) but was not able to detect the faintest dwarf galaxies. We also need the 12 and 22 \micÊ\ constraints to match the IRS spectra to the photometry (Section \ref{sssec:irs}). WISE flux densities at 3.4 and 4.6 \mic \ are also given for completeness. We present here WISE data for the DGS {\rev galaxies only as \spitz\ and \irasÊ\ already provide the equivalent spectral coverage for the brighter KINGFISH galaxies}.

The WISE maps were retrieved from the NASA/IPAC ISA database and the AllWISE database\footnote{The query form is available at: http://sha.ipac.caltech.edu.}. Most sources in the DGS are resolved by WISE as the resolution is $\sim$ 6\arcsecÊ\ for the first three bands and 12\arcsecÊ\ for the WISE 22 \micÊ\ band. Profile-fit and standard aperture photometry provided by the WISE database can underestimate the brightness of resolved sources and can suffer from confusion with nearby objects (Section 2.2 of the AllWISE explanatory supplement\footnote{This document is available at http://wise2.ipac.caltech.edu/docs/\\ release/allwise/expsup/index.html.}). 

We {\rev therefore} perform {\rev our own} aperture photometry on the maps following the method outlined in Section 4.3 of the All WISE explanatory supplement. The apertures are the same as for IRAC and \hersc\ photometry whenever possible. In some cases, the morphology of the source in the first two bands is quite different from the morphology of the source in the last two bands. These two sets of wavelengths are not tracing the same physical component, old stellar population at 3.4 \mic\ and 4.6 \mic\ on one side, warm dust at 12 \micÊ\ and 22 \mic\ on the other side, and the near-infrared (NIR) morphology may differ from the MIR morphology. In this case we have to use two different apertures if we want to encompass the whole emission of the source and optimise the S/N ratio at all four wavelengths. For the unresolved sources, we use the profile-fit photometry provided by the All WISE database. 

The uncertainty comes from the flux determination, from the background estimation, from correlated noise in the maps, and from the calibration and measurement of the zero-point magnitudes (see the AllWISE explanatory supplement). The calibration of WISE have been tied to that of \spitz\ IRAC 3.6 \mic\ for WISE1, IRAC 4.5 \mic\ for WISE2, IRS for WISE3, and MIPS 24 \micÊ\ for WISE4 \citep[see][]{Jarrett2011}, thus we adopt a calibration error of 10 \% at 3.4 \mic\ and 4.6 \mic, 5\% at 12 \mic, and 7\% at 22 \mic. 
As for IRAC photometry, we report upper limits if the detection is below 3$\sigma$. The final WISE flux densities and apertures used are given in Table \ref{t:wise}.

WISE measurements compare well with other instruments, IRAC at 3.6 \micÊ\ and 4.5 \mic, \iras \ at 12 \mic, and MIPS at 24 \mic. The details of the comparison are given in Appendix \ref{app:compMIR}.

\subsection{All-sky survey data from 2MASS and IRAS}\label{ssec:allsky}

We complete our set of \spitz\, WISE and \hersc\ data by searching the literature for data from 2MASS \citep[Two Microns All Sky Survey,][]{Skrutskie2006} and \iras.

The 2MASS data in the J (1.24 \mic), H (1.66 \mic) and K$_s$ (2.16 \mic) bands, for the DGS sample is given in Table \ref{t:jhkiras}, and has been compiled from the literature: the NASA/IPAC ISA 2MASS Point Source Catalog\footnote{The catalog is available at: http://irsa.ipac.caltech.edu.
}, the 2MASS Extended Objects Final Release, the 2MASS Large Galaxy Atlas \citep{Jarrett2003}, \cite{Engelbracht2008, Dale2009}. In some cases, the original data is given in magnitudes. To convert this into flux densities, we use the zero-magnitude flux values from \cite{Cohen2003}. 2MASS data for the KINGFISH galaxies is presented in \cite{Dale2007}. 

We also compile \iras\ data for the DGS at 12, 25, 60 and 100 \mic\ from the literature: the NASA/IPAC ISA IRAS Faint Source (v2.0 1990)$^9$ 
and Point Source (v2.1) Catalogs$^9$, 
\cite{Rice1988, Sanders2003, Engelbracht2008} and is given in Table \ref{t:jhkiras}. Note that \iras\ is not very sensitive for the low-luminosity dwarf galaxies and that the resolution varies from $\sim$ 30\arcsec\ at 12 \mic\ to $\sim$ 120\arcsec at 100 \mic. Even for the detected sources, such low resolution implies that several sources may be mixed in the beam and indistinguishable. That is why some of the \iras\ fluxes of Table \ref{t:jhkiras} have been noted as unreliable (with the note $^g$). 25 DGS galaxies have \iras\ measurements and only 15 are detected in all of the IRAS bands. KINGFISH \iras\ data have been extracted from \cite{Sanders2003} and the IRAS Faint Source (v2.0 1990) and Point Source (v2.1) Catalogs.

\subsection{Star formation activity}\label{ssec:sfa}

\subsubsection{Star-formation rates}\label{sssec:sfr}

To estimate the SFR for our sources, we consider two of the most widely used SFR tracers, the far-ultraviolet (FUV) and the H$\alpha$ luminosities. The observed stellar emission traced either by FUV or H$\alpha$ has to be corrected for dust attenuation, and this correction is usually made using observational dust tracers such as the TIR or 24 \mic\ luminosities \citep{Kennicutt2009, Calzetti2010, Hao2011, Kennicutt2012}. 

In our sample, we {\rev can derive SFR} for 90\% of the galaxies using H$\alpha$+TIR, 88\% using H$\alpha$+24 \mic, 77\% using FUV+TIR and 78\% using FUV+24 \mic. {\rev We thus use the diagnostic based on the observed H$\alpha$ luminosity, corrected for attenuation using \ltir}, and the formula by \cite{Kennicutt2009} for a Kroupa initial mass function (IMF). Moreover, \cite{Kennicutt2009} showed for the \spitz-SINGS sample that this composite tracer gave the most robust SFR measurements compared to H$\alpha$ measurements corrected for dust attenuation using the Balmer decrement. 

The {\rev integrated} H$\alpha$ luminosities for the sample have been taken from the literature \citep[mostly from][]{Moustakas2006a, Kennicutt2008, Kennicutt2009}, and are listed in Table \ref{t:sfr}. Care has been taken in providing H$\alpha$ luminosities corrected for underlying stellar absorption, for N{\sc II} line contamination and foreground Galactic extinction. The same extinction curve was used to correct our sources for foreground Galactic extinction: A(H$\alpha$)=0.6 A$_B$ \citep{ODonnell1994}. The B-band extinction, A$_B$, is taken from the NASA/IPAC Extragalactic Database (NED\footnote{This database is available at http://ned.ipac.caltech.edu/.}) and the measurements of \cite{Schlafly2011}. {\rev For the five galaxies missing either H$\alpha$ or TIR data, we use the mean of the other SFR estimates.} The error on the SFR is derived by taking the standard deviation of all of the SFR estimates when more than one is available, {\rev otherwise we use the 20\% median error on the SFR in the sample (reported in Table \ref{t:SEDparam})}.

\begin{figure*}[h!tbp]
\begin{center}
\includegraphics[width=8.8cm]{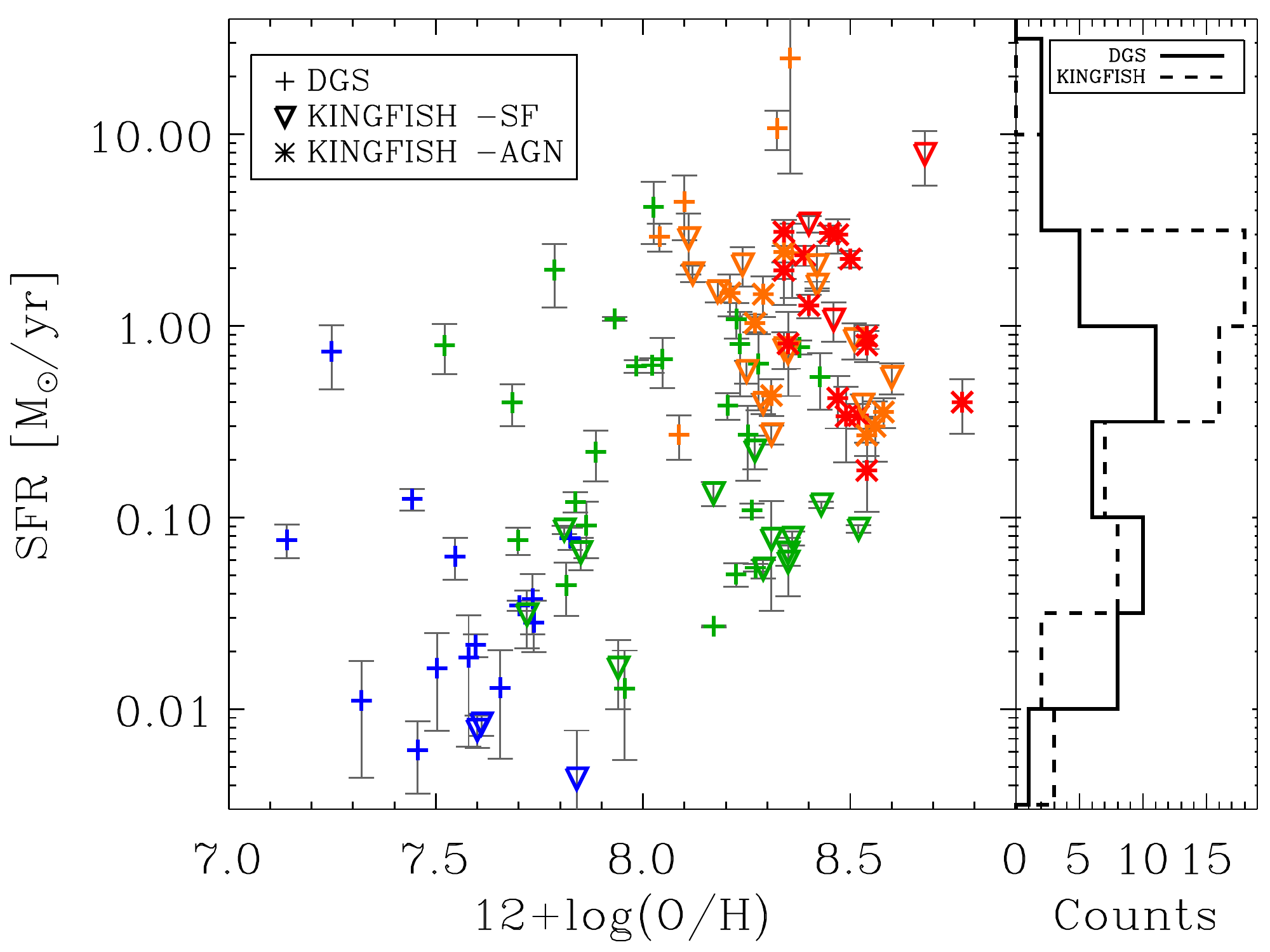}
\includegraphics[width=8.8cm]{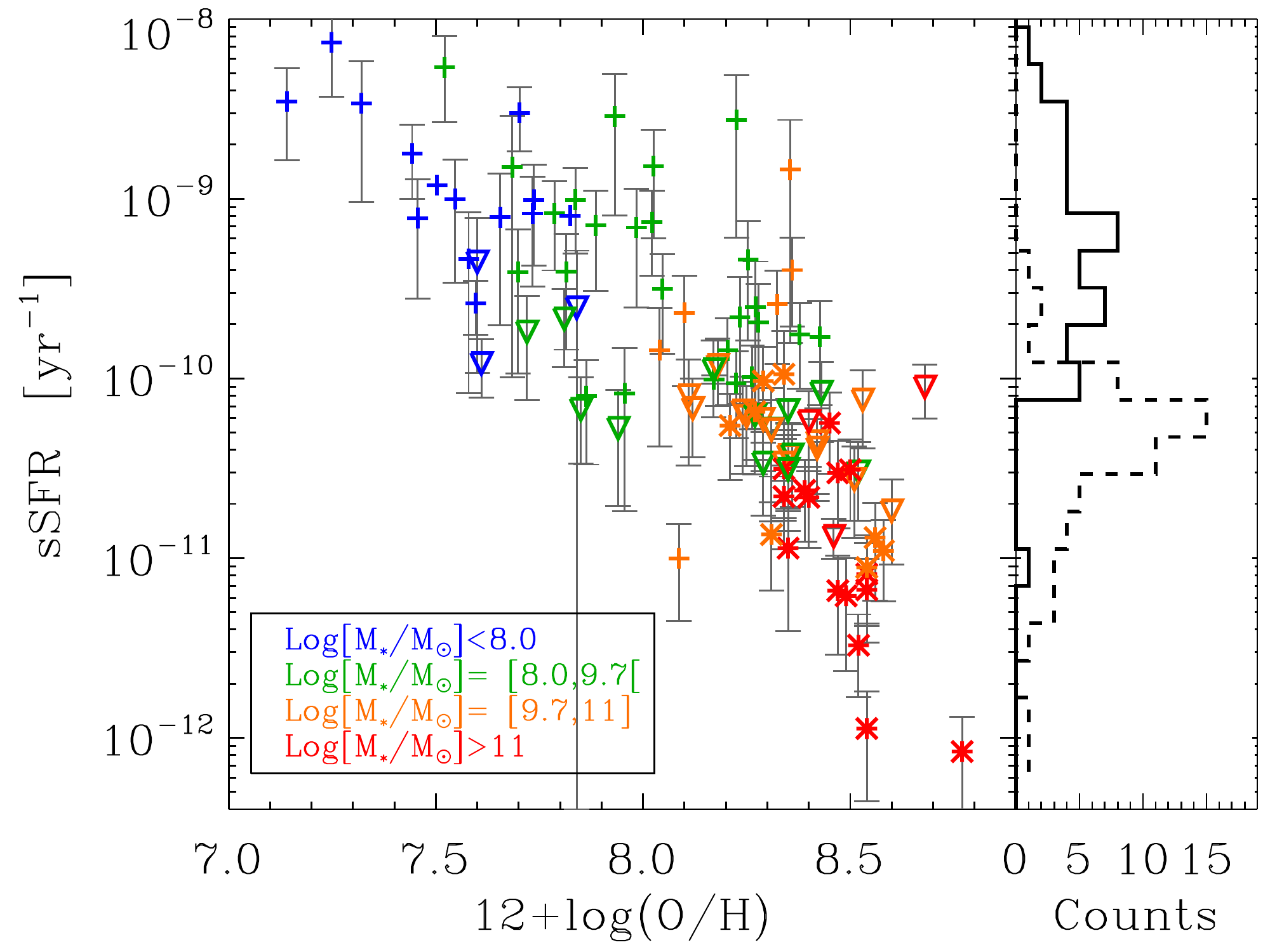}
\caption{Star-formation rates, SFR, {\it (left)}, specific star-formation rates, \ssfr, {\it (right)}, for the DGS (crosses) and KINGFISH samples as a function of metallicity, colour coded with \mstar\ (colour legend in right panel). For the KINGFISH sample, we distinguish the sources that have been classified as ``star-forming'' (SF, downward triangles) or as ``AGN'' (stars) by \cite{Kennicutt2011}.
The distribution of each parameter is indicated on the side of each panel for both samples: solid line for DGS and dashed line for KINGFISH.}
\label{f:ssfr}
\end{center}
\end{figure*}

\cite{Lee2009} cautioned the use of H$\alpha$ to derive SFR for galaxies with integrated L$_{H\alpha}$ $\leq$ 2.5 $\times$ 10$^{39}$ erg.s$^{-1}$ or equivalently SFR $\lesssim$ 0.01 \sfr. Ten galaxies in our sample are below these limits in L$_{H\alpha}$ or SFR. For these very faint galaxies, H$\alpha$-SFR will be underestimated, even after correction for dust attenuation {\rev \citep{Lee2009, Meurer2009}}. For these ten galaxies, we use the calibration provided by \cite{Lee2009} (specifically their equation 10), which converts the non-dust corrected H$\alpha$ luminosities into dust-corrected SFR, based on the \cite{Kennicutt1998} FUV-SFR. These obtained SFRs are from a calibration assuming a Salpeter IMF \citep{Kennicutt1998} and we thus convert them into Kroupa IMF-based SFRs using the correction factors tabulated in Table 1 of \cite{Kennicutt2012}. The error on the SFR for these ten galaxies is taken from Table 2 in \cite{Lee2009}. The resulting SFRs for these ten galaxies are a factor of 2 larger than the previously estimated SFRs from H$\alpha$ and TIR. The H$\alpha$ luminosities and final SFR are given in Table \ref{t:sfr}, and shown in Fig. \ref{f:ssfr}.

The SFRs in our sample cover four orders of magnitude, with a median SFR of 0.27~\sfr\ and 0.54~\sfr\ for the DGS and KINGFISH samples respectively (see Table \ref{t:SEDparam}). We see from Fig. \ref{f:ssfr} that at a given stellar mass, galaxies with the highest SFRs have the lowest metallicities. This is consistent with the findings of \cite{LaraLopez2010, Mannucci2010, Hunt2012} on the ``fundamental metallicity relation'' (FMR) describing the various scaling relations between metallicity, stellar mass and star-formation rate.

\subsubsection{Specific star-formation rate}\label{sssec:ssfr}

We also consider the specific star-formation rate, \ssfr, i.e., SFR normalised by the stellar mass, to remove scaling effects in our analysis of such a wide variety of environments. 
The \ssfr\ values are given in Table \ref{t:sfr} and plotted as a function of metallicity in Fig.~\ref{f:ssfr}.

The median \ssfr\ is one order of magnitude higher in the DGS than in KINGFISH (0.51 Gyr$^{-1}$ versus 0.04 Gyr$^{-1}$). In the KINGFISH sample, we distinguish the galaxies according to their nuclear types as defined by \cite{Kennicutt2011}: star-forming (SF) versus non-thermal emission (AGN). The KINGFISH ``AGN''  galaxies have the lowest \ssfr\ in Fig. \ref{f:ssfr}, with a median \ssfr\ of 0.01 Gyr$^{-1}$, versus 0.06 Gyr$^{-1}$ for the KINGFISH ``SF'' sample.  
The stellar masses were estimated on the IRAC 3.6 \mic\ and 4.5 \mic\ flux densities, which might not be appropriate for AGN-type galaxies, possibly overestimating \mstar\ and thus yielding underestimated \ssfr\ values.

The clear correlation between \ssfr\ and metallicity seen in Fig. \ref{f:ssfr} follows directly from the FMR. 
The \ssfrÊ\ can also be interpreted as the amount of star formation occurring today compared to the output of the cumulated past star-formation, i.e, the mass of stars already formed, tracing the integrated star-formation history. The \ssfr\ value can thus be boosted up by the low stellar mass of the dwarf galaxies even if the present star-formation rate is comparable to that in more metal-rich environments. 
This is consistent with the so-called ``downsizing'' effect, i.e.,  
more massive galaxies are more efficient in converting gas into stars than low-mass galaxies \citep[e.g.][]{Sandage1986, Cowie1996, Gavazzi1996, Boselli2001, Gavazzi2002, Brinchmann2004, Hughes2013}. The less massive galaxies form the bulk of their stellar mass later than more massive systems, and thus undergo a slower metal enrichment. This is consistent with our picture of low-mass, low-metallicity galaxies having a higher \ssfr\ today than their massive metal-rich counterparts.

\section{Dust modelling in the DGS and KINGFISH galaxies}\label{sec:model}

In this section, we describe the dust SED model, the parameters and the derivation of the errors on the best-fit parameters (Sects. \ref{ssec:descmodel}, \ref{ssec:modelparam}, \ref{ssec:errors}). In Sect. \ref{ssec:compBB}, we compare our results with the dust properties derived with a modified blackbody model (namely the dust mass and the temperature). In Sect. {\rev \ref{ssec:isrf}} and \ref{ssec:amcar}, we investigate the effect of a different {\rev description for the radiation field} and of a different dust composition on the dust properties, {\rev respectively}. And finally in Sect. \ref{ssec:submm} we show that the 500 \mic\ data point is a necessary constraint to accurately estimate the dust mass, and investigate the influence of a potential submm excess. 

\subsection{Description of the SED model}\label{ssec:descmodel}

We adopt the semi-empirical dust SED model presented in \cite{Galliano2011}.  The model adopts the Galactic grain composition made of silicate grains, carbon grains in the form of graphite, and PAHs. The optical properties  are taken from \cite{WeingartnerDraine2001, LaorDraine1993} and \cite{DraineLi2007} respectively. The assumed size distribution is that determined by \cite{Zubko2004} for their BARE-GR-S model, but we allow the relative normalisation of the PAH component to vary for more flexibility. {\rev The graphite-to-silicate ratio is kept fixed to the \cite{Zubko2004} value.}
We assume that the ISRF illuminating the dust grains has the spectral shape of the solar neighbourhood ISRF \citep{Mathis1983}. Only its intensity varies, controlled by the parameter $U$\footnote{{\rev $U$ is expressed relative to the solar neighbourhood value U$_\odot$~=~2.2 $\times$ 10$^{-5}$ W.m$^{-2}$}}. {\rev The temperature fluctuations of stochastically heated grains are computed using the transition matrix method \citep{GuhathakurtaDraine1989}.}

We assume that the dust properties are uniform, i.e., the size distribution and the grain optical properties are constant within the galaxy, and that only the starlight intensity varies. We distribute the mass into different mass-elements of uniform illumination with the empirical recommendation from \cite{Dale2001} : the distribution of dust mass per starlight intensity can be approximated by a power law of index $\alpha$, $dM/dU~\propto$~U$^{-\alpha}$, with U varying between \umin\ and \umin + \DU. This simply expresses that most of the dust mass should reside in the coldest components, with low starlight intensities $U$. In most cases, this is flexible enough to reproduce dense and diffuse media, and provides a simple parameterization of the physical conditions in the ISM.

Emission from old stars can also contribute to the IR emission and especially in the NIR. Thus we add a stellar continuum to the dust emission, parameterised by the stellar mass of the galaxy $M_{\star}$. This parameter is not designed to estimate the stellar mass, but rather to have a good fit in the NIR \citep[see][for details]{Galliano2011}. 

For each band at a wavelength $\lambda_i$ the synthetic luminosity 
is computed by convolving the model with the spectral response of each band and using the appropriate spectral convention of each instrument. This is done for all of the observational constraints except at SPIRE wavelengths, as the input flux densities are already colour-corrected (see Sect. \ref{ssec:hersc}). 

The fit is performed with the {\sc mpcurvefit} iterative procedure in {\sc IDL} based on the Levenberg-Marquardt method.
This model has been used to model galaxies before by \cite{Galametz2009,Galametz2010,Galametz2011,Galametz2013a, Hony2010, Meixner2010, OHalloran2010}.

\subsection{Setting the parameters}\label{ssec:modelparam}

The model can be described by {\rev six main} free parameters:

\begin{itemize}
\item \mdust: the total dust mass,
\item \umin: the minimum of the starlight intensity distribution,
\item \DU: the difference between the maximum and minimum of the starlight intensity distribution,
\item $\alpha$: the index of the power law describing the starlight intensity distribution,
\item \fpah: the PAH mass fraction, normalised to the Galactic PAH mass fraction\footnote{The Galactic PAH mass fraction is \fpah$_\odot$=4.57\% from \cite{Zubko2004}.}, and 
\item $M_\star$: the stellar mass.
\end{itemize}

{\rev Units and bounds used for the fitting are described in \cite{Galliano2011}. In addition, we vary the mass fraction of very small grains (non-PAH grains with radius {\it a}~$\leq$~10 nm), \fvsg, relative to the Galactic value\footnote{We compute the Galactic mass fraction of non-PAH grains with radius~$\leq$~10 nm in \cite{Zubko2004} model: \fvsg$_\odot$ = 16\%.}. It is constrained by the 8 - 30 \mic\ photometry, plus IRS spectroscopy for the DGS sources, and controls the MIR continuum. We also vary the ionised-to-total PAH mass ratio in the DGS sources, f$_{{\rm ion}}$, as we have the IRS spectra to constrain it. These two parameters are used in order to allow for a more flexible and a better fit in the MIR, especially to reproduce the steeply rising continuum of the dwarf galaxies, but will not be discussed any further.
} 

From the final best-fit models, we derive the TIR luminosity, \ltir, by integrating the total dust SED between 1 and 1000 \mic\ for each galaxy.
We compute as well the first and second moments of the starlight intensity distribution, \uav\ and (\sigU)$^2$, corresponding to the mass-averaged starlight intensity and the variance in the starlight intensity distribution respectively. \uav\ and (\sigU)$^2$ are given by \citep[see Eqs. 9 to 12 from][]{Galliano2011}: 

\begin{equation}
\langle U\rangle =\frac{1}{M_{{\rm dust}}} {\int_{U_{{\rm min}}}^{U_{{\rm min}}+\Delta U} U \times \frac{\rm dM_{{\rm dust}}}{\rm dU} \, \mathrm dU}
\label{eq:U}
\end{equation}

\begin{equation}
(\sigma U)^2 =\frac{1}{M_{{\rm dust}}} {\int_{U_{{\rm min}}}^{U_{{\rm min}}+\Delta U} (U-\langle U\rangle)^2 \times \frac{\rm dM_{{\rm dust}}}{\rm dU} \, \mathrm dU}
\label{eq:sigU}
\end{equation}

\uav\ can be seen as an indirect measure of the average dust temperature and \sigUÊ\ of the broadness of the FIR peak of the SED. 

\begin{figure}[h!tbp]
\begin{center}
\includegraphics[width=9.0cm]{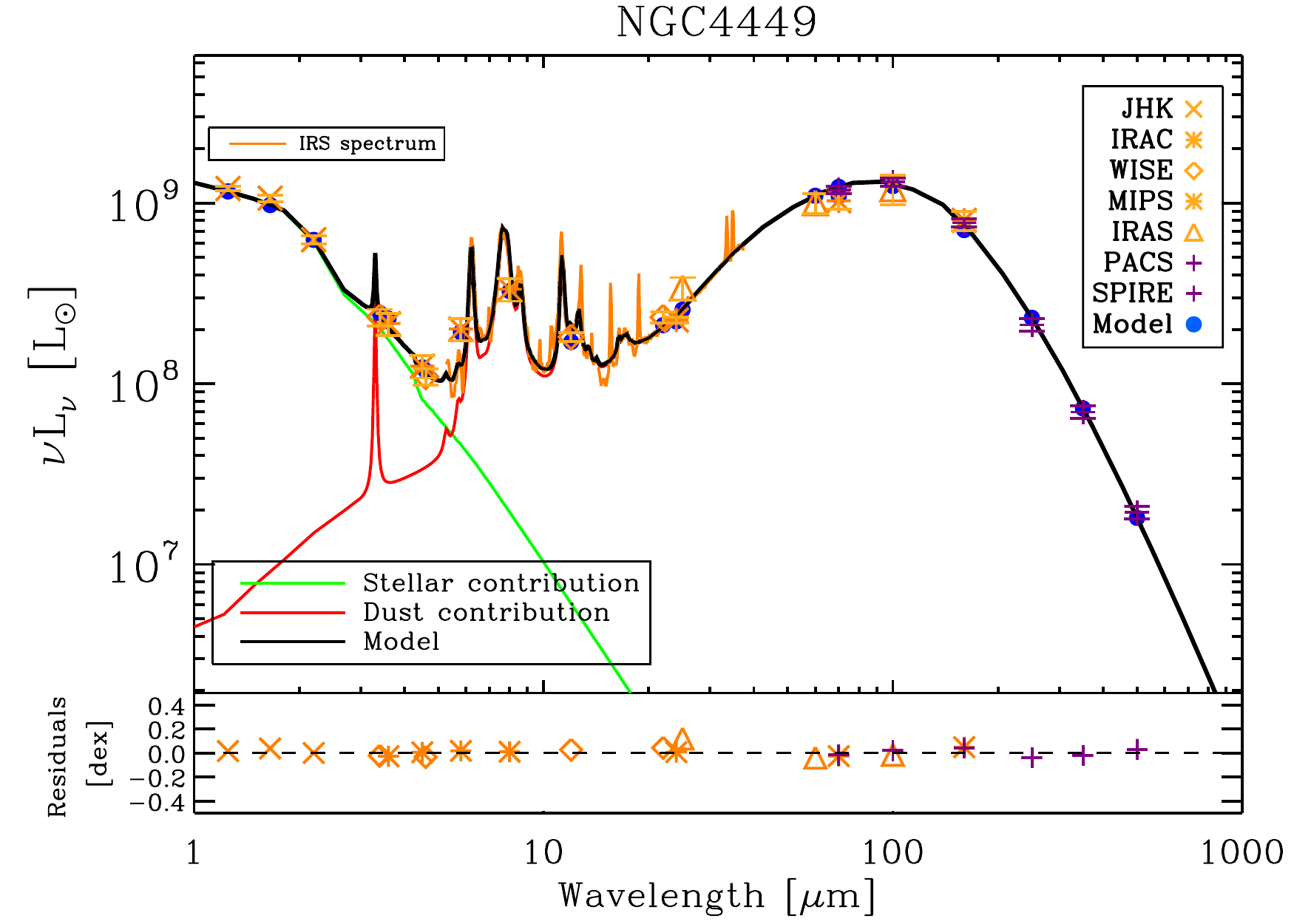}
\caption{SEDs for a DGS source NGC~4449. The observed SED includes the \hersc\ data (purple crosses) as well as any available ancillary data (in orange). The different symbols code for the different instruments: Xs for 2MASS bands, stars for \spitz\ IRAC and MIPS, diamonds for WISE and triangles for IRAS. The IRS spectrum is displayed in 
orange. The total modelled SED in black is the sum of the stellar (green) and dust (red) contributions. The modelled points in the different bands are the filled blue circles. Residuals are shown on the bottom panel.}
\label{f:exSED}
\end{center}
\end{figure}

This model is applied to the observed SEDs of DGS and KINGFISH galaxies. An example of fit is given in Fig. \ref{f:exSED}, the others are shown in Figs. \ref{f:SED_DGSall} and \ref{f:SED_KFall}. For four DGS galaxies, there are not enough constraints to properly fit a SED: HS~1236+3937, HS~2352+2733, Tol~0618-402 and UGCA~20.
We do not consider KINGFISH galaxies without \hersc\ detections either: NGC~0584, NGC~1404, DDO~154 and DDO~165. We also remove from the subsequent analysis the three KINGFISH galaxies for which the AGN emission is dominant: NGC~1316, NGC~4725 and NGC~4736 (see Section \ref{ssec:samples}). Indeed, \cite{Ciesla2015} showed that the emission of an AGN can significantly impact the total IR emission from contribution $\gtrsim$40\%. For the other KINGFISH ``AGN" sources, we do not expect the total IR emission to be affected by the AGN contribution on global galaxy scales, and due to the difficulty of accurately constraining such low AGN fractions \citep{Ciesla2015} we do not apply any specific modelling. 
We have a total of 98 galaxies to which our SED model is applied.

\subsubsection{Wavelength coverage}

The wavelength coverage is not entirely the same for all of the galaxies. For the DGS galaxies, the observational constraints come from the 2MASS survey and the \spitz, \wise, \iras\ and \hersc\ instruments (Section \ref{sec:data}). 
For the KINGFISH galaxies, the observational constraints originate from the same instruments (except \wise\ and \spitz/IRS).
We have complete spectral coverage for 93\% of the sources in the 1 - 5 \micÊ\ NIR range (five wavelengths\footnote{We consider only one datapoint when several observations are available for one wavelength (e.g., either WISE 3.4 \mic\ or IRAC 3.6 \mic, either IRAS 100 \micÊ\ or PACS 100 \mic, etc.) to avoid redundancy.}), 87\% in the 5 - 50 \mic\ MIR range (four wavelengths), 90\% in the 50 - 200 \mic\ FIR range (four wavelengths), and 76\% in the 200 - 500 \mic\ submm range (three wavelengths). 
\cite{Ciesla2014} 
showed that SPIRE constraints were particularly important to account for the cold dust (see also Section \ref{ssec:submm}). The 24\% of our sample without SPIRE detections are all dwarf galaxies, harbouring particularly warm dust, with a peak of the FIR SED at wavelengths $\sim$ 30 - 70 \mic, well sampled by constraints until 160 \mic. Thus, although heterogeneous, the wavelength coverage achieved in our sample is excellent and we are confident in the parameters we derive with these sets of constraints.

{\rev When several observations are available at the same wavelength we favour the fluxes obtained with higher resolution data}.
When IRS constraints are available, we weight the IRS data points for them to contribute equally as the other MIR constraints in the fit. Whenever PAH features are absent from the IRS spectrum, we fix \fpah\ = 0, and fix $f_{{\rm ion}}$ = 0.5, {\rev to reduce the number of free parameters.}
Some galaxies are not detected at one or several wavelengths, and we {\rev impose to the best-fit model to be consistent with the upper limits.} 

\subsubsection{Additional features}\label{sssec:addfeat}

\begin{figure}[ht!bp]
\begin{center}
\includegraphics[width=9.0cm]{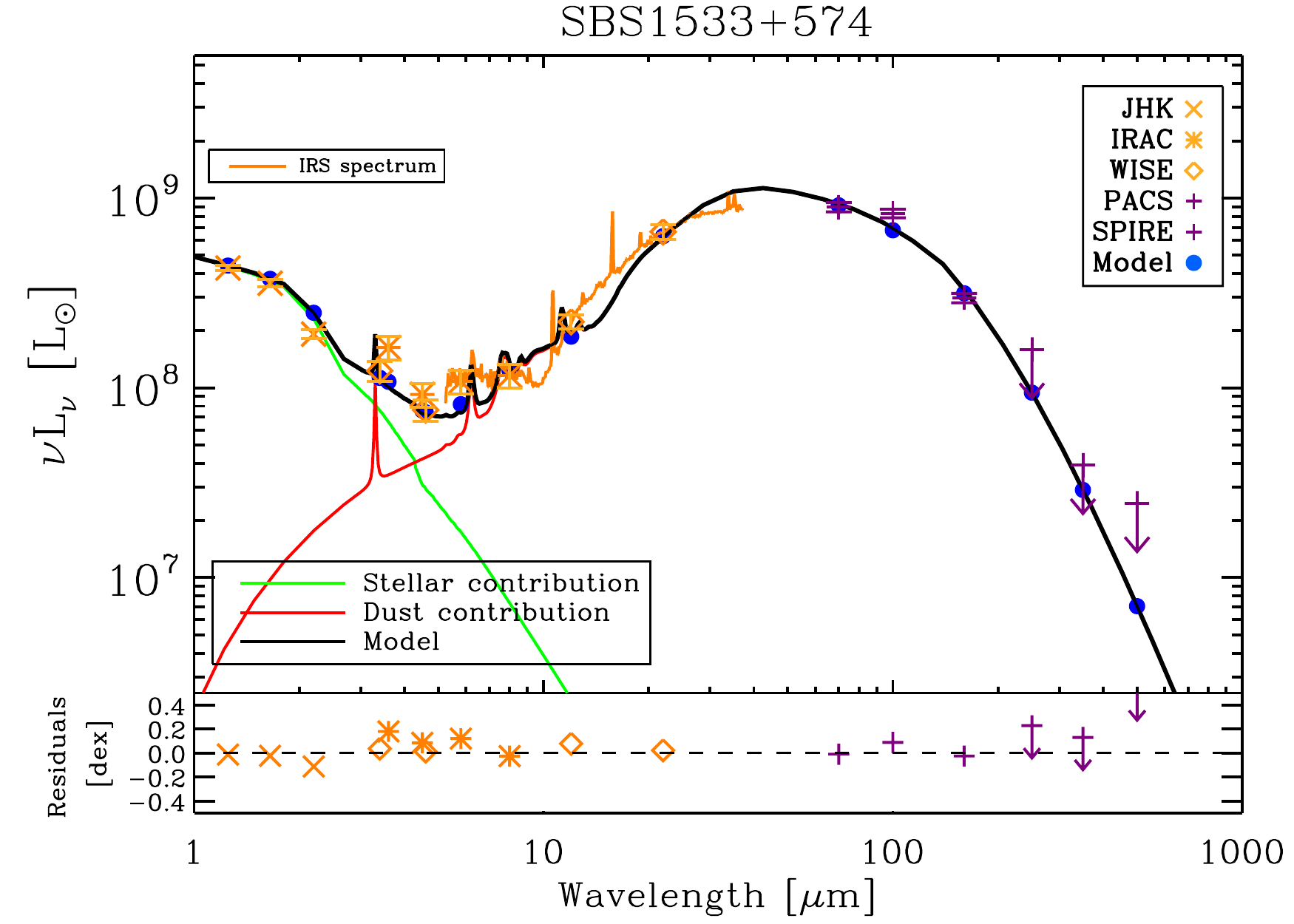}
\includegraphics[width=9.0cm]{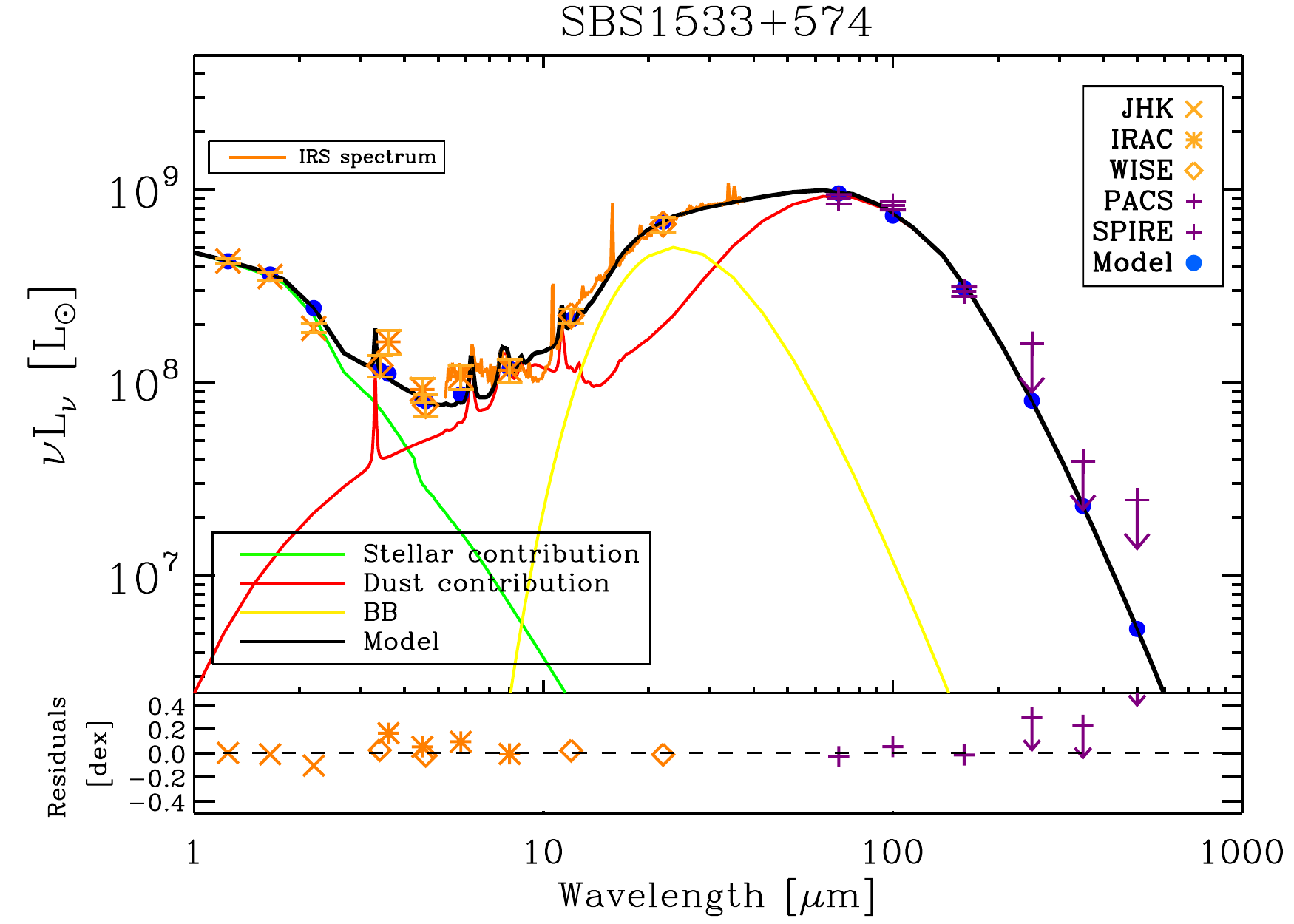}
\caption{SED for SBS1533+574 without {\it (top)} and with {\it (bottom)} the extra MIR modified blackbody, shown in yellow. The other colours and the symbols are the same as for Fig.Ê\ref{f:exSED}.}
\label{f:MIRmodBB}
\end{center}
\end{figure}

For eleven DGS galaxies the observed MIR continuum shape cannot be well reproduced by our model. 
An example is given in the top panel of Fig. \ref{f:MIRmodBB} for SBS~1533+574. For these eleven sources, we add to the model an extra modified blackbody (modBB) component at MIR wavelengths, with a fixed {\it $\beta$} = 2.0 and a temperature varying between 80 and 300 K. It can be physically interpreted as a contribution from hot \HII\ regions to the total emission of the galaxy. As dwarf galaxies have small physical sizes and low dust attenuation, the emission from the energetic \HII\ regions can indeed affect the total emission from the whole galaxy {\rev \citep{daCunha2008, Galametz2010, Hermelo2013}}. This is observed also for FIR-line emission that is often dominated on galaxy-wide scales by the emission from star-forming regions or diffuse ionised gas in dwarf galaxies \citep[e.g.,][]{Cormier2015}. In these cases, the \cite{Dale2001} empirical recipe of $dM/dU \propto U^{-\alpha}$ may not be flexible enough to provide a satisfactory fit. Of course, a modBB is not the optimal way of modelling the emission from hot grains in \HII\ regions \citep[see][]{Galliano2008, Groves2008} as these grains are likely not in thermal equilibrium, but we adopt this approach as a first approximation. 

This additional component does not impact significantly the total dust mass determination: for each of the eleven cases, dust mass estimates with and without the extra MIR modBB are consistent within errors except for SBS~1533+574. However, in this case the addition of a warm modBB dramatically changes the modelled MIR-to-submm shape of the SED (see Fig. \ref{f:MIRmodBB}). The best fit MIR modBB temperatures are reported in Table \ref{t:Dustparam}. The fits without the MIR modBB are shown in Appendix \ref{ap:default}, Fig. \ref{f:Noteseds} for comparison.

Specific modelling was required for five DGS galaxies and is detailed in Appendix \ref{ap:default}.

\subsection{Estimating the errors on the best-fit parameters}\label{ssec:errors}

The errors on the best-fit parameters are estimated by generating 300 random realisations of the observed SEDs within the observational errors, taking special care for errors which are correlated between different bands (see below). For each galaxy, we perform fits of the 300 random realisations of the SEDs in order to obtain a distribution for the parameters. Following the method outlined in \cite{RemyRuyer2013}, the perturbation to the observed fluxes is the sum of two components:

\begin{itemize}
 \item{A normal random independent variable representing the measurement errors at each wavelength.}
 \item{A normal random variable describing the calibration errors that takes into account the correlation between the wavebands for each instrument (as detailed below).}
\end{itemize}

The errors are taken as the 66.67\% confidence level. The best-fit parameters with errors are given in Table \ref{t:Dustparam} for all of the galaxies.

We detail below the decomposition of the calibration errors and eventual correlations between the different bands for all of the instruments, except \hersc\ as this decomposition has already been presented in \cite{RemyRuyer2013}. {\rev Taking into account the correlations in the observationnal errors allows for smaller error bars on the dust parameters.} \\ 

\paragraph{2MASS:} \cite{Jarrett2003} quote a 2 - 3\% uncertainty on the zero-magnitude flux values. To be conservative, we assume an independent error of 3\% in each band. 

\paragraph{IRAC:} The total calibration uncertainty used for the DGS and KINGFISH IRAC fluxes is $\sim$~10\%. This can be decomposed into two parts:
\begin{itemize}
\item \cite{Reach2005} give a 2\% uncertainty in all of the IRAC bands. This error is correlated between the four bands.
\item The IRAC Instrument Handbook (Section 4.3) also gives a 10\% error to account for several systematic effects in the calibration. This error is independent from band to band.
\end{itemize}

\paragraph{IRS:} The IRS spectra extracted from the CASSIS database provide a decomposition of the total error on the flux densities into three parts:
\begin{itemize}
\item Part of the error is the statistical error on the flux determination, 
independent for the different wavelengths.
\item Part of the error is systematic and due to the flux difference between the two nod spectra and is correlated for all of the wavelengths over the SL range on one side and the LL range on the other side.
\item The third component of the error in IRS spectra is the calibration error, and this error is correlated between the two SL and LL modules since the same calibrator star was used for both modules. 
According to \cite{Lebouteiller2011}, the global IRS calibration is better than the 2\% level.
\end{itemize}

\paragraph{MIPS:} For the DGS MIPS photometry, \cite{Bendo2012} used a 4\% calibration error at 24 \mic\ \citep{Engelbracht2007}, 10\% at 70 \mic\ \citep{Gordon2007} and 12\% at 160 \mic\ \citep{Stansberry2007}. For KINGFISH, the same uncertainties were used by \cite{Dale2007} except for MIPS 70 \mic\ where they adopted a 7\% calibration error. According to the MIPS Instrument Handbook, the calibration of the MIPS 160 \mic\ band has been done using the 24 \mic\ and 70 \mic\ observations of asteroids. We can thus consider that the calibration errors for MIPS 24 \mic\ and MIPS 70 \mic\ are both correlated with MIPS 160 \mic.

\paragraph{WISE:} \wise\ calibration has been performed on stars and is tied to \spitz\ calibration according to \cite{Jarrett2011}. For each wavelength, the correlations between the bands can be summarised this way:
\begin{itemize}
\item The \wise\ 3.4 \mic\ calibration error is decomposed in an independent part, proper to \wise, of 2.4 \%, and is correlated with the IRAC 3.6 \mic\ band. \item Similarly, the \wise\ 4.6 \mic\ error has an independent part of 2.8\% and is correlated with the IRAC 4.5 \mic\ band.
\item The \wise\ 12 \mic\ error has an independent part of 4.5\% and is correlated with the IRS-SL/LL modules. 
\item And finally, the \wise\ 22 \mic\ error has an independent part of 5.7\% and is correlated with the MIPS 24 \mic\ band.
\end{itemize}

\paragraph{IRAS:} According to the IRAS explanatory supplement\footnote{This document is available at: http://lambda.gsfc.nasa.gov/\\ product/iras/docs/exp.sup/.}, the calibration of \iras\ has been tied to the \cite{Rieke1984} ground-based photometric system at 12 \mic. The three \iras\ bands at 12, 25 and 60 \mic\ have been calibrated using stellar models, and the \iras\ 100 \mic\ calibration used asteroids. The relative uncertainties relative to the ground-based 12 \mic\ are 2\%, 5\% and 5\% for \iras\ 12, 25 and 60 \mic\ respectively, independent from band to band. The absolute uncertainty on the 12 \mic\ flux density is 4\%, in common and correlated between the three bands. The uncertainty at 100 \mic\ is 10\%, and is not correlated with any of the other \iras\ bands.

\subsection{Comparison with a single modified blackbody}\label{ssec:compBB}
 
\begin{figure*}[h!tbp]
\begin{center}
\includegraphics[width=16cm]{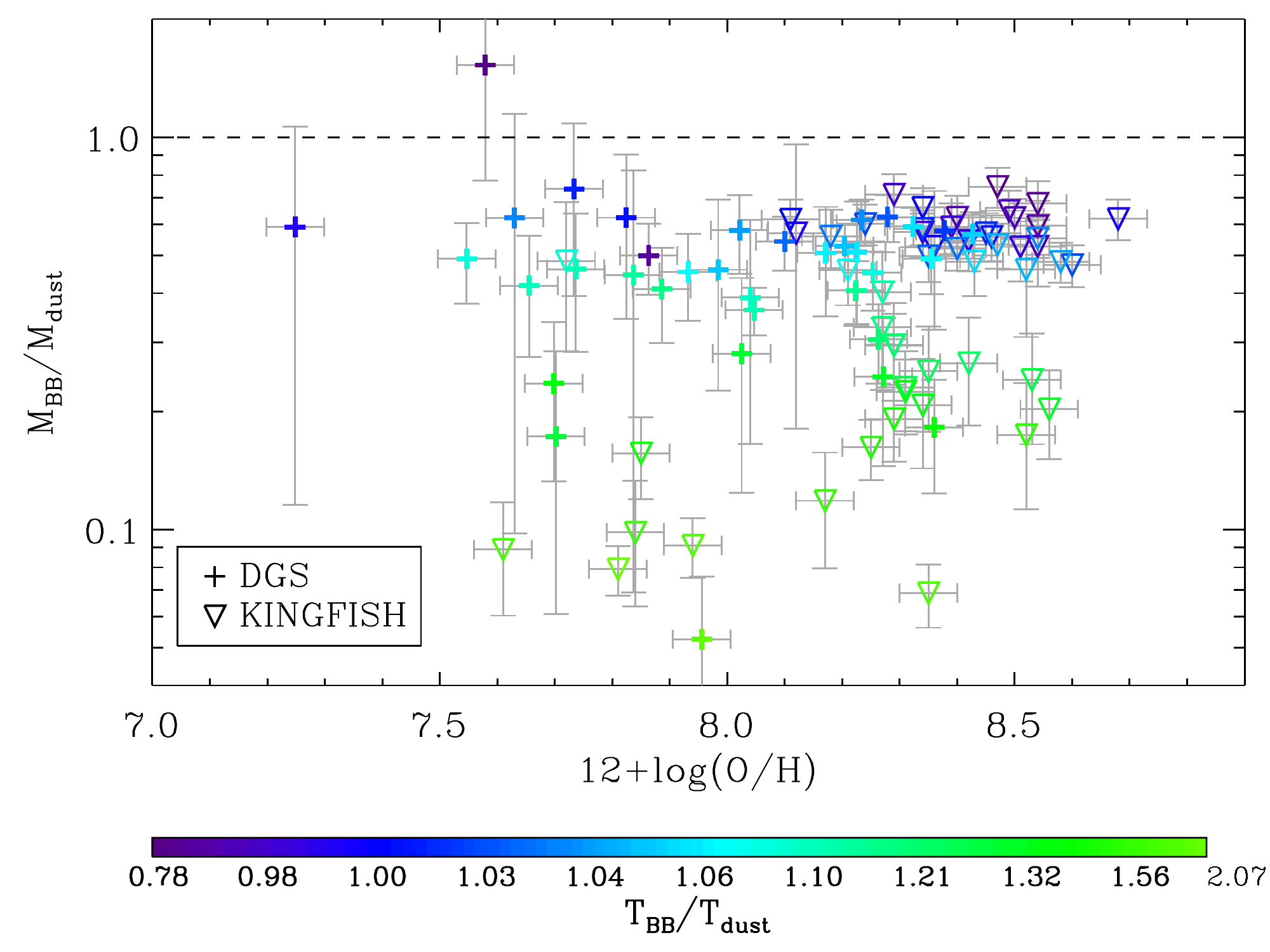}
\caption{Ratios of the dust masses estimated with a modBB model and with a semi-empirical SED model, $M_{{\rm BB}}/M_{{\rm dust}}$, for the DGS (crosses) and KINGFISH (downward triangles) samples, as a function of the metallicity. The colour codes the ratio between the dust temperatures estimated from modBB fits and the SED model \citep[assuming T$_{MW}$ = 19.7 K,][see text for details]{PlanckCollaboration2014XI} $T_{{\rm BB}}$/$T_{{\rm dust}}$.
}
\label{f:compBBsed}
\end{center}
\end{figure*}

We conducted a first study of the dust properties in the DGS using a modBB model to describe the dust emission in \cite{RemyRuyer2013}. Although very popular in the literature especially when limited data is available, a modBB model assumes a single temperature for the dust grains, and this affects the resulting dust properties. 
Fig. \ref{f:compBBsed} shows the ratios between the dust masses estimated with a modBB model, $M_{{\rm BB}}$, and with our semi-empirical SED model, \mdust, for both DGS and KINGFISH samples, as a function of metallicity. 

For this comparison, the modBB masses have been updated from \cite{RemyRuyer2013} to be consistent with the updated \hersc\ data. We also fixed the emissivity index $\beta$ =2.0 to be consistent with the effective emissivity index of our dust model. {\rev We otherwise use the same wavelength range and fitting procedure as in \cite{RemyRuyer2013}}. The opacity at the reference wavelength, $\kappa(\lambda_0)$ = 4.5 m$^2$kg$^{-1}$ at 100 \mic, used to normalise the modBB model is derived from the opacities of the dust composition adopted in the model \citep{Galliano2011}. We find that the modBB model almost systematically underestimates the dust mass compared to a semi-empirical SED model, with a median ratio of 0.48 but also ratios sometimes as low as 0.1 (Fig. \ref{f:compBBsed}), and that this underestimation does not depend on the metallicity of the source. 

{\rev We also compare the temperatures derived from a modBB fit, $T_{{\rm BB}}$, to the average dust temperatures derived from the SED fits, $T_{{\rm dust}}$. We can estimate $T_{{\rm dust}}$ directly by integrating over 
$T=T_{{\rm MW}} \times$ U$^{1/(4+\beta)}$ in Eq \ref{eq:U}, with $\beta$=2.0 for our model. We adopt T$_{{\rm MW}}$=19.7 K from \cite{PlanckCollaboration2014XI}. We see from the colour coding on Fig. \ref{f:compBBsed}, that the more $T_{{\rm BB}}$ is overestimated compared to $T_{{\rm dust}}$, the more the modBB underestimates the dust mass compared to a more complex dust modelling.}

{\rev When modelling a galaxy with a broad distribution of equilibrium temperatures, the modBB model will tend to average over all temperatures to best fit the peak of the SED. The modBB will get biased towards higher dust temperatures, and thus lower dust masses, as seen in Fig. \ref{f:compBBsed}. This also applies, to a smaller extent, in the case where the dust emission is dominated by dust heated by the diffuse ISRF because the two big grain populations (graphite and silicate) do not reach the same equilibrium temperatures. 
}

Other studies have compared dust masses from modBB to those from more complex dust modelling {\rev using the \cite{DraineLi2007} dust SED model (hereafter DL07), and report an underestimation of 10 - 20\% \citep{Magrini2011, Bianchi2013}. The grain mixture from \cite{Zubko2004} used in our modelling have a slightly different emissivity compared to the DL07 model (i.e., size distribution effect), causing the difference with the results obtained here.}

\subsection{Influence of the ISRF description}\label{ssec:isrf}

{\rev
We describe the starlight intensity distribution by a power-law component only \citep{Dale2001} and not with an additional diffuse component as is sometimes done by other studies \citep{Draine2007, Aniano2012, Dale2012}. These studies were dedicated to the modelling of metal-rich spirals. Our study is motivated by the modelling of the low-metallicity sources and adding a diffuse component to the starlight intensity distribution would not be appropriate. Moreover, \cite{Galliano2011} showed on the low-metallicity LMC that the additional diffuse component was not necessary. Additionally, this diffuse component would add yet another free parameter, which would be difficult to constrain for the few dwarf galaxies not detected beyond 160 \mic. We apply this starlight intensity parametrisation to the KINGFIHSH galaxies as well, to have a consistent and homogeneous modelling approach for the whole sample.
Nonetheless, we check that the two descriptions for the starlight intensity distribution (power-law only or power-law + diffuse) give the same \mdust\ and \uav\ for the KINGFISH galaxies. We have a median ratio of \mdust(power-law+diffuse)/\mdust(power-law only) = 1.03 $\pm$ 0.15, and \uav(power-law+diffuse)/\uav(power-law only) = 1.00 $\pm$ 0.12. Given that the average error is 20\% on \mdust\ and 21\% on \uav, we have a good agreement between the two parametrisations of the starlight intensity.  
}
 
Note that we use the same ISRF for all of the galaxies for consistency, not taking into account the fact that the ISRF is expected to be harder in low-metallicity galaxies \citep{Galliano2003, Galliano2005, Madden2006}. {\rev At fixed energy density, i.e. $U$, and given that the absorption efficiency, Q$_{{\rm abs}}$, is not a strongly varying function of wavelength in the UV/visible regime, the hardness of the radiation field would not greatly impact the emission from the big grains in thermal equilibrium. Under these assumptions,} increasing the hardness of the ISRF in dwarf galaxies would only affect the emission spectrum of the stochastically heated grains. 

\subsection{Influence of the dust composition: amorphous carbons}\label{ssec:amcar}

Several studies suggested that amorphous carbon grains provide a better description of carbonaceous dust in the ISM than graphite. For example, \cite{SerraDiaz-Cano2008} showed that the erosion of amorphous carbon grains in shocks is more efficient than for graphite grains and matches better the observations of high fractional abundances of carbon in the gas phase of shocked regions. They conclude that even if the carbon dust is in form of graphite when injected in the ISM, it is unlikely that it will remain graphitic as it evolves in the ISM and is subject to erosion or ion irradiation. 

We compute another set of dust masses for our sample using amorphous carbons instead of graphite grains in our full SED model, {\rev and keeping the same carbon mass budget in the grains}. The amorphous carbons optical properties are taken from \cite{Zubko1996} \citep[see][]{Galliano2011}. We use exactly the same procedure and options (e.g., additional MIR modBB or not) to be able to directly compare the graphite (noted with $_{[Gr]}$) and amorphous carbon (noted with $_{[Ac]}$) dust masses (reported in Table \ref{t:SEDparam}). 

We find that the amorphous carbon dust masses are about 2.5 times lower than the graphite dust masses. This is because the amorphous carbon dust is more emissive in the submm domain and needs less dust mass to account for the same luminosity.
\cite{Galliano2011} showed that this dust composition was better suited for the low-metallicity LMC. However, we do not find any dependence of \mdust$_{[Ac]}$/\mdust$_{[Gr]}$ on metallicity, nor with any other galaxy property. This is because we fixed the dust composition {\it a priori} in our models.
Regarding the other parameters of the fit, \ltir, \sigU, and \fpah\ do not vary significantly between the two dust compositions, while \uav$_{[Ac]}$/\uav$_{[Gr]}$ $\sim$ 2.0. 
{\rev At fixed SED shape, i.e. fixed average equilibrium temperature, the radiation field intensity required to heat the amorphous carbon grains is higher as they have a higher emissivity in the submm and absorb more in the UV/visible regime.}

In most studies similar to ours where dust SED modelling is used to derive and interpret dust properties \citep[e.g.,][]{Draine2007, Galliano2008, Galametz2011, Dale2012, Cortese2012b, Ciesla2014, RemyRuyer2014} the models mostly use graphite for the carbonaceous component. So to ease the comparison with these various studies, we use \mdust$_{[Gr]}$ in the rest of this paper. As the ratio \mdust$_{[Ac]}$/\mdust$_{[Gr]}$ is mostly independent of any of the galaxy properties tested here, using \mdust$_{[Ac]}$ instead would systematically shift the trends presented in the following sections without affecting the general conclusions.

\subsection{Influence of the wavelength coverage and submm excess}\label{ssec:submm}
{\rev
The importance of submm observations longwards of 160 \mic\ has been shown by \cite{Gordon2010, Galametz2011, Dale2012, Ciesla2014}.
We find that the dust masses estimated without \hersc\ constraints are overestimated for high-metallicity galaxies (i.e., 12+log(O/H) $\gtrsim$ 8.2) and underestimated for lower metallicity galaxies in agreement with the conclusions of the previously mentioned studies. 

Excess emission at submm wavelengths (850 - 870 \mic), appearing around 400 - 500 \mic, and presently unaccounted for by the standard dust emission models, has been observed in some low-metallicity systems or low-mass spirals \citep[e.g.,][]{Galliano2003, Galliano2005, Dumke2004, Bendo2006, Galametz2009, Galametz2012, Zhu2009, Bot2010, Grossi2010, Dale2012, RemyRuyer2013, Ciesla2014, Galametz2014, Gordon2014, Grossi2015}. Several hypotheses have been made to explain this excess emission, but so far remain inconclusive. As the submm excess is not the main focus of this paper, we do not discuss them here. The presence of this submm excess can impact the dust mass, as a large amount of cold dust could be contributing to this submm emission. 

We adopt the same definition of the excess as that in \cite{RemyRuyer2013}, i.e., an excess is present if the residual at 500 \mic\ is greater than its error bar (1$\sigma$ criteria). Note that the model used in \cite{RemyRuyer2013}, against which the 500 \mic\ emission is compared, is a modBB with a fixed $\beta$ of 2.0, and that the SPIRE flux densities have been updated to match the latest SPIRE calibration and beam areas for this study (see Sect \ref{ssec:hersc}). The findings of \cite{RemyRuyer2013} (identification of the galaxies with excess and number of excess sources) are thus not directly comparable to our results. 

Eight galaxies exhibit an excess at 500 \mic: Haro~11, HS~0052+2536, Mrk~930, NGC~1569, NGC~625, NGC~337, NGC~3049 and NGC~4631. For these eight excess galaxies (except one, NGC3049) including or omitting the 500 \mic\ point in the fit yields dust mass estimates that are consistent within the errors, because the 500 \mic\ excess is quite small ($\leq$ 1.5 $\sigma$).
However, omitting the 500 \mic\ point in the fit for all of the sample leads to underestimated dust masses (by a factor 2 to 4) for 15\% of the 78 galaxies detected at 500 \mic, because part of the coldest component of the dust is not fully taken into account. 

For the purpose of our study, we thus need to include the 500 \micÊ\ point in the model to get an accurate estimation of the dust mass, and this estimation will not be biased for the eight sources with this small submm excess at 500 \mic. 
}

\section{Dust properties}\label{sec:dustprop}

In this section, we scrutinise the dust properties derived from our realistic dust modelling over the whole IR wavelength range, and present results for \mdust\footnote{From now on, we consider the dust masses estimated with the model using graphite carbon grains, and including the 500 \mic\ point in the fit.}, \ltir, \fpah, \uav\ and \sigU. 
The median error on each parameter together with the spanned range of values and the median values are shown in Table \ref{t:SEDparam} for both samples.

\begin{table*}[h!tbp]                                                                               
\begin{center}                                                                                      
\caption{DGS and KINGFISH SED parameters.}                                                                                                                                                              
\label{t:SEDparam}                                                                                  
\begin{tabular}{lc | ccc | ccc}                                                                     
\hline                                                                                              
\hline                                                                                              
Parameter                                  & Median err& \multicolumn{3}{c}{DGS}                 & \multicolumn{3}{c}{KINGFISH} \\     
                                                &  [\%] & min & median & max                   & min & median & max \\                 
\hline                                                                                              
12+log(O/H)                               &      26$^*$  &  7.14     &  7.93     &  8.43     &  7.54     &  8.35     &  8.77   \\ 
  &                                                        &            IZw18  &                   &           He2-10  &           DDO154  &                   &        NGC1316  \\ 
\hline                                  
Log(\mstar) [Log(\msun)]                  &      47  &  6.51     &  8.58     & 10.62     &  7.24     & 10.44     & 11.68   \\ 
  &                                                        &      HS0822+3542  &                   &            UM448  &           DDO053  &                   &        NGC1316  \\ 
\hline                                  
Log(SFR) [Log(\sfr)]                      &      22  & -2.21     & -0.66     &  1.40     & -2.36     & -0.38     &  0.90   \\ 
  &                                                        &          UGC4483  &                   &           Haro11  &           M81dwB  &                   &        NGC2146  \\ 
\hline                                  
Log(\ssfr) [Log(yr$^{-1}$)]               &      52  &-11.00     & -9.16     & -8.13     &-12.08     &-10.35     & -9.35   \\ 
  &                                                        &      Tol0618-402  &                   &      SBS0335-052  &          NGC1316  &                   &         DDO053  \\ 
\hline                                  
\hline                                  
Log(\mdust) [Log(\msun)]                  &      19  &  2.44     &  5.69     &  7.70     &  4.04     &  7.28     &  8.16   \\ 
  &                                                        &          UGC4483  &                   &            UM311  &           DDO053  &                   &        NGC5457  \\ 
\hline                                  
Log(\ltir) [Log(\lsun)]                   &       3  &  6.42     &  8.98     & 11.30     &  6.65     &  9.81     & 11.13   \\ 
  &                                                        &          UGC4483  &                   &           Haro11  &           M81dwB  &                   &        NGC2146  \\ 
\hline                                  
\fpah [\fpah$_\odot$]                     &      17  &  0$^{**}$&  0.21     &  0.66     &  0$^{**}$&  0.90     &  1.64   \\ 
  &                                                        &                   &                   &          NGC4449  &                   &                   &        NGC4254  \\ 
\hline                                  
Log(\uav) [Log(U$_\odot$)]                &      21  & -0.51     &  1.30     &  3.44     & -0.62     &  0.29     &  1.41   \\ 
  &                                                        &          NGC6822  &                   &      SBS0335-052  &          NGC4236  &                   &        NGC1377  \\ 
\hline                                  
Log(\sigU) [Log(U$_\odot$)]               &      40  &  0.17     &  2.50     &  4.77     & -0.59     &  0.58     &  3.14   \\ 
  &                                                        &            UM311  &                   &      SBS0335-052  &          NGC2841  &                   &        NGC2798  \\ 
\hline                                  
Log(\mdust/\mstar)                        &      51  & -4.53     & -3.08     & -1.79     & -4.02     & -2.96     & -2.02   \\ 
  &                                                        &      SBS0335-052  &                   &           Pox186  &          NGC5866  &                   &        NGC4236  \\ 
\hline                                  
\end{tabular}                                                                                       
\end{center}                                                                                        
{\footnotesize                                                                                      
{\rev $^*$: corresponding to 0.1 dex, see \cite{Madden2013}}.\\
$^{**}$: 0 for the minimum $f_{PAH}$ value means that PAHs are not detected in at least one galaxy in the sample.\\                                                                                        
Note: \fpah$_\odot$ = 4.57 \% in our model and U$_\odot$ = 2.2 $\times$ 10$^{-5}$ W.m$^{-2}$.}                                                                                                          
\end{table*}

For each parameter we examine the relation with metallicity expressed as 12+log(O/H), the star-formation activity  traced by the \ssfr, estimated in Sect. \ref{sssec:ssfr}, and the stellar mass \mstar. For simplicity in the notations and to ease the reading, we use Z for 12+log(O/H), even if Z is not strictly equivalent to the oxygen abundance.

We compute Spearman rank correlation coefficients\footnote{The Spearman rank coefficient, $\rho$, indicates how well the relationship between X and Y can be described by a monotonic function: monotonically increasing: $\rho$ $>$ 0, or monotonically decreasing: $\rho$ $<$ 0. They are computed with the {\sc r\_correlate} function in IDL.} for every pair of parameters and present them in Table \ref{t:corr} to determine the most significant relations to investigate. For the 98 galaxies, we determine that the correlation (or anti-correlation) is significant if $|\rho| >$ 0.37 at a significance level of 0.01\% (i.e., the probability that the two variables are motonically correlated is greater than 99.99\%). 

\begin{table}[h!tbp]                                                                                
\begin{center}                                                                                      
\caption{Dust properties: correlations from Spearman rank coefficients.}                                                                                                                                
\label{t:corr}                                                                                      
\begin{tabular}{lcccc}                                                                              
\hline                                                                                              
\hline                                                                                              
Param &          12+log(O/H) &     \ssfr &   \mstar \\
\hline                                                                                              
12+log(O/H) &          1.00 &     -0.78 &      0.84 \\
\mdust &               0.78 &     -0.72 &      0.92 \\
\ltir &                0.64 &     -0.44 &      0.87 \\
\fpah &                0.67 &     -0.77 &      0.70 \\
\uav &                -0.47 &      0.66 &     -0.35 \\
\sigU &               -0.52 &      0.71 &     -0.44 \\
\mdust/\mstar &        0.02 &     -0.04 &     -0.06 \\
\hline                                                                                              
\end{tabular}                                                                                       
\end{center}                                                                                        
\end{table}

\subsection{Dust mass}

The dust masses in our sample cover a range of more than five orders of magnitude. 
The median value is 4.4 $\times$ 10$^5$ \msun\ for the DGS sample and 1.9 $\times$ 10$^7$ \msun\ for the KINGFISH sample, about two orders of magnitude above the median dust mass for the DGS, in agreement with the findings of \cite{RemyRuyer2013}.

The median error on the dust mass is $\sim$20\% (Table \ref{t:SEDparam}). This error only takes into account the uncertainties in the observations. No systematic modelling uncertainty is included due to the difficulty of determining it. However, with the different tests performed in Sects. \ref{ssec:compBB}, \ref{ssec:amcar} and \ref{ssec:submm} where we changed the model, the dust composition or the submm wavelength coverage, we can say that we have a conservative systematic modelling uncertainty of a factor 2 to 3 on the dust mass estimate. 

\begin{figure}[h!tbp]
\begin{center}
\includegraphics[width=9.0cm]{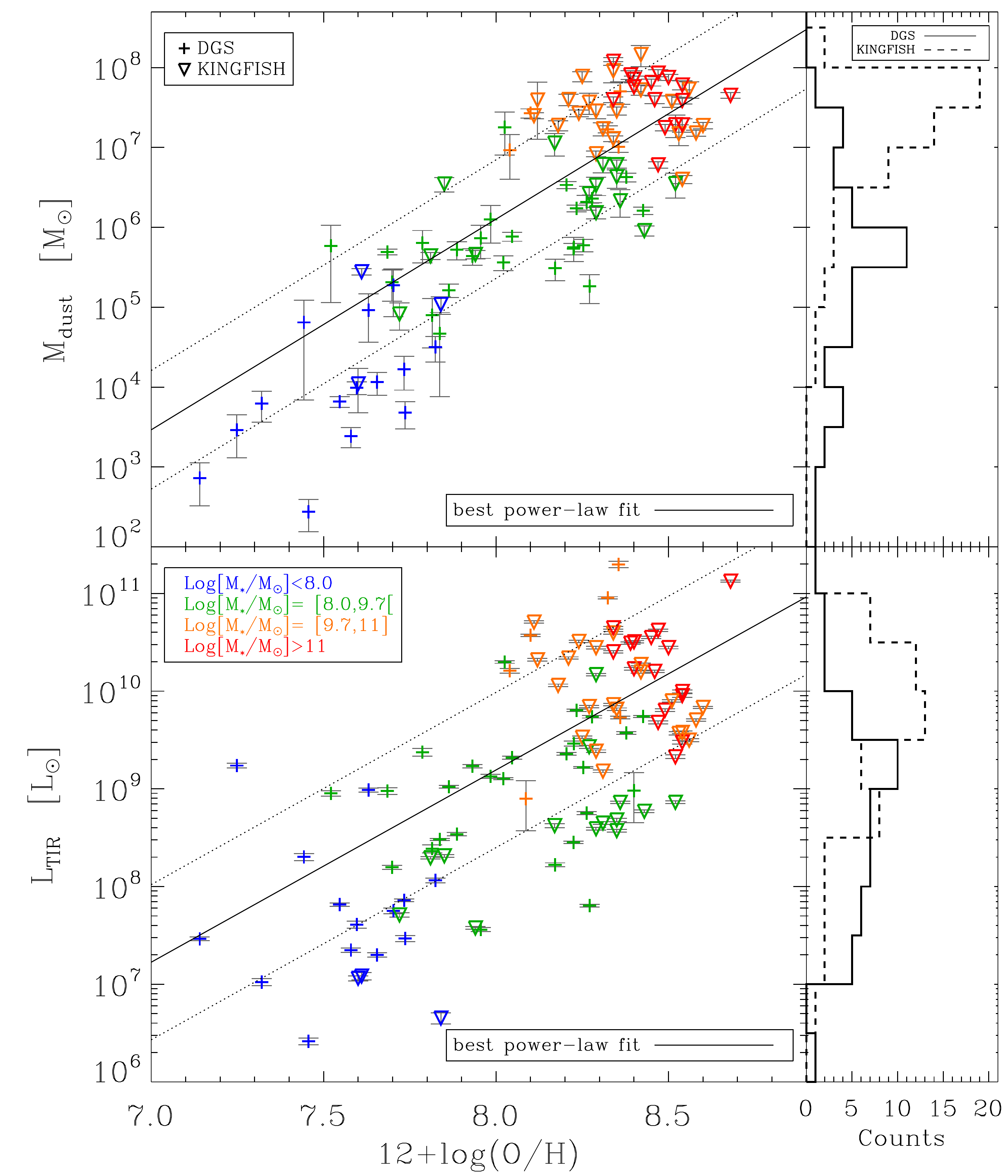}
\caption{Dust masses {\it (top)}, TIR luminosities {\it (bottom)} 
for the DGS (crosses) and KINGFISH (downward triangles) samples as a function of metallicity, colour coded with \mstar. 
The best power-law fit is indicated as a black line, together with the 1$\sigma$ dispersion as dotted lines for the two panels. The distribution of each parameter is indicated on the side of each panel for both samples: solid line for DGS and dashed line for KINGFISH.}
\label{f:Zcorr}
\end{center}
\end{figure}

As seen in Table \ref{t:corr}, the stellar mass gives the strongest correlation with the dust mass ($\rho$(\mdust, \mstar)=0.92), and not the metallicity: $\rho$(\mdust, \Ao) = 0.78. The correlation between stellar mass and metallicity is also stronger than between dust mass and metallicity, $\rho$(\mstar, \Ao) = 0.84. This holds whether we consider the full sample, low-metallicity sources only or metal-rich sources only. This simply translates a scaling effect where a more massive galaxy also contains more dust \citep[see also][]{Ciesla2014}. The correlation between the dust mass and metallicity is thus a direct consequence of the mass-metallicity relation \citep{Tremonti2004}.
This is illustrated in Figure \ref{f:Zcorr}. The best fit relation gives: 

\begin{equation}
{\rm log}(M_{{\rm dust}}) = (-15.0 \pm 0.9) + (2.6 \pm 0.1) \times (12+{\rm log(O/H)})
\end{equation}

\noindent with a dispersion of 0.74 dex from the relation. 
                                                                                      
\subsection{Total infrared luminosity}\label{ssec:ltir}

\subsubsection{Estimating \ltir}

Several definitions exist to derive \ltir. We chose to estimate \ltir\ by integrating the best-fit modelled dust SED, L$_{TIR [SED]}$, (i.e., not including the stellar continuum) between 1 and 1000 \mic. Some studies use [8-1100] \mic\ or [3-1100] \mic\ intervals for the integration, but using any of the two other intervals gives equivalent results (see Table \ref{t:ltircomp}). 

{\it Comparing \lfir\ and \ltir:} We also compare the FIR luminosities, \lfir\ over the interval [50-650] \mic\ used in \cite{RemyRuyer2013}, with \ltir. We find that the \lfir\ accounts for about 65$\pm$13\% of the TIR luminosity, and gets lower as the peak of the SED broadens (($\rho$(\lfir/\ltir, \sigU) =  -0.83)). 
As the peak of the dust SED shifts to shorter wavelengths and broadens, the [50-650] \micÊ\ FIR interval does not capture the bulk of the emitted dust luminosity anymore. For very broad SEDs, we can miss up to 70\% of the TIR luminosity. The most extreme case is SBS~0335-052 where \lfir/\ltir = 10\%, consequence of its very peculiar SED peaking around 30 \mic. \\

\begin{table}[h!tbp]                                                                               
\begin{center}                                                                                      
\caption{\ltir\ comparisons.}                                                                                                                                                              
\label{t:ltircomp}                                                                                  
\begin{tabular}{lc}                                                                         
\hline                                                                                              
\hline
x					& Median ratio \\ 
\hline                                                                                              
Definitions  			&		x / \ltir$_{[1-1000]}$ \\
\hline
TIR: \ltir$_{[8-1100]}$		&		1.00 $\pm$ 0.02	\\
TIR: \ltir$_{[3-1100]}$		&		1.05 $\pm$ 0.08	\\
FIR: \lfir$_{[50-650]}$		&		0.65 $\pm$ 0.13	\\
\hline                                                                                              
\cite{Galametz2013b}		&     		x / \ltir$_{[3-1100]}$ \\
\hline
70					&		1.13 $\pm$ 0.29	\\
100					&		0.72 $\pm$ 0.27	\\
24 - 160				&		1.01	$\pm$ 0.35	\\
24 - 70 - 160			&		1.01	$\pm$ 0.14	\\
24 - 70 - 100 - 160		&		1.00	$\pm$ 0.13	\\
\hline
\cite{Madden2013, Kennicutt2011}				& 		x / L$_{{\rm TIR [24-70-160, G13]}}$ \\
\hline
$L_{{\rm TIR [pre-\hersc]}}$ &		1.13 $\pm$ 0.29	\\
\hline                                  
\end{tabular}                                                                                       
\end{center}                                                                                        
\end{table}

{\it Prescriptions for the DGS:} \cite{Galametz2013b} explored various \ltir\ calibrations from \spitz\ and \hersc\ bands, using the KINGFISH sample. 
We test all of the possible relations using 24, 70, 100 and 160 \mic, and look at the ratio of the estimated \ltir\ over the \ltir\ derived from our best-fit SED\footnote{For this comparison we use \ltirÊ\ integrated over [3-1100] \mic\ to match the interval used in \cite{Galametz2013b}. } to determine which relations are the most appropriate for the DGS sample. 

We find that for the DGS galaxies, the best monochromatic relation is using 70 \mic\ with a median ratio of 1.13~$\pm$~0.29 (Table \ref{t:ltircomp}). Using 70 \mic\ tends to overestimate the \ltir\ for the DGS, while using 100 \mic\ tends to underestimate the \ltir\ (Table \ref{t:ltircomp}). This is due to the higher $F_{70}/F_{100}$ colour in dwarf galaxies than in more-metal rich galaxies over which these relations have been calibrated.
The best relation using two bands is with 24 and 160 \mic. Increasing the number of bands gives similar results, with decreasing scatter. The smallest dispersion using three bands is achieved by combining 24, 70 and 160 \mic. 
Thus to estimate \ltirÊ\ for galaxies with similar metallicity and star-formation activity, we recommend using the 70 \mic, the 24-160 \mic, or the 24-70-160 \mic\ calibration depending on the number of available constraints.

Pre-\hersc\ estimates of the \ltir\ derived with \spitz\ fluxes and the \cite{DaleHelou2002} formula ($L_{{\rm TIR [pre-\hersc]}}$) from \cite{Madden2013} and \cite{Kennicutt2011} are consistent with the \ltir\ derived here. 
All of these \ltir\ comparisons are summarised in Table \ref{t:ltircomp}, and show that the TIR luminosity is quite a robust parameter. \\

\subsubsection{\ltir\ and metallicity}

\ltir\ is presented in Fig. \ref{f:Zcorr} (bottom panel) as a function of metallicity. The \ltir\ in both samples cover five orders of magnitude, and the low-metallicity dwarf galaxies are about 6.5 times less luminous in the IR than the more metal-rich environments. 
As for the dust mass, \ltirÊ\ is strongly correlated with the stellar mass, $\rho$(\ltir, \mstar)=0.87, due to scaling effects. The correlation between \ltirÊ\ and metallicity ($\rho$(\ltir, \Ao)=0.64) is also a consequence of the mass-metallicity relation. 
As seen in Fig. \ref{f:Zcorr}, the relation of \ltir\ with metallicity is more dispersed than between the dust mass and the metallicity. The best fit relation gives: 

\begin{equation}
{\rm log}(L_{{\rm TIR}}) = (-6.6 \pm 0.2) + (2.0 \pm 0.1) \times (12+{\rm log(O/H)})
\end{equation}

\noindent with a dispersion of 0.79 dex from the relation. From the construction of the model, \ltirÊ\ is directly linked to the product \mdust$\times$\uav\ {\rev at first order}. And while \mdust\ is correlated with metallicity, \uav\ is anti-correlated with metallicity, which results in a flattened correlation with larger scatter between \ltir\ and metallicity.  
\ltir \ is not correlated with \uav\ ($\rho$(\uav,~\ltir)~=~-0.05), indicating that it is the amount of dust that primarily drives the luminosity. 

\subsection{PAHs}\label{ssec:fpah}

\begin{figure*}[h!tbp]
\begin{center}
\includegraphics[width=17cm]{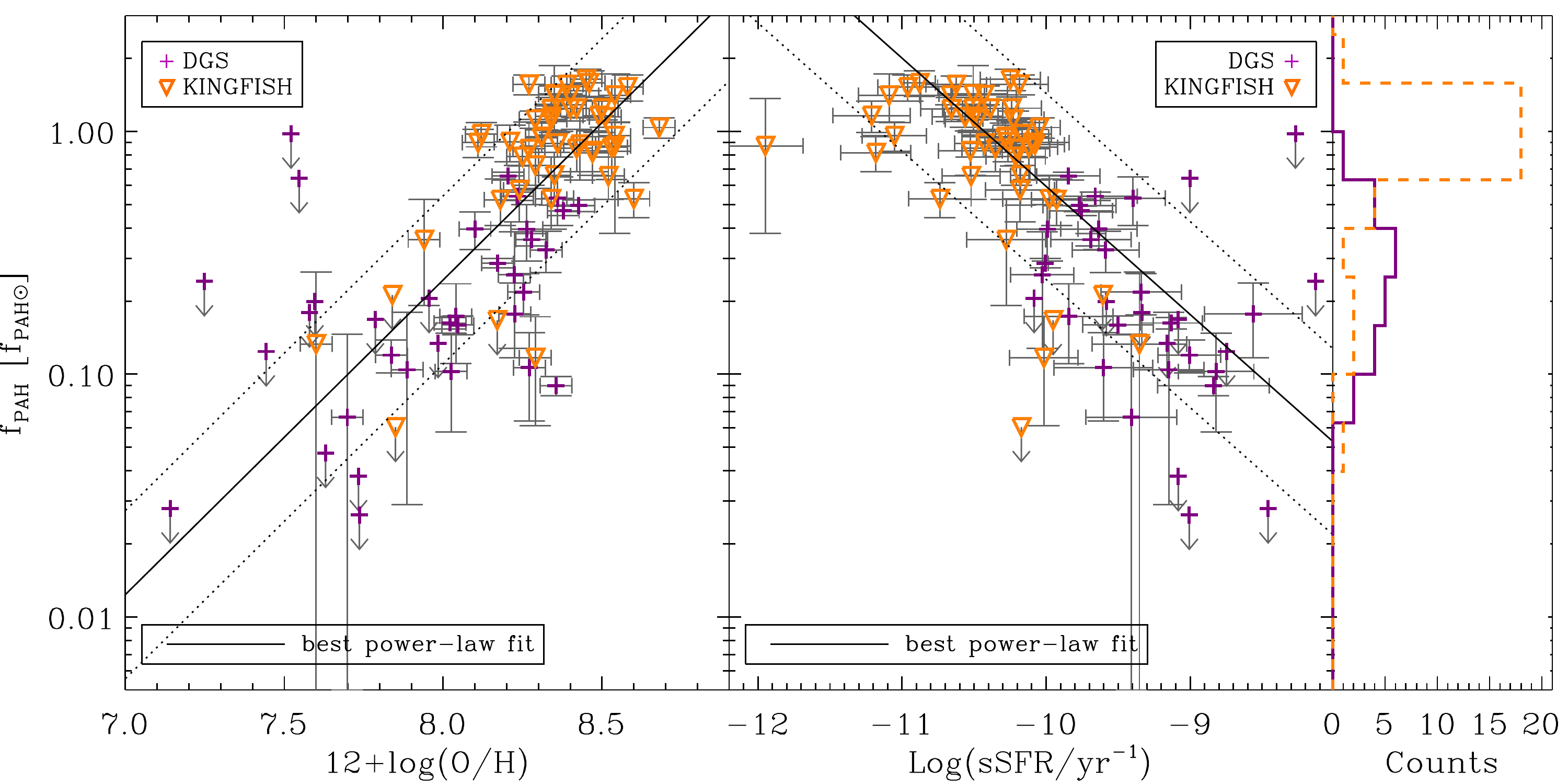}
\caption{PAH mass fractions for the DGS (purple crosses) and KINGFISH (orange downward triangles) samples as a function of metallicity {\it (left panel)} and \ssfr\ {\it (right panel)}. \fpahÊ\ is expressed in units of \fpah$_\odot$, with \fpah$_\odot$ = 4.57\%. The distribution of \fpah\ is indicated on the side for both samples: plain purple line for DGS galaxies and dashed orange line for the KINGFISH sample. In each panel, the best power-law fit is indicated as a solid line, together with the 1$\sigma$ dispersion as dotted lines.}
\label{f:fpah}
\end{center}
\end{figure*}

Lower PAH abundances in low-metallicity systems than in more metal-rich galaxies have previously been observed \citep[e.g.][]{Madden2000, Boselli2004, Engelbracht2005, Wu2006, OHalloran2006, Draine2007, Smith2007, Gordon2008, Galliano2008, Wu2011, Ciesla2014}. 
From our model we obtain an estimate of the mass fraction of PAHs {\rev for our sources, including the 19 low-metallicity galaxies with new IRAC and IRS data. 
We have an average error of 17\% on \fpah, and} a very good agreement between our \fpah\ to the PAH mass fraction derived by \cite{Dale2012} for the KINGFISH sample using the \cite{DraineLi2007} model (median ratio of 1.03 $\pm$ 0.24).
{\rev For the DGS galaxies with featureless IRS continuum,} we fit the model to the observations again, leaving \fpah\ free, and use this value as an upper limit, reported in Table \ref{t:Dustparam}. 

In our sample, the mass fraction of PAHs covers two orders of magnitude, and are detected over 1.1 dex in metallicity down to $\sim$0.08 \zsun, {\rev expanding the range explored by previous works}. From Table \ref{t:corr}, \fpah\ correlates very well with \ssfr\ and metallicity ($\rho$(\fpah, \ssfr) = -0.77, and $\rho$(\fpah, \Ao) = 0.67, {\rev not including the upper limits}), with a PAH mass fraction decreasing with decreasing metallicity and increasing \ssfr, {\rev confirming results from the previously mentioned studies.} This is shown in Fig. \ref{f:fpah}. The best-fit relations of \fpah\ with metallicity and \ssfr\ give: 

{\small
\begin{equation}
\left  \{
 \begin{array}{rcl}
{\rm log}(f_{{\rm PAH}}) & = &  (-11.0 \pm 0.3) + (1.30 \pm 0.04) \times (12+{\rm log(O/H)}) \\
{\rm log}(f_{{\rm PAH}}) & = & (-5.5 \pm 0.1) - (0.53 \pm 0.01) \times {\rm log}({\rm sSFR})  \end{array} 
 \right .
\end{equation}
}

\noindent with a dispersion of 0.35 dex around the relation with metallicity, and of 0.38 dex around the relation with \ssfr. 

{\rev The lower abundance of PAHs in dwarf galaxies is not controlled by a single parameter but rather arises from a joint effect of both low metallicity and high \ssfr. 
The higher \ssfr\ results in a harder and more intense galaxy-wide ISRF. Combined with the lower dust attenuation, PAHs are efficiently destroyed by hard UV photons but also shocks and cosmic rays \citep[e.g.,][]{Madden2006, Engelbracht2008, Micelotta2011}. Due to the small physical size of dwarf galaxies, PAHs are also subject to destruction by SN shock waves on galaxy wide scales \citep{OHalloran2006, Micelotta2010}. 
In addition, lower C/O ratios in dwarf galaxies \citep{Garnett1995, Garnett1999} mean that less material is available in the ISM to form the PAHs than to form the oxygen-rich silicate grains which make the bulk of the dust mass. The deficiency of PAHs in low-metallicity galaxies can also be explained by the delayed injection of carbon dust into the ISM by AGB stars \citep{Galliano2008}.
}

\subsection{Temperature distribution}\label{ssec:temp}

\cite{RemyRuyer2013} showed that dwarf galaxies harbour warmer dust and present a potentially broad dust SED peak. 
Warmer dust in dwarf galaxies had been discovered first with \iras\ \citep[e.g.][]{Helou1986, Hunter1989, MelisseIsrael1994, Galliano2003}, and this was confirmed later with \spitz\ \citep[e.g.][]{Galliano2005, Rosenberg2006, Cannon2006, Galametz2009}. More recent studies have confirmed that low-mass galaxies have broader IR SED peaks \citep[e.g.][]{Boselli2012, SmithD2012, Ciesla2014}, or show a flattening of their FIR SED slope compared to more massive systems \citep{Boselli2010b, Cortese2014a}. These studies also showed that the most actively star forming galaxies were the ones with the warmest dust and the broadest SED peak \citep{Boselli2010b, Ciesla2014}.

\begin{figure}[h!tbp]
\begin{center}
\includegraphics[width=8.8cm]{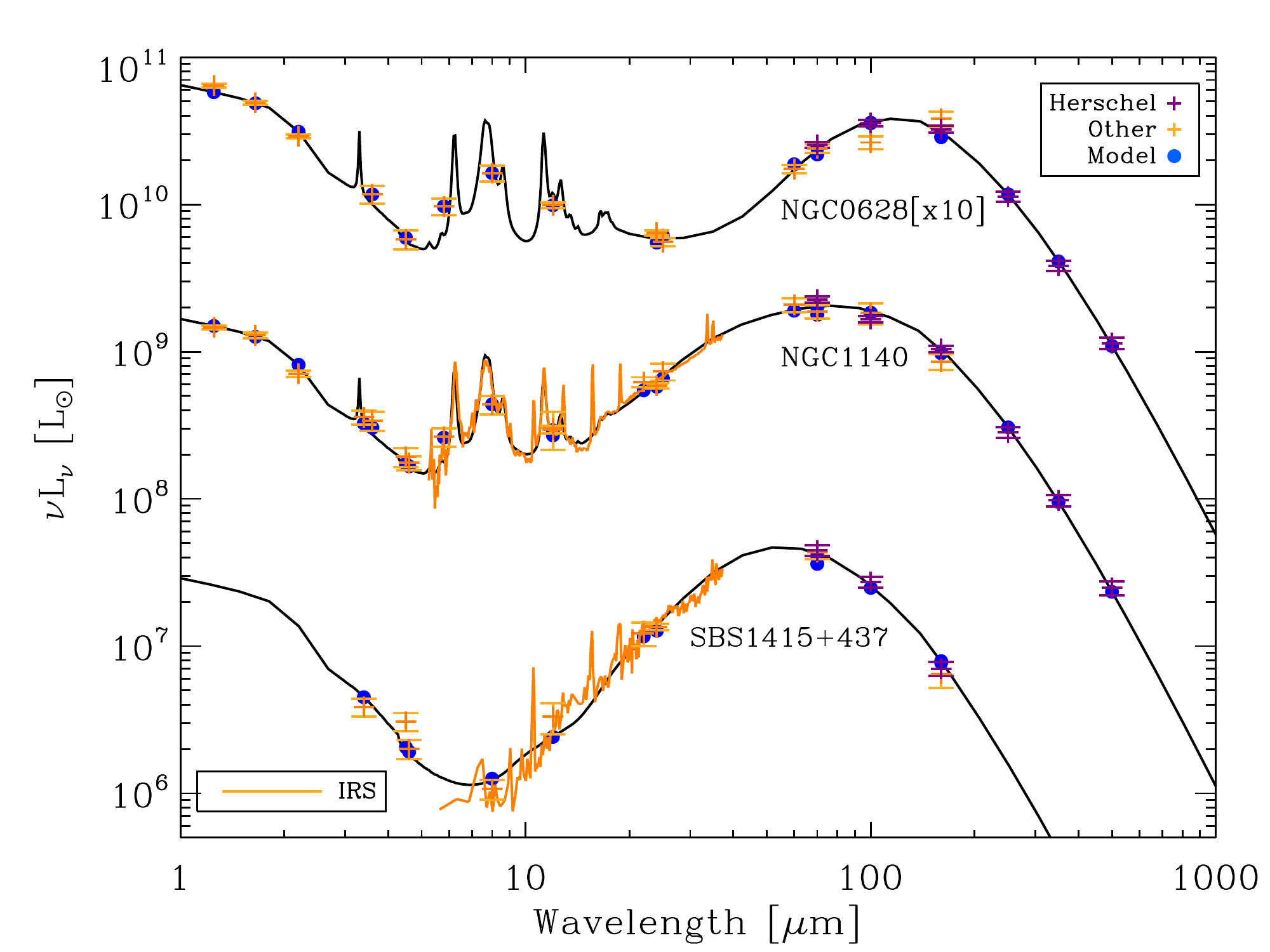}
\caption{Example of different SED shapes, with increasing broadness around the FIR peak for increasing \sigU. NGC~0628, NGC~1140 and SBS~1415+437 have log(\sigU/U$_\odot$)Ê= 0.4, 2.1 and 3.0, respectively. The location of these three galaxies is indicated in Fig. \ref{f:UavsU}.}
\label{f:sigUSEDs}
\end{center}
\end{figure}

\begin{figure*}[h!tbp]
\begin{center}
\includegraphics[width=16cm]{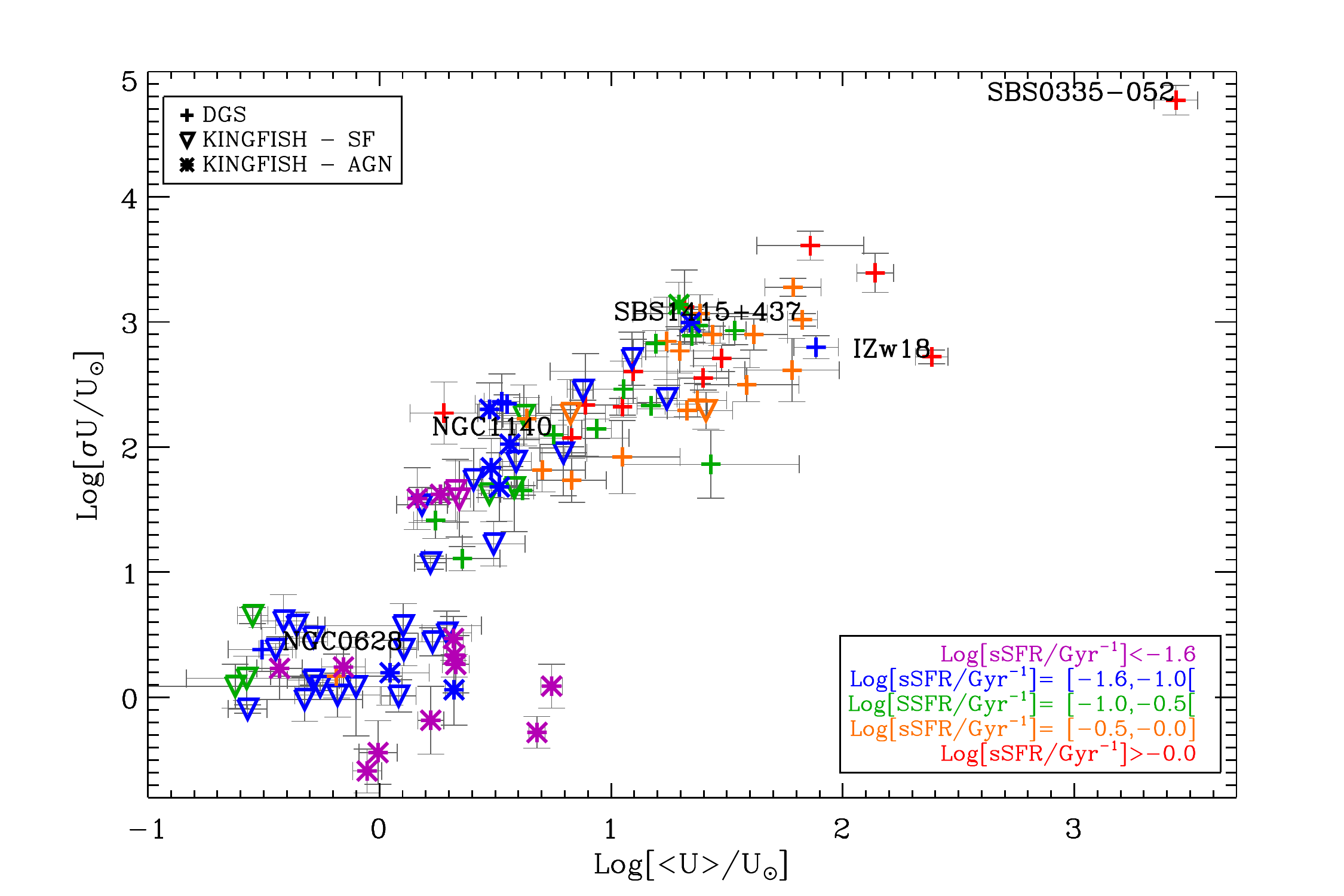}
\caption{\sigU\ as a function of the average starlight intensity \uav\ for the DGS (crosses) and KINGFISH samples. For the KINGFISH sample, we distinguish the sources that have been classified as ``star-forming'' (SF, downward triangles) or as ``AGN'' (stars) by \cite{Kennicutt2011}. The average dust equilibrium temperature is shown on the top axis. 
The colour codes the \ssfr\ of the sources, in units of Gyr$^{-1}$. 
We indicate the location of the two most metal-poor galaxies, I~Zw~18 and SBS~0335-052, and of the three galaxies of Fig. \ref{f:sigUSEDs}.}
\label{f:UavsU}
\end{center}
\end{figure*}

A temperature distribution is necessary to properly describe the dust emission (Sect. \ref{ssec:compBB}), and this dust temperature distribution is directly linked to the range of starlight intensities, $\Delta U$, to which the dust is exposed. However, to avoid being affected by potential degeneracies between the starlight intensity distribution parameters ($\Delta U$, \umin\ and $\alpha$) we consider the standard deviation of the starlight intensity distribution, \sigU, to describe the range of starlight intensities to which the dust is exposed. Fig. \ref{f:sigUSEDs} illustrates how \sigU\ traces the width of the SED peak on three examples. Fig. \ref{f:UavsU} shows the range of starlight intensities to which the dust is exposed, \sigU, as a function of the average starlight intensity \uav.

As explained in Sect. \ref{ssec:modelparam}, we added a MIR modBB to get a better match to the observed MIR SED for eleven galaxies. 
We add a delta-function describing this additional single temperature component to the U distribution in Eqs \ref{eq:U} and \ref{eq:sigU}: $\delta(U~-~U_{{\rm MIR,BB}}$), with $U_{{\rm MIR,BB}}$ = (T$_{{\rm MIR, BB}}$/T$_{{\rm MW}}$)$^6$. 
\uav\ covers a range of four orders of magnitude equivalent to $\sim$ 50 K in dust temperature, from 12 K to 64 K in the DGS; and from 12 K to 30 K in the KINGFISH sample. 
The median T$_{{\rm dust}}$ is 26 K for the DGS and 20 K for the KINGFISH sample, consistent with low-metallicity galaxies harbouring warmer dust than more metal-rich environments \citep{RemyRuyer2013}. The median \sigU\ is 290 U$_\odot$ for the DGS and 4 U$_\odot$ for the KINGFISH sample, reflecting the broader SED peak for low-metallicity environments compared to more metal-rich sources. 

Both \uav\ and \sigU\ strongly correlates with \ssfr\ (Table \ref{t:corr}), although the correlation is stronger with \sigU. The correlation of both parameters with metallicity is weaker than with \ssfr\ ($\rho$(\Ao,~\sigU)=-0.52, $\rho$(\Ao, \uav)=-0.47), and is a side effect of the strong correlation between \ssfr\ and metallicity ($\rho$(\Ao,~\ssfr)~=~-0.78). This holds whether we consider the full sample or low-metallicity sources on one side and metal-rich sources on the other side.
The \ssfr \ thus seems to be the parameter driving the dust SED shape.

In Fig. \ref{f:UavsU}, we clearly distinguish two clusters of points, one in the lower left corner of the plot, mostly metal-rich KINGFISH galaxies, with low \uav\ and low \sigU\ (``cold and narrow'' SEDs), and one mainly composed of low-metallicity galaxies, with higher \uav\ and \sigU\ (``hot and broad'' SEDs). However some high-metallicity sources can be found in the ``hot and broad" group, and vice versa. 
This is due to their higher (or lower) \ssfr\ as shown by the colours on Fig. \ref{f:UavsU}.
For example, the spiral galaxy NGC~0628 has the same metallicity than the star-forming dwarf galaxy NGC~1140 (Z = 0.5 \zsun) but very different SED shapes (Fig. \ref{f:sigUSEDs}) due to their different star-formation activity: log(\ssfr/Gyr$^{-1}$) = -1.4 for NGC~0628 and -0.8 for NGC~1140.
This confirms that the \ssfr\ is the parameter determining the dust SED shape, and that metallicity only plays a secondary role {\rev as also noted by \cite{daCunha2008, SmithD2012}}. Low-metallicity sources extend the trend outlined by metal-rich galaxies to warmer temperatures and broader temperature distributions due to their higher star-formation activity in a smooth transition rather than in a sharp change. This has been noted by \cite{Cortese2014a} on a smaller range of metallicities (0.6 dex).

In more active galaxies, the dust spans a wider temperature range translating into a broader dust SED peak. 
In a galaxy undergoing an active phase of star formation (e.g., with high \ssfr), the ISM will be clumpier as {\rev a large number of embedded star-forming clumps} are spread all over the galaxy. This clumpier ISM allows for a wider equilibrium temperature distribution of the dust grains, {\rev skewed towards higher dust temperatures} (and thus higher \sigU\ and \uav). Evidence for this clumpier structure of the ISM in dwarf galaxies can be directly seen from the resolved UV-to-mm observations of the LMC and SMC, and has also been suggested by \cite{Cormier2015} from a detailed study of the DGS FIR fine-structure cooling lines. This irregular ISM structure due to feedback processes related to star formation and supernov\ae\ events result in an ISM in which the dust grains are exposed to a range of stellar populations and thus show a larger distribution of dust temperatures.
The dwarf galaxies in our sample contain warmer dust primarily because they are intensively forming stars. Metallicity has a secondary impact on the dust temperature, as the low dust attenuation allows to heat the dust deeper within the molecular clouds.

In Fig. \ref{f:UavsU} we have also split the KINGFISH sample in the ``SF'' and ``low-luminosity AGN'' groups. The warmest KINGFISH galaxies (i.e., with high \uav) are mostly the KINGFISH low-luminosity AGNs. {\rev It might be possible that}, despite a small contribution to the total luminosity, the emission from the central AGN is powerful enough to impact the dust heating on global galaxy scales and to increase the average dust temperature, thus to have an impact on the global shape of the dust SED. Warmer dust in the presence of an AGN had already been seen in \cite{Kirkpatrick2012}. 

\section{Towards a comprehensive view of the dust properties in low-metallicity environments}\label{sec:compview}

After studying the different relationships between the dust properties and fundamental galaxy parameters, we now analyse our results in the context of galaxy evolution, and attempt to draw a consistent picture of the dust in low-metallicity environments.

\begin{figure*}[h!tbp]
\begin{center}
\includegraphics[width=14cm]{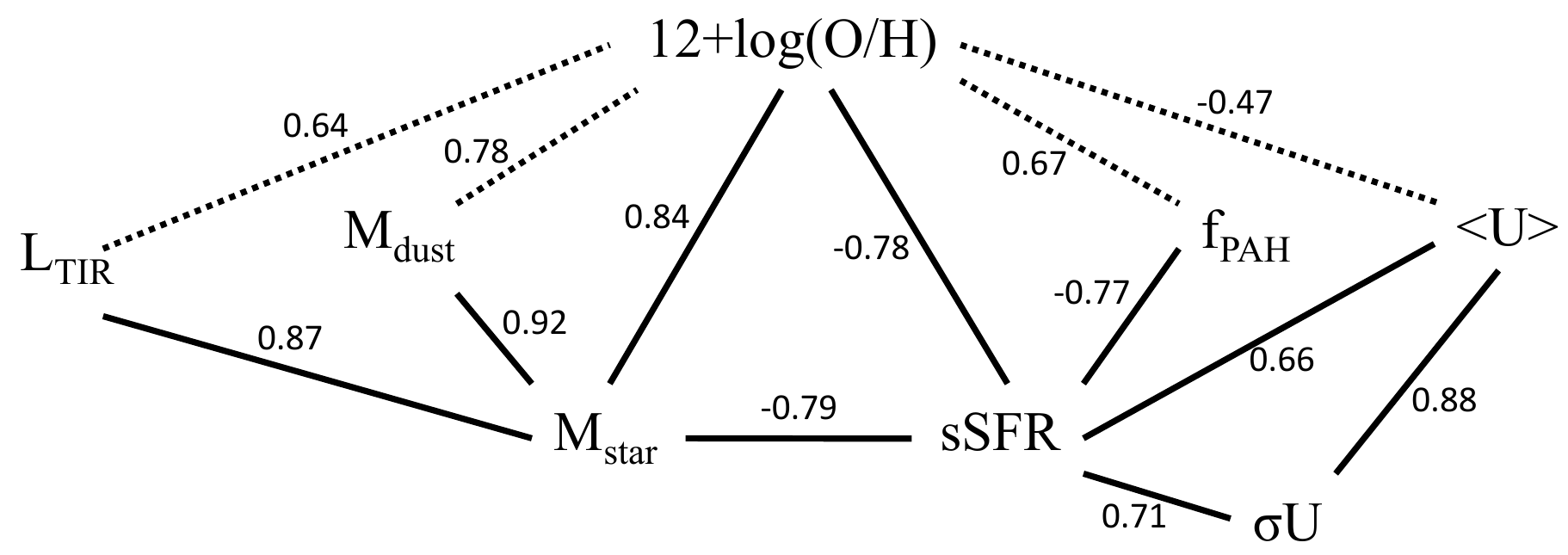}
\caption{Schematic view of the various correlations between parameters in the DGS and KINGFISH samples. The numbers on the lines are the Spearman correlation coefficients reported from Table \ref{t:corr}. The solid lines indicates the dominating correlations between parameters and the dashed lines indicates several secondary correlations of interest.}
\label{f:corr}
\end{center}
\end{figure*}

\begin{figure*}[h!tbp]
\begin{center}
\includegraphics[width=14cm]{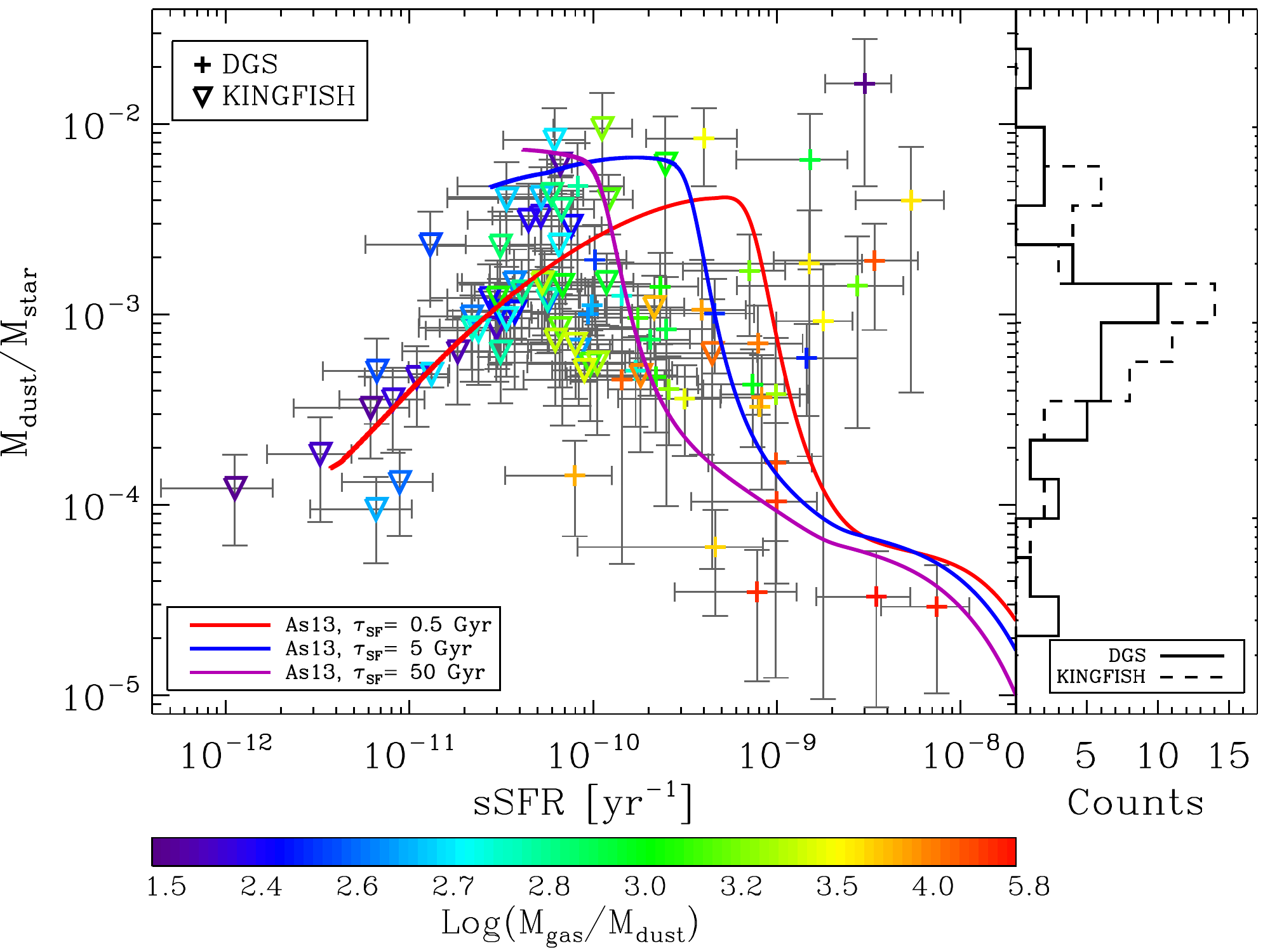}
\caption{Dust-to-stellar mass ratios for the DGS (crosses) and KINGFISH (downward triangles) samples as a function of \ssfr, colour coded by the gas-to-dust mass ratio (G/D), \mgas/\mdust, from \cite{RemyRuyer2014}. The distribution of \mdust/\mstar\ is indicated on the side for both samples: solid line for DGS and dashed line for KINGFISH. Modelled evolutionary tracks from \cite{Asano2013} are shown for different star-formation timescales ($\tau_{{\rm SF}}$ = 0.5, 5 and 50 Gyr) by the red, blue and purple lines.}
\label{f:dustar}
\end{center}
\end{figure*}

The most famous picture of galaxy evolution is the widely studied mass-metallicity relation. The ISM matter lifecycle implies that the metallicity of a galaxy increases as the galaxy evolves through several cycles of star formation, building up its stellar mass \citep{Lequeux1979, Tremonti2004, Kirby2013, Cook2014, Goncalves2014}. 
We saw that the mass-metallicity relation drives all of the extensive quantities, \mdustÊ\ and \ltir, while the intensive quantities \fpah, \uav\ and \sigU, are mostly driven by \ssfr. This is shown schematically in Fig. \ref{f:corr}. From this figure, is is clear that metallicity only plays a secondary role in shaping the dust properties. Remember however, that our sample is dominated at high metallicities by spiral galaxies and by star-forming, gas-rich dwarf galaxies at low metallicities. The correlations presented here thus may suffer from some selection bias. 
The dust-to-stellar mass ratio being a normalised (and thus intensive) quantity enables us to look at the dust mass build up with respect to the stellar mass. Fig. \ref{f:dustar} shows the dust-to-stellar mass ratios for our sample as a function of \ssfr.

The dust-to-stellar mass ratio has been extensively studied previously {\rev \citep[e.g.][]{daCunha2010, Skibba2011, Bourne2012, Cortese2012a, Smith2012a}}. These studies have shown that the dust-to-stellar mass ratio decreases for increasing stellar mass (or metallicity) and decreasing \ssfr. We do not find any correlation for the dust-to-stellar mass ratio with metallicity, stellar mass or \ssfr. However, considering each sample separately, we find a rather weak correlation between the dust-to-stellar mass ratio and metallicity ($\rho$(\mdust/\mstar, \Ao)=-0.44) or \ssfr\ ($\rho$(\mdust/\mstar, \ssfr)=0.47) for the KINGFISH sample, and still no correlations for the DGS sources : $\rho$(\mdust/\mstar, \Ao)=0.18 and $\rho$(\mdust/\mstar, \ssfr)=-0.04.
Thus the results for the metal-rich KINGFISH galaxies are in agreement with the findings of the previously mentioned studies. However, the low-metallicity sample does not extend the observed behaviour of the metal-rich galaxies to higher \ssfr.

The peculiar behaviour of the dust-to-stellar mass ratios for the low-metallicity DGS galaxies is due to their chemical evolutionary stage. This can be traced by the gas-to-dust-mass ratio \citep[G/D,][]{RemyRuyer2014}, and is shown by the colours in Fig. \ref{f:dustar}.
Modelled evolutionary tracks from the chemical evolution model of \cite{Asano2013} are also shown in Fig. \ref{f:dustar}, for different star-formation timescales ($\tau_{{\rm SF}}$ = 0.5, 5 and 50 Gyr). This chemical evolution model, based on models from \cite{Hirashita1999, Inoue2011}, includes dust production by stellar sources (AGB stars and Type II SNe) and by dust growth processes in the ISM. Dust is destroyed by SN shocks. \cite{RemyRuyer2014} showed that this model can successfully reproduce the observed trend between G/D and metallicity, thanks to the dust growth mechanism.

At high \ssfr, thus low-metallicity and low stellar masses, the dust content is very low giving very low \mdust/\mstar\ and high G/D. The dust production is controlled there by stellar sources only. Then when a critical metallicity is reached in the ISM, dust growth by metal accretion on the dust grains in the ISM becomes the major process for building up the dust mass \citep{Asano2013, Zhukovska2014} and the dust mass rapidly increases without significant consumption of the gas reservoir or star formation. This rapid increase of the dust mass results in an increase of the dust-to-stellar mass ratio and a decrease of the G/D \citep{RemyRuyer2014}. The large scatter in the dust-to-stellar mass ratio for the highly star forming galaxies in Fig. \ref{f:dustar} is thus due to their different evolutionary stage, as traced by their G/D.
Then dust growth processes saturate when all the available metals are locked up in the grains. In the meantime, star formation continues, consuming the gas reservoir and increasing the stellar mass. This results in a decreasing \ssfr, decreasing dust-to-stellar mass ratio and decreasing G/D.

The nucleus type (star-forming or AGN) also has an impact in shaping the dust properties \citep{Fritz2006, Magdis2014, Ciesla2015}, meaning that the picture drawn from Figs. \ref{f:corr} and \ref{f:dustar} is far from being that simple. \ssfr\ and metallicity are not the only parameters involved in the evolution of the dust properties in galaxies. All of these fundamental parameters can also be affected by inflows and outflows, rendering more complex in practice the relations between stellar mass, star formation, metallicity and dust properties, and can change significantly the position of a galaxy in the \mdust/\mstar - \ssfrÊ\ plot.

Moreover, the dust masses are estimated here with the same dust mixture, using silicates, graphite and PAH grains, for all of the galaxies. However, amorphous carbon grains could be used instead of graphite grains to model the carbonaceous dust \citep[][and see Section \ref{ssec:amcar}]{Galliano2011, Galametz2013a, Jones2013}. The silicate grain properties used in our model are derived from empirically-normalised astrophysical data, and more realistic types of silicate grains could also be used \citep{Kohler2014}. Changing the dust composition results in different dust properties, e.g., the dust mass or dust temperature as we saw with amorphous carbons in Section \ref{ssec:amcar}. The dust composition can also vary from galaxy to galaxy, which would in turn impact the derived dust masses, and the observed trends. 
Thanks to the wealth of spatially resolved dust studies with \hersc, we know now that the dust composition and size distribution also vary {\it within} each galaxy \citep{Smith2012b, Galametz2012, Mattsson2014b, Viaene2014, Ysard2015}. Dust evolves from diffuse to denser regions where bigger grains or aggregates can eventually form \citep{Kohler2015}, thus impacting the SED shape, and the dust parameters. Studying the impact of a varying dust composition and grain size distribution within and between galaxies on the SED and the derived dust properties is now the next step.


\section{Conclusion}

In this paper, we provided a coherent picture of the evolution of the dust properties from metal-poor to metal-rich environments. 
Our sample comprises 109 galaxies, spanning almost 2 dex in metallicity, dominated by spiral galaxies at high metallicities and by star-forming gas-rich dwarf galaxies at low metallicities. Observed SEDs are gathered over the whole IR-to-submm wavelength range, with constraints from 2MASS, \spitz, \wise, \iras, and \hersc. The full data set is presented here for the DGS sample of dwarf galaxies. 
The dust properties (namely, dust mass, TIR luminosity, PAH mass fraction and dust temperature distribution) are derived in a systematic way using a realistic semi-empirical dust SED model from \cite{Galliano2011}, and then compared to fundamental galaxy parameters: stellar mass, metallicity and \ssfr\ for a final sample of 98 galaxies. 

The dust mass is a critical parameter for constraining chemical evolution models. We showed here that different model assumptions could greatly impact the estimated dust mass. 
A single-temperature modBB model underestimates the dust mass by a factor of 2 compared to a semi-empirical SED model, even with careful matching of the effective optical properties in the modBB. The modBB model overestimates the effective average dust temperature
thus leading to the underestimation of the dust mass, as also shown by \cite{Bendo2015}.
Changing the carbonaceous component in the dust mixture from graphite grains to amorphous carbon grains results in a decrease of the dust mass estimate by a factor 2.5. Amorphous carbon grains are more emissive than graphite grains, so less dust is needed to account for the same luminosity.

We find an excess at 500 \micÊ\ for eight galaxies in our sample. The excess is rather small ($\leq$ 1.5 $\sigma$) as the transition from thermal dust and submm excess emission observed at longer wavelengths occurs around 500 \mic. Including the 500 \micÊ\ point during the fitting procedure does not result in a drastic overestimation of the dust mass for the excess galaxies. However, leaving the 500 \micÊ\ data point out of the fit results in an underestimation of the dust mass by a factor 2 to 4 for 15\% of the galaxies in our sample, because the cold dust component is then not properly constrained.
Estimating dust masses is thus subject to non-negligible systematic modelling uncertainties.

We present various ways of estimating \ltir\ in the DGS sample with mono/polychromatic indicators, using the calibrations presented in \cite{Galametz2013b}. For galaxies with similar metallicity and star-formation activity, we recommend using the 70, 24-160, or 24-70-160 calibrations of \cite{Galametz2013b}.

The dust temperature distribution and PAH mass fraction are primarily driven by the \ssfr, with a second order effect from metallicity. In our sample, PAHs are detected down to Z $\sim$ 1/12 \zsun. Low PAH abundances in dwarf galaxies are a consequence of the high star-formation activity and low dust attenuation. 
This combined effect of \ssfr\ and metallicity is also responsible for the higher average dust temperatures in starbursting low-metallicity dwarf galaxies. We find a median dust temperature (derived from the average starlight intensity \uav) for the DGS sample of $\sim$26~K, and $\sim$20~K in KINGFISH galaxies. The higher star-formation activity results in a clumpier ISM allowing for a larger range of dust equilibrium temperatures. 

The extensive dust properties, \mdust\ and \ltir, are driven by the mass-metallicity relation, reflecting a scaling effect: the more massive galaxies contain more dust, are more luminous, and are also more metal-rich. However, the dust mass build up with respect to stellar content is not the same in highly star-forming, low-metallicity sources than in more metal rich systems. The dust-to-stellar mass ratios of metal-rich sources follow an increasing trend of \mdust/\mstar\ with \ssfr, previously observed on other samples \citep{daCunha2010, Skibba2011, Cortese2012a}. On the other hand, for the more actively star forming galaxies (\ssfr\ $>$ 0.1 Gyr$^{-1}$) the trend is far less clear, with increasing scatter. The peculiar behaviour of the low-metallicity sources is driven by their chemical evolutionary stage: at low metallicity and high star-formation activity , the dust production is dominated by stellar sources only. After a critical metallicity is reached, the dust-to-stellar mass ratio rapidly increases because dust growth processes in the ISM dominate the dust production. Then dust growth saturates while star formation goes on, forming stars from the gas reservoir, resulting in a decreasing \ssfr\ and \mdust/\mstar. This completes and confirms our results on the gas-to-dust mass ratios derived in \cite{RemyRuyer2014}. Others effects such as inflows or outflows, presence of an AGN, or varying dust composition between and within galaxies have not been considered here but can also affect our picture of the evolution of the dust properties in galaxies.


\begin{acknowledgements}

The authors would like to thank R. Asano for providing the evolutionary tracks for his model for Fig. \ref{f:dustar}, and H. Hirashita and T. Takeuchi for interesting discussions on the chemical evolution models.
This research was made possible through the financial support of the Agence Nationale de la Recherche (ANR) through the programme SYMPATICO (Program Blanc Projet ANR-11-BS56-0023) and also through the EU FP7 funded project DustPedia (Grant No.606847). IDL is a postdoctoral researcher of the FWO-Vlaanderen (Belgium). \\
This research has made use of the NASA/IPAC Extragalactic Database (NED) and of the NASA/ IPAC Infrared Science Archive which are operated by the Jet Propulsion Laboratory, California Institute of Technology, under contract with the National Aeronautics and Space Administration.

PACS has been developed by MPE (Germany); UVIE (Austria); KU Leuven, CSL, IMEC (Belgium); CEA, LAM (France); MPIA (Germany); INAF-IFSI/OAA/OAP/OAT, LENS, SISSA (Italy); IAC (Spain). This development has been supported by BMVIT (Austria), ESA-PRODEX (Belgium), CEA/CNES (France), DLR (Germany), ASI/INAF (Italy), and CICYT/MCYT (Spain). SPIRE has been developed by Cardiff University (UK); Univ. Lethbridge (Canada); NAOC (China); CEA, LAM (France); IFSI, Univ. Padua (Italy); IAC (Spain); SNSB (Sweden); Imperial College London, RAL, UCL-MSSL, UKATC, Univ. Sussex (UK) and Caltech, JPL, NHSC, Univ. Colorado (USA). This development has been supported by CSA (Canada); NAOC (China); CEA, CNES, CNRS (France); ASI (Italy); MCINN (Spain); Stockholm Observatory (Sweden); STFC (UK); and NASA (USA).

SPIRE has been developed by a consortium of institutes led by Cardiff Univ. (UK) and including: Univ. Lethbridge (Canada); NAOC (China); CEA, LAM (France); IFSI, Univ. Padua (Italy); IAC (Spain); Stockholm Observatory (Sweden); Imperial College London, RAL, UCL-MSSL, UKATC, Univ. Sussex (UK); and Caltech, JPL, NHSC, Univ. Colorado (USA). This development has been supported by national funding agencies: CSA (Canada); NAOC (China); CEA, CNES, CNRS (France); ASI (Italy); MCINN (Spain); SNSB (Sweden); STFC, UKSA (UK); and NASA (USA).

\end{acknowledgements}


\bibliographystyle{aa}
\bibliography{mybiblio.bib}

\Online

\section{SED fits}\label{sec:SEDall}

\begin{figure*}[h!tbp]
\begin{center}
\includegraphics[width=17cm]{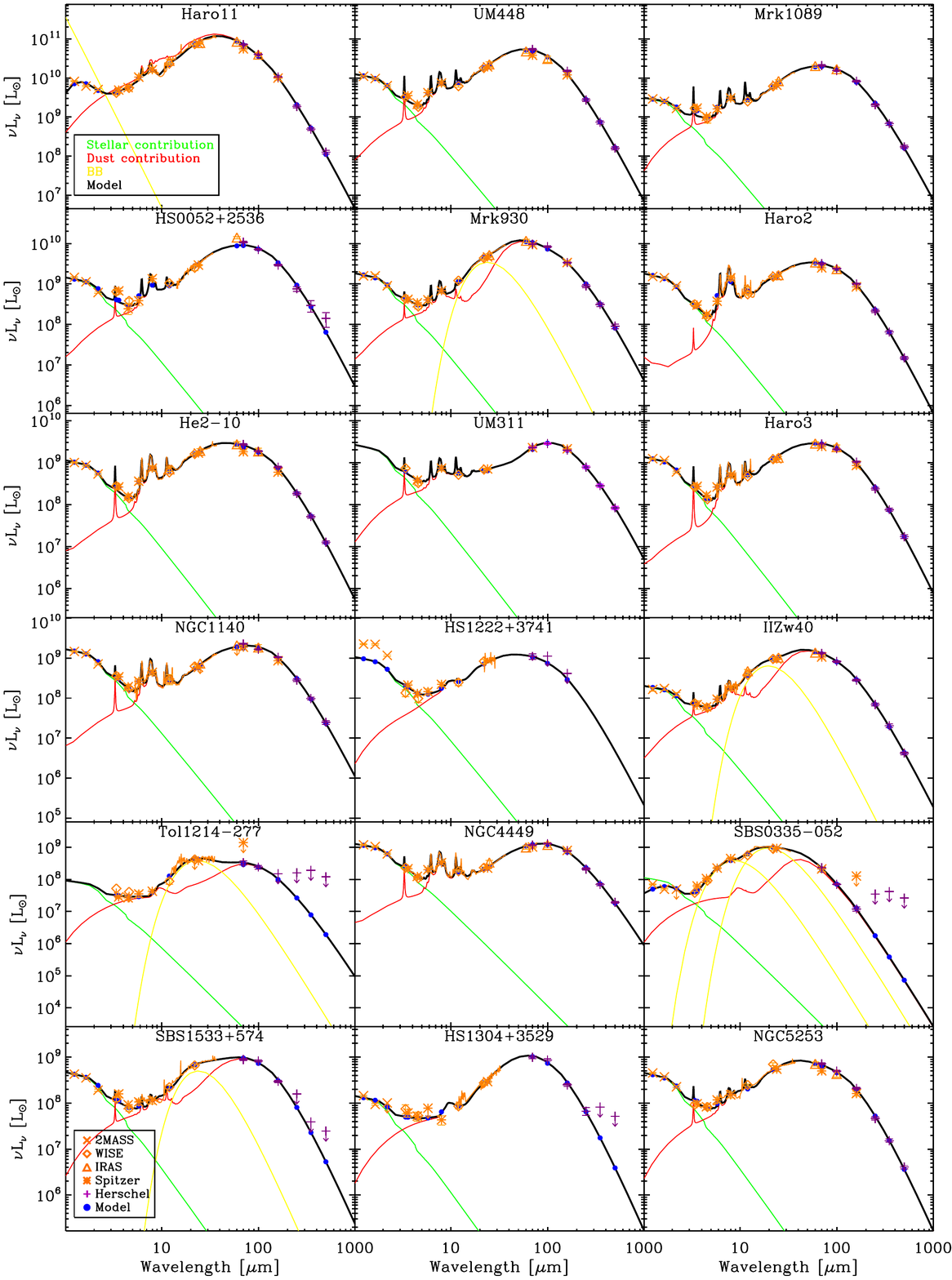}
\caption{DGS SEDs: The observed SEDs include the \hersc\ data (purple crosses) as well as any available ancillary data (in orange). The different symbols refer to the different instruments: Xs for 2MASS bands, stars for \spitz\ IRAC and MIPS, diamonds for WISE and triangles for IRAS. The IRS spectrum is displayed in orange. The total modelled SED in black is the sum of the stellar (green) and dust (red) contributions, and eventually of the modified blackbody contribution (yellow). The modelled points in the different bands are the filled blue circles.}
\label{f:SED_DGSall}
\end{center}
\end{figure*}

\addtocounter{figure}{-1}
\begin{figure*}[h!tbp]
\begin{center}
\includegraphics[width=18cm]{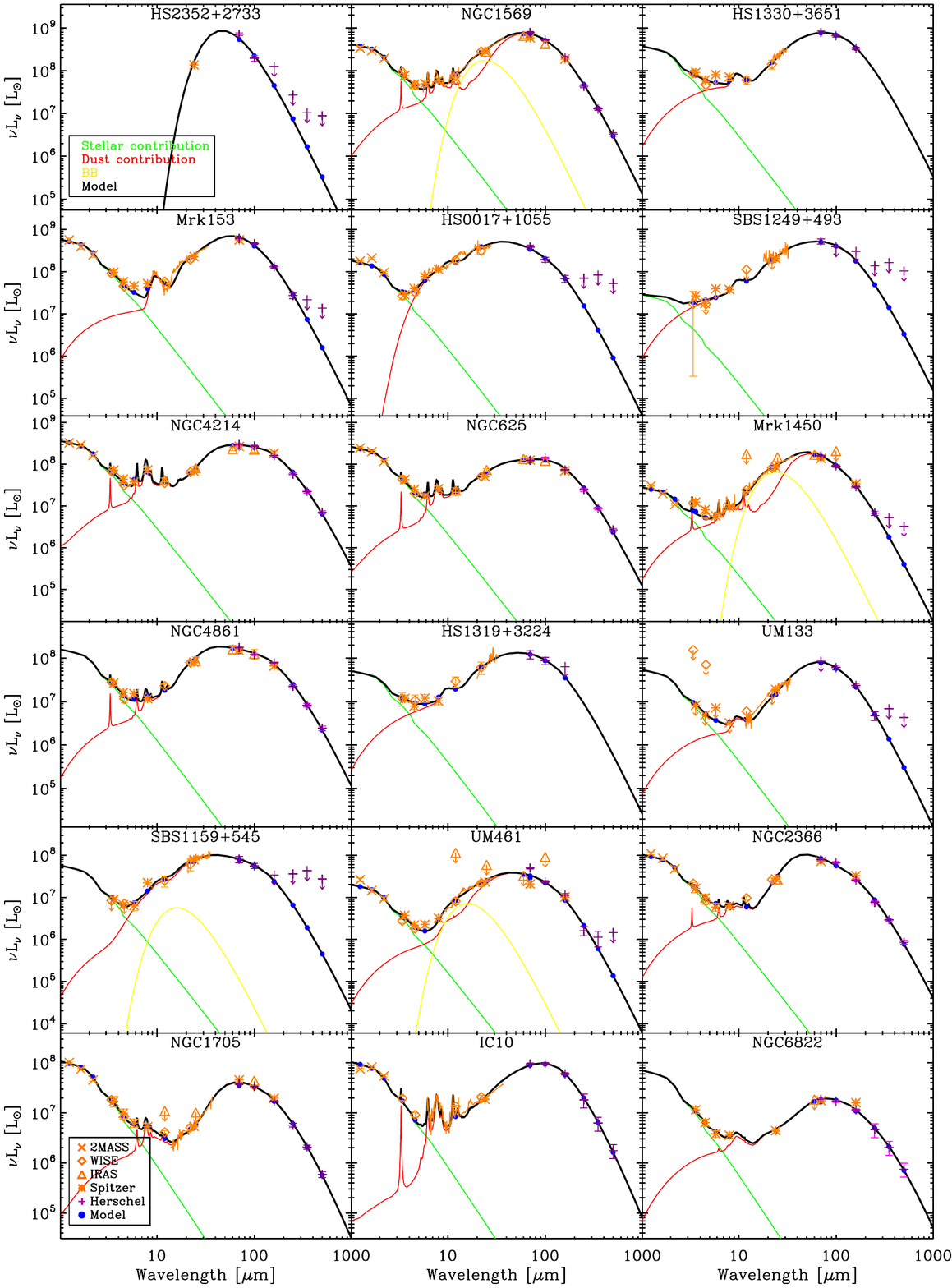}
\caption{DGS SEDs (continued).}
\end{center}
\end{figure*}

\addtocounter{figure}{-1}
\begin{figure*}[h!tbp]
\begin{center}
\includegraphics[width=18cm]{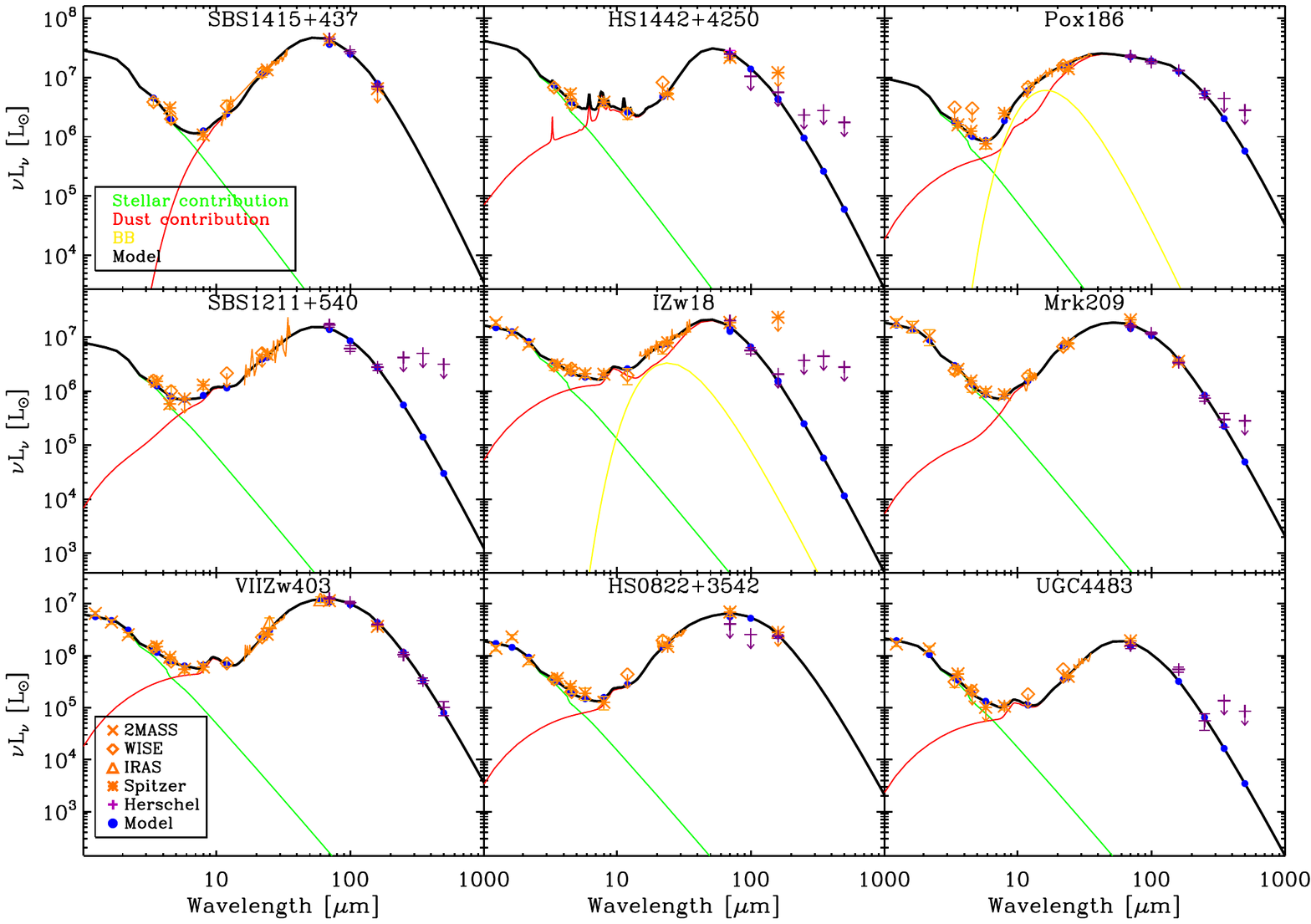}
\caption{DGS SEDs (continued).}
\end{center}
\end{figure*}

\begin{figure*}[h!tbp]
\begin{center}
\includegraphics[width=18cm]{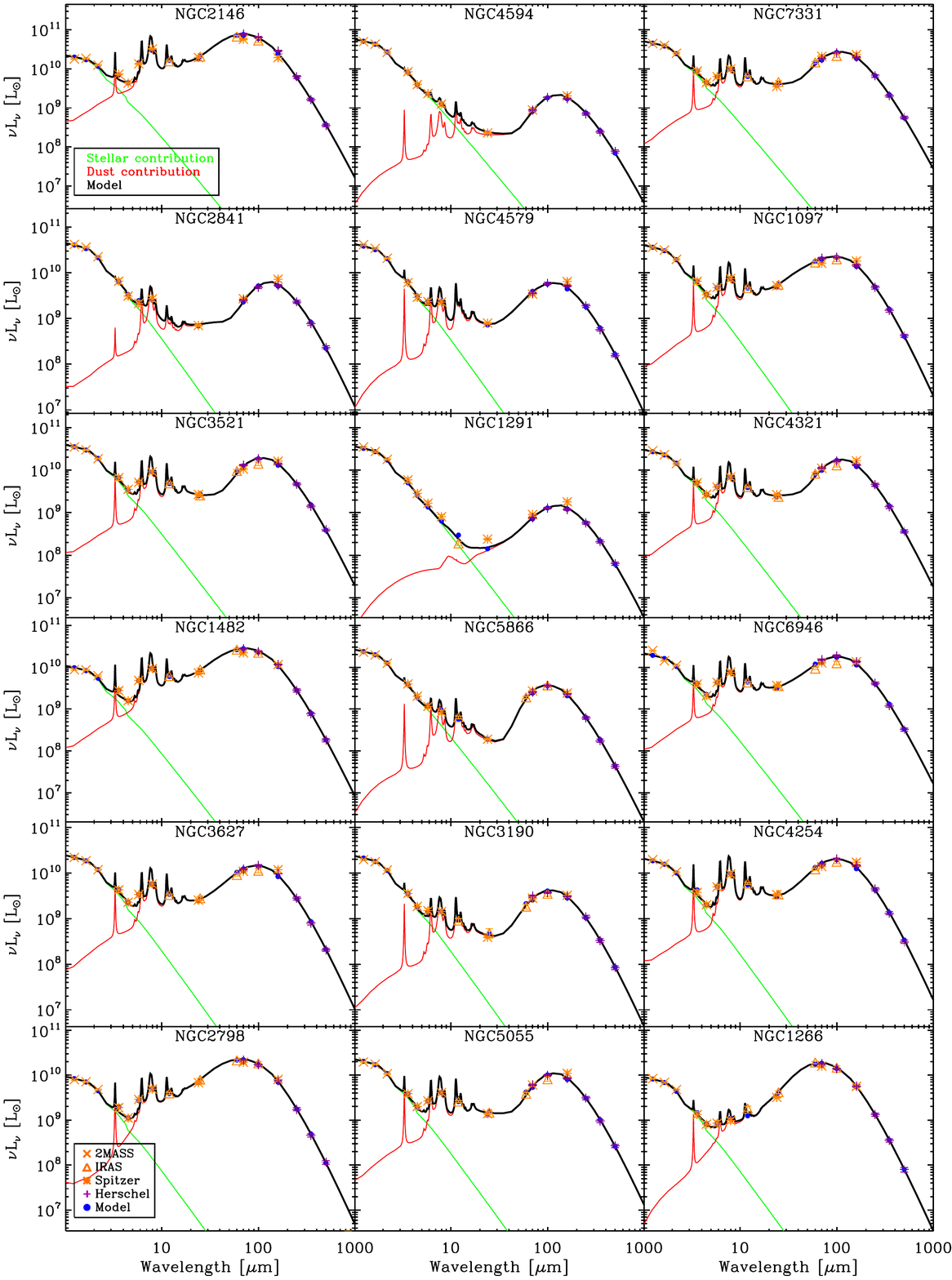}
\caption{KINGFISH SEDs: The colors and symbols are the same as in Fig. \ref{f:SED_DGSall}.}
\label{f:SED_KFall}
\end{center}
\end{figure*}

\addtocounter{figure}{-1}
\begin{figure*}[h!tbp]
\begin{center}
\includegraphics[width=18cm]{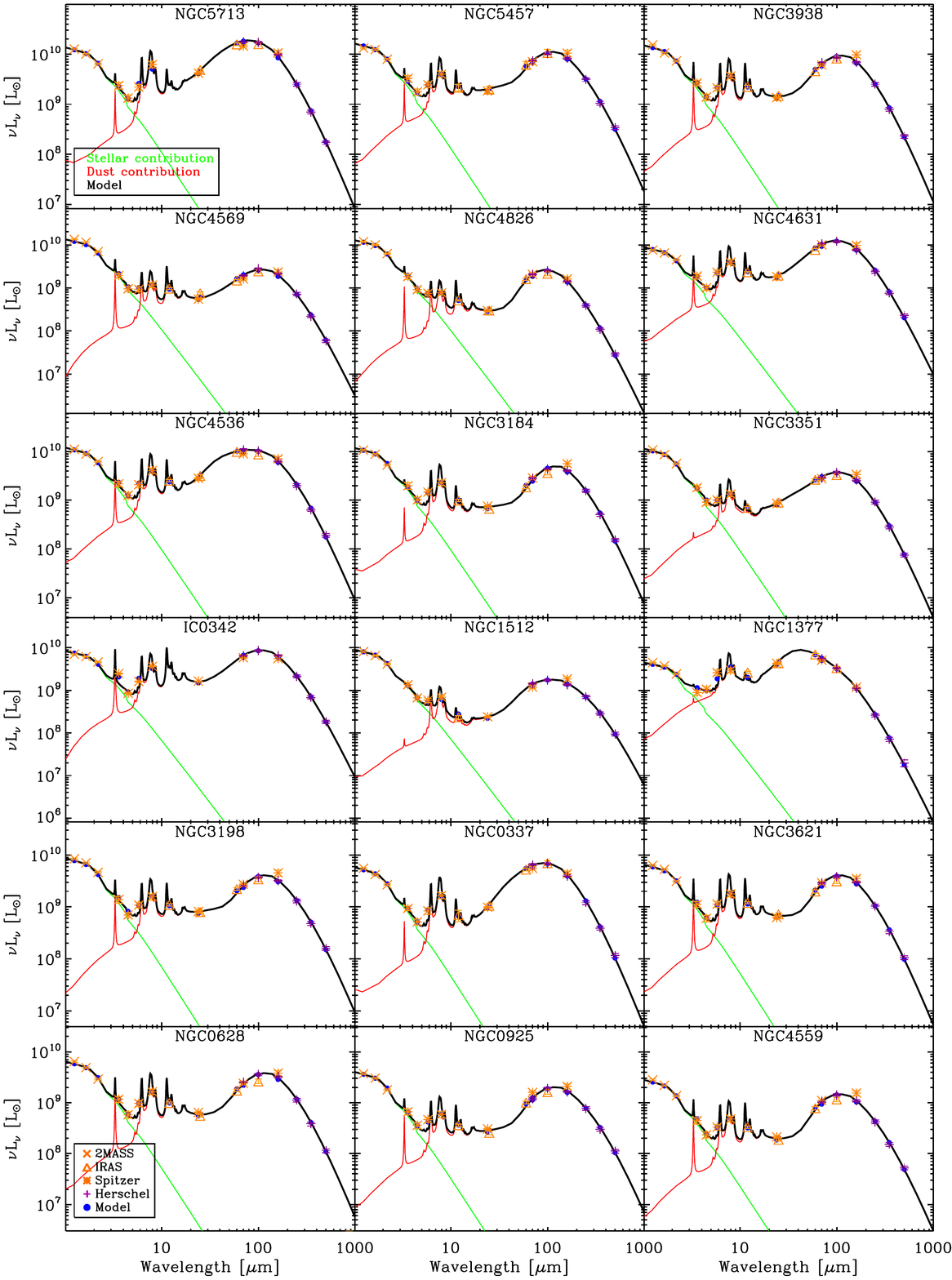}
\caption{KINGFISH SEDs (continued).}
\end{center}
\end{figure*}

\addtocounter{figure}{-1}
\begin{figure*}[h!tbp]
\begin{center}
\includegraphics[width=18cm]{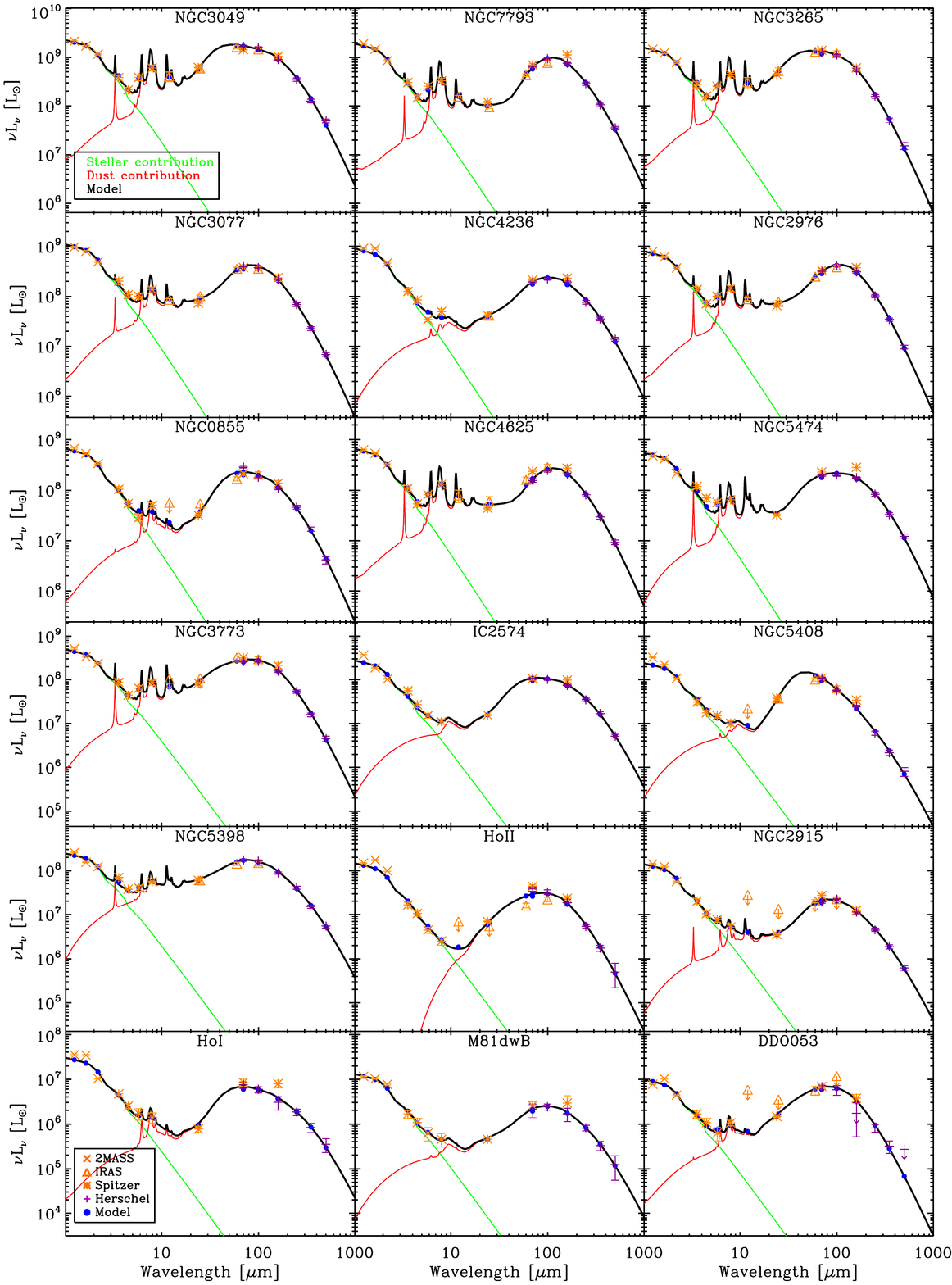}
\caption{KINGFISH SEDs (continued).}
\end{center}
\end{figure*}

\section{Data Tables}

\include{SFRparam_DGS_KF_fin}

\include{Herschel_fluxes_fin}
\include{IRAC_fluxes_fin}
\include{WISE_fluxes_fin}
\include{Literature_fluxes_fin}

\include{Dustparam_DGS_KF_fin}


\appendix

\section{Update of the \hersc\ data}\label{app:updateHersc}

\subsection{PACS}
In \cite{RemyRuyer2013}, the PACS data for the DGS galaxies were processed either with {\sc PhotProject} (provided by HIPE) for point sources or with {\sc Scanamorphos} \citep[version 14,][]{Roussel2013} for extended sources. The update of {\sc Scanamorphos} to version 23 implies a significant update on the red band at 160 \mic\ due to the PACS distortion flat-field that is now taken into account. This  effect was already included in {\sc PhotProject}. 
For the DGS data, the {\sc Scanamorphos} maps are thus reprocessed with v23 of the algorithm for the three PACS bands for consistency. The photometry is done with the same method as in \cite{RemyRuyer2013}, and using the exact same apertures. There are three changes to note : \\

{\it - NGC~1705 -} The v14 maps of {\sc Scanamorphos} for this galaxy presented a non-uniform background which compelled us to choose the {\sc PhotProject} maps in \cite{RemyRuyer2013} (see note $^f$ in their Table 2). This has been corrected in the new version of the maps and so we now consider the {\sc Scanamorphos} maps for this galaxy. 

{\it - UGC~4483 -} The new {\sc Scanamorphos} maps now manage to reconstruct better the emission for this galaxy at 160 \mic\ yielding a strong detection (9$\sigma$) in the new map. Moreover, this detection at 160 \mic\ is more coherent with the rest of the FIR/submm photometry (detection at 250 \mic\ by SPIRE). So we now consider the {\sc Scanamorphos} maps for this galaxy and not the {\sc PhotProject} maps as in \cite{RemyRuyer2013}. 

{\it - UM~133 -} For this galaxy \cite{RemyRuyer2013} noted that the 70 \mic\ flux density might present some discrepancies with others FIR measurements. Using {\sc PhotProject} maps instead of {\sc Scanamorphos} maps yields a 70 \mic\ flux density which is more consistent with the FIR shape of the SED. So we now consider the {\sc PhotProject} maps for this galaxy. \\

The new flux densities for the DGS are presented in Table \ref{t:herschel}. For the KINGFISH PACS data, we applied a corrective factor of 0.925 to the 160 \mic\ flux densities of \cite{Dale2012} (H. Roussel, priv. com.). 

\subsection{SPIRE}
There have been two different updates for the SPIRE data : the calibration and the beam model used to compute the beam areas\footnote{The beam areas are needed to convert the original data in Jy/beam to Jy/pix for the aperture photometry.}.

The SPIRE maps for the DGS have been reprocessed with HIPE v12, as in \cite{RemyRuyer2013}. 
We {\rev apply the same method as in \cite{RemyRuyer2013} to redo the photometry on the new data, depending if the source is extended or point like}. For the extended sources, there are two steps to apply before performing aperture photometry: convert the data from point-source calibration to extended-source calibration, via the $K_{PtoE}$ factor\footnote{This factor was denoted $K_{4e}/K_{4p}$ in \cite{RemyRuyer2013}.}, and convert the data from Jy/beam to Jy/pix, using the beam area. 
The values for the $K_{PtoE}$ factor, given the SPIRE Data Reduction Guide\footnote{The SPIRE Data Reduction Guide, v3.0, is available at: http://herschel.esac.esa.int/hcss-doc-12.0/print/spire\_drg/spire\_drg.pdf\\ \#spire\_drg.}, are 0.99858, 1.00151 and 0.99933 at 250, 350 and 500 \mic\ respectively. For the ``slightly extended'' sources \citep[see][]{RemyRuyer2013} we do not apply this $K_{PtoE}$ factor.
The beam areas also depend on the spectral shape of the considered source, which means that the beam areas will be {\it different} for each galaxy. To produce the final maps, the SPIRE pipeline assumes a spectral shape with a dependence $F_\nu \propto \nu^{-1}$, which corresponds to beam areas of 465, 823 and 1769 arcsec$^2$ at 250 \mic, 350 \mic\ and 500 \mic\ respectively. These are the beam areas we used for the second step \cite[identical to those used in][]{RemyRuyer2013}, before performing aperture photometry.

To account for the different FIR spectral shapes and thus beam areas, we need to apply a colour correction factor. 
The colour correction factors, $K_{col}$, have been tabulated in the SPIRE Data Reduction Guide for spectral shapes with $F_\nu \propto \nu^{\alpha}$. For point sources and extended sources, the $K_{col}$ are given in Table 6.13 of the SPIRE Data Reduction Guide. For slightly extended sources, we apply the point source colour correction times the effective beam area ratios tabulated in Table 6.8 of the SPIRE Data Reduction Guide. Note that this effective beam area ratio is {\it already} taken into account in the colour correction for extended sources.
The method we use to colour correct our flux densities is the following: we fit a line in logarithmic space to the three SPIRE flux densities to find the spectral slope $\alpha$. We include PACS 160 \mic\ data if there are some SPIRE non-detections. Once $\alpha$ is determined, we apply the corresponding colour correction factors to the flux densities. We iterate these few steps until $\alpha$ does not vary by more than 0.1\%. Usually two iterations are enough.

In Table \ref{t:herschel}, we give the updated fluxes densities {\it before} the colour correction step. 
On the other hand, the colour-corrected flux densities are the ones used in the fitting procedure. The colour corrections are of the order of 6\%, 4\% and 4\% at 250 \mic, 350 \mic, and 500 \mic\ respectively for the DGS. \\

The SPIRE maps of \cite{Dale2012} for the KINGFISH sample were reduced with HIPE v5, and aperture photometry was performed on the maps to extract the flux densities. 
Since then, there have been two updates of the calibration with the releases of v7 and v11 of HIPE. To account for the update at v7 we multiply the flux densities from \cite{Dale2012} by 1.0, 1.0067 and 1.0\footnote{G. J. Bendo, priv. com.} at 250, 350 and 500~\mic\ respectively. To account for v11 update, we multiply the flux densities by 1.0253, 1.0250 and 1.0125$^{23}$ at 250, 350 and 500 \mic\ respectively. 
No $K_{PtoE}$ factor had previously been applied to the flux densities (D. Dale, priv. com.). Assuming that the KINGFISH sources are all extended, we also apply the $K_{PtoE}$ factor to all of the galaxies. 
We also convert the fluxes to match the beam areas mentioned previously (465, 823 and 1769 arcsec$^2$). Then these updated flux densities are colour corrected following the method described for the DGS galaxies above. The colour corrections are of the order of $\sim$ 5\% for the three SPIRE wavelengths for the KINGFISH galaxies.

\section{DGS IRAC and IRS Observing Logs}\label{ap:ObsLogs}

\begin{table}[h!tbp]
\begin{center}
\caption{DGS \spitz/IRAC and IRS Observing Log.}
\label{t:IRSObsLog}
{\small
 \begin{tabular}{lc | cc}
\hline
\hline
 & {\bf IRAC} & \multicolumn{2}{c}{{\bf IRS}} \\
\hline
Sources & AOR key & AOR key & Extraction  \\
\hline
Haro 11        & 4326400 &  9007104  & Optimal  \\
Haro 2        & 5539840 & 9489920  &   Map  \\
Haro 3         &  11180288 &   12556288 & Tapered  \\
He 2-10        & 4329472 &  4340480 & Tapered  \\
HS 0017+1055   & 26387200 &   26393344 & Optimal  \\
HS 0052+2536   & 26387456 &  17463040  & Optimal  \\
HS 0822+3542   & 4328960 & 1763808  & Optimal   \\
HS 1222+3741   & 17564928 & 26393600 & Optimal  \\
HS 1236+3937   &  26387712 & 26393856   & Optimal  \\
HS 1304+3529   &   26387968 & 26394112  & Optimal  \\
HS 1319+3224   & 26388480 & 26394624  & Optimal    \\
HS 1330+3651   & 26388736 & 26394880  & Optimal  \\
HS 1442+4250   & 10388480 & 12562944 & Optimal   \\
HS 2352+2733$^a$ & 26388992 & 26395136  & -   \\
I Zw 18         & 4330752 &  16205568  & Optimal \\
IC 10          & 4424960 & 26396672 &  Map  \\
II Zw 40        &4327936 & 9007616 & Optimal  \\
Mrk 1089    & 11250432 & 26395392 & Tapered  \\
Mrk 1450       & 4334336 & 16206080 (SL),  & Optimal  \\
			&		& 9011712 (LL) 	& \\
Mrk 153        & 4333056 & 4342272  & Optimal \\
Mrk 209        & 22556672 & 12557568  & Optimal  \\
Mrk 930        & 4338944 & 4344320  & Tapered  \\
NGC 1140       & 4327168 & 4830976 & Tapered  \\
NGC 1569       &4434944 &  3856640 & Tapered   \\
NGC 1705   & 5535744 & 9513216  & Map  \\
NGC 2366       & 4436480 & 21920768 & Map  \\
NGC 4214$^b$       & 4457984 &  -   & -  \\
NGC 4449       & 4467456 & 26396928 &  Map \\
NGC 4861$^c$       & 4337408 &   -  & -  \\
NGC 5253       & 4386048 & 4386304   & Map  \\
NGC 625$^c$        &  22520064 &  -   &- \\
NGC 6822$^{b,c}$       & 5507072 &  -  &-  \\
Pox 186        & 26389248 &  12629760 & Optimal   \\
SBS 0335-052   & 4327424 & 8986880  & Optimal   \\
SBS 1159+545   & 26389504 & 9008896  & Optimal  \\
SBS 1211+540   &26389760 & 26395392 & Optimal   \\
SBS 1249+493   & 26390016 &  26395904 & Optimal  \\
SBS 1415+437   & 10392832 & 12562432 (SL),  & Optimal  \\
			&		& 8990464 (LL)	& \\
SBS 1533+574   & 17563904 & 8996352 & Optimal  \\
Tol 0618-402   & 4328448 & 8090624 & Optimal    \\
Tol 1214-277   & 4336384 & 9008128 & Optimal  \\
UGC 4483       &4329728 & 26396160  & Optimal   \\
UGCA 20$^a$        &26390272 &  26396416  &-   \\
UM 133        &26390528 &  21922304  & Map  \\
UM 311$^b$         & 10392576 &  - &-  \\
UM 448         & 4334592 & 4342784 & Tapered    \\
UM 461         & 4335104 & 16204032 (SL),   & Optimal   \\
			&		& 9006336 (LL)	& \\
VII Zw 403      &4334080 & 9005824   & Tapered  \\
\hline
\end{tabular}
}
\end{center}
\footnotesize{
{\bf Notes : } \\
$^a$: For these galaxies, the IRS slits are not centred on the source position and thus we do not present any IRS spectrum. \\
$^b$: For these galaxies, only local pointings were performed and we cannot present an IRS spectrum for the total galaxy. \\
$^c$: Only high resolution IRS spectra (SH and/or LH) are available. \\
}
 \end{table}

\section{Quality checks for IRAC and WISE DGS photometry}\label{app:compMIR}

\subsection{IRAC: Comparison to previous literature measurements}

\begin{figure*}[h!tbp]
\begin{center}
\includegraphics[width=9.0cm]{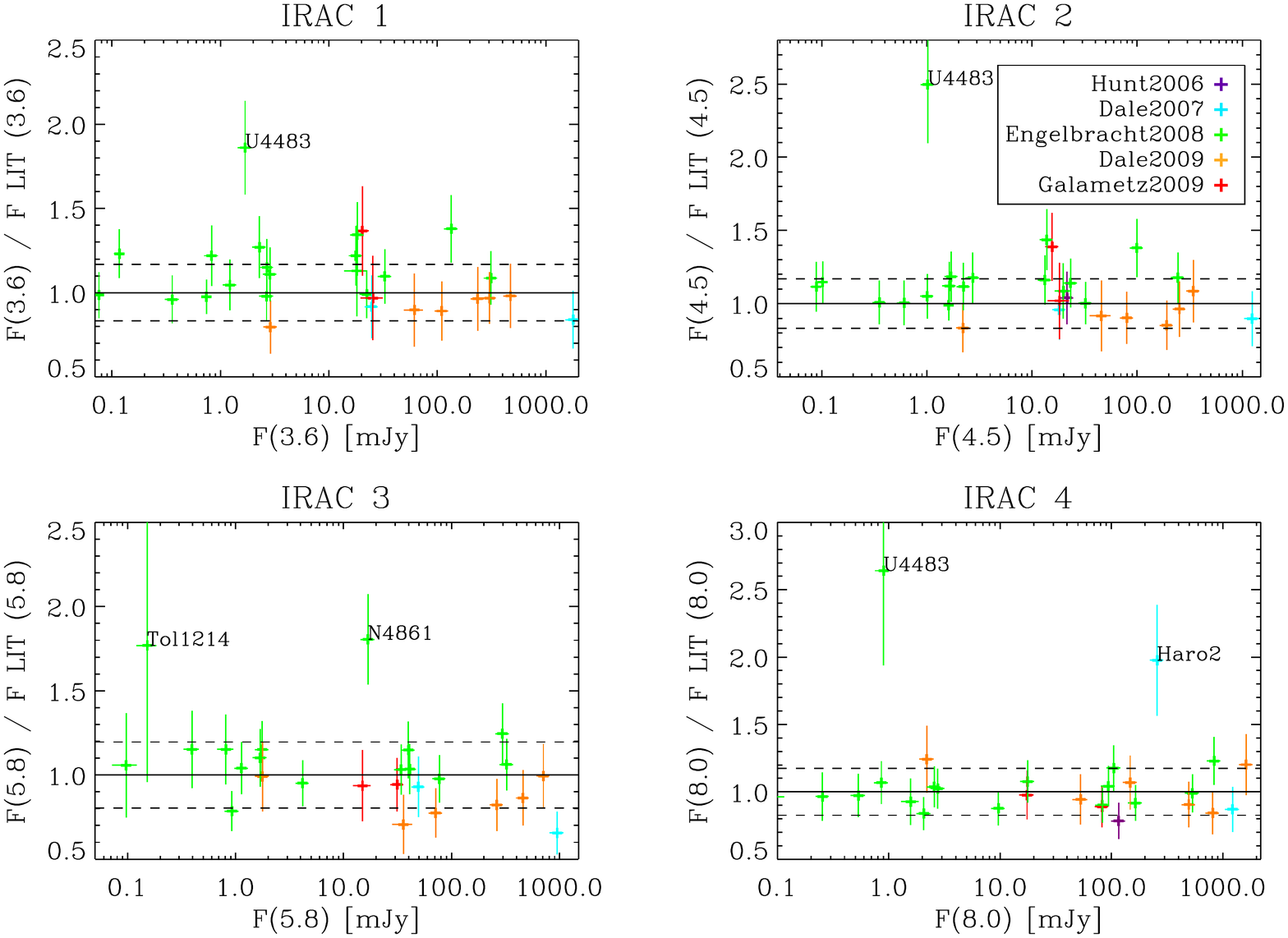}
\includegraphics[width=9.0cm]{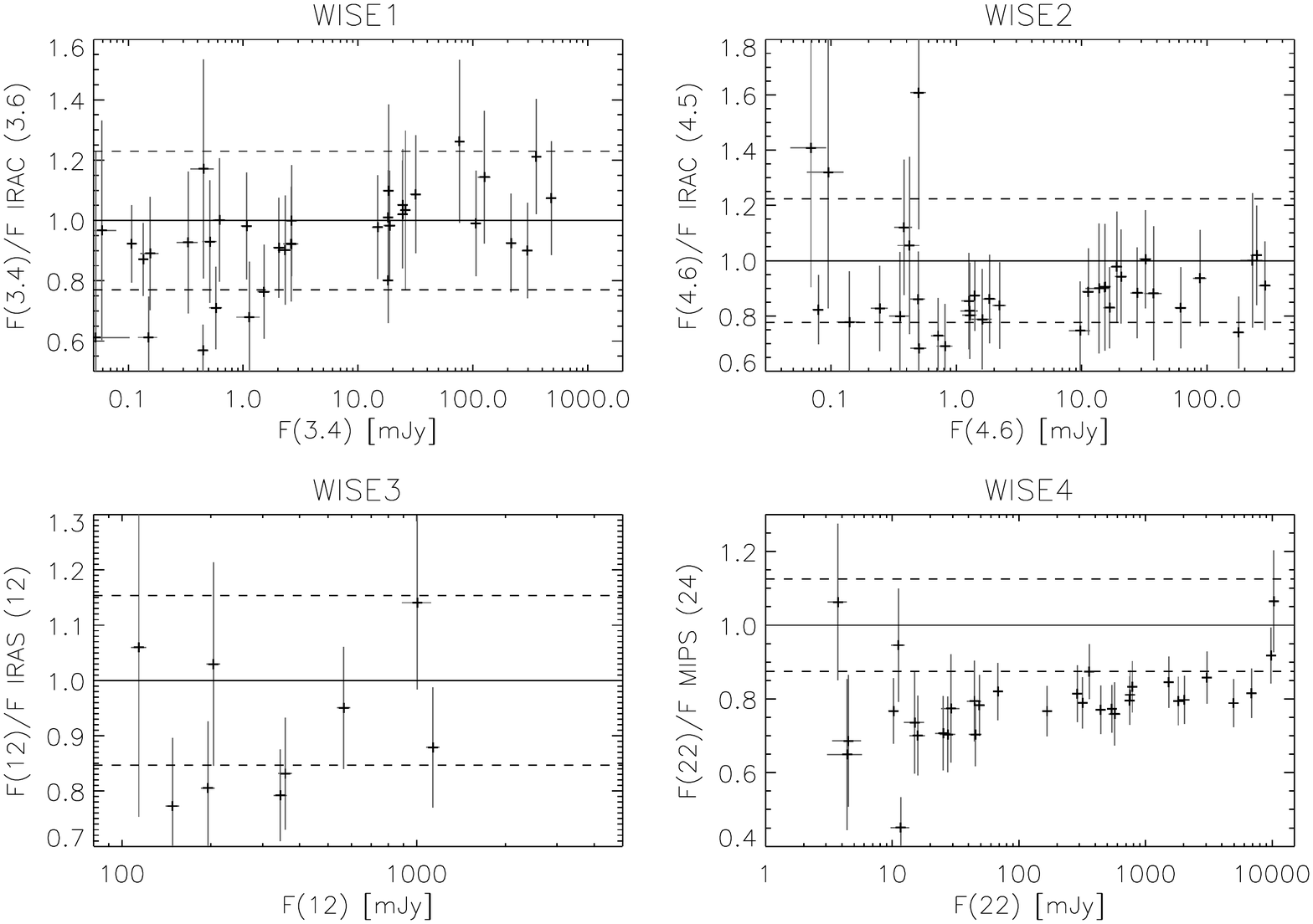}
\caption{
{\it (left panels)} Comparison of our IRAC flux densities and literature IRAC flux densities: $F(\lambda_0)/F LIT(\lambda_0)$ flux density ratios as a function of our IRAC flux density, $F(\lambda_0)$, at 3.6 \mic, 4.5 \mic, 5.8 \mic, and 8.0 \mic. The average uncertainties are 17\%, 17\%, 19\% and 17\% at 3.6, 4.5, 5.8 and 8.0 \mic. Colours distinguish the reference for the literature measurement.
{\it (right panels)} Ratios of WISE flux densities to other MIR measurements (either IRAC, IRAS or MIPS) as a function of WISE flux density at 3.4 \mic, 4.6 \mic, 12 \mic\ and 22 \mic. The average uncertainties are 20\%, 22\%, 15\% and 12\% at 3.4, 4.6, 12 and 22 \mic\ respectively.
On all panels, we show the unit ratio as a solid line as well as the average uncertainties on the ratio in each bands as dashed lines.
}
\label{f:IRACcomp}
\end{center}
\end{figure*}

We compare IRAC flux densities previously available in the literature and the flux densities derived in this work. We use the ratios of our IRAC measurement to the literature IRAC flux densities; a ratio of 1 corresponds to a very good agreement. The comparison is presented in Fig. \ref{f:IRACcomp} and shows good agreement between the two measurements, even if some outliers are present:

{\it - Haro~2 -} The comparison to the literature values from \cite{Dale2007} agrees for the first three IRAC bands (ratios of 0.92, 0.96 and 0.93 at 3.6 \mic, 4.5 \mic, and 5.8 \mic\ respectively) but not for the IRAC 8.0 \mic\ band where the ratio goes to $\sim$ 2. 
Recomputing the flux density with the aperture given in \cite{Dale2007} gives a flux in very close agreement to the one we find with our own aperture (ratio of 1.009 at 8.0 \mic), and also for the three other IRAC bands. Additionally, we find that \cite{Marble2010} performed photometry for Haro~2 in an IRS-matched aperture and their result is in good agreement with ours (ratio = 1.036 at 8.0 \mic). 

{\it - NGC~4861 - UGC 4483 - Tol 1214-277 -} Our measurements are systematically larger than the flux densities from \cite{Engelbracht2008}. No precise information is given in \cite{Engelbracht2008} about the size of the apertures used for IRAC photometry, so it is difficult to asses that we are really comparing similar measurements. Nonetheless, a possible explanation for NGC~4861 could be that we include Mrk~39 with which NGC~4861 is interacting in our aperture and that \cite{Engelbracht2008} did not. 
 
Excluding points where one of the two measurements (ours or literature values) is an upper limit, we get a mean ratio of our IRAC measurements to those in the literature of 1.05 $\pm$ 0.11 for IRAC 3.6 \mic, 1.06 $\pm$ 0.11 for IRAC 4.5 \mic, 1.02 $\pm$ 0.14 for IRAC 5.8 \mic\ and 1.01 $\pm$ 0.11 for IRAC 8.0 \mic. This is to be compared to an average uncertainty on the ratios of 17\%, 17\%, 19\% and 17\% at 3.6, 4.5, 5.8 and 8.0 \mic. The error on the mean ratio is lower than the average uncertainty on the ratios for all of the bands indicating a good photometric agreement between the two measurements.

\subsection{WISE: Comparison to other MIR measurements}

WISE photometry can be compared to \spitz\ IRAC measurements for the first two bands, to \irasÊ\ 12 \mic\ and to MIPS 24 \mic. WISE, IRAC, IRAS and MIPS are not calibrated on the same reference spectral shape, so we first apply colour corrections to our flux densities before the comparison. These colour corrections are of the order of $<$ 1\%, 3\%, 4\%, 2\%, 4\%, 5\%, $<$ 1\% and 9\% for WISE 3.4 \mic, WISE 4.6 \mic, WISE 12 \mic, WISE 22 \mic, IRAC 3.6 \mic, IRAC 4.5 \mic, IRAS 12 \mic, and MIPS 24 \mic\ respectively. 
We remove from the comparison all of the upper limits in any of the bands and we are left with 35 galaxies for WISE1, 37 galaxies for WISE2, 9 galaxies for WISE3 and 32 galaxies for WISE4. 

The ratios of WISE flux densities to the corresponding MIR flux densities are shown in Fig. \ref{f:IRACcomp}. The median ratios are F$_{{\rm WISE 3.4}}$/F$_{{\rm IRAC 3.6}}$ = 0.97 $\pm$ 0.12, F$_{{\rm WISE 4.6}}$/F$_{{\rm IRAC 4.5}}$ = 0.87 $\pm$ 0.17, F$_{{\rm WISE 12}}$/F$_{{\rm IRAS 12}}$ = 0.88 $\pm$ 0.13, and F$_{{\rm WISE 22}}$/F$_{{\rm MIPS 24}}$ = 0.79 $\pm$ 0.22. 
We note that the dispersion is larger for the first two WISE bands for WISE flux densities less than $\sim$ 1 mJy. WISE 22 \micÊ\ flux densities are also systematically lower than MIPS 24 \micÊ\ flux densities, reflecting the difference in wavelength between the two bands and the steeply rising MIR continuum observed in the SED of dwarf galaxies.

For consistency, we also compare the WISE photometry from the AllWISE database to our measurements on the maps. We confirm that for the brightest and most extended sources, the photometry provided by the database underestimates the emission, and especially at 12 and 22 \mic.

\section{DGS IRS spectra}\label{ap:IRSdata}

\subsection{IRS data reduction}

Two versions of the spectra are available in the CASSIS database \citep{Lebouteiller2011} for the staring observations: an ``optimal'' extraction better suited for point sources and a ``tapered column'' extraction, better suited for partially extended sources. A message advises the user if the ``tapered column'' extraction or the ``optimal'' extraction method should be chosen. The chosen extraction for each of the DGS targets, as well as AORkeys, are summarised in Appendix \ref{ap:ObsLogs}. CASSIS uses the {\sc AdOpt} algorithm \citep[Advanced Optimal extraction,][]{Lebouteiller2010} within the Spectroscopic Modeling Analysis and Reduction Tool \cite[SMART,][]{Higdon2004}. {\sc AdOpt} enables optimal extraction of spectra using a super-sampled PSF. The pipeline includes such steps as image cleaning, individual exposure combination and background subtraction. Specific attention is given to identification and removal of bad pixels and outlier rejection at the image and spectra levels.

Some discrepant fluxes between orders at some wavelengths have to be removed from the spectra. These cut wavelengths are given in Table \ref{t:IRSchar}. 
For the extended galaxies, there is a conversion step from MJy/sr to Jy where the IRS spectrum is multiplied by the area over which it has been extracted. The spectra are then rescaled to match the observed photometry.

\begin{table}[h!tbp]
\begin{center}
\caption{IRS cut-off wavelengths.}
{\small
\begin{tabular}{lc}
\hline
\hline
Module & Cut-offs (\mic) \\
\hline
SL1 &   [$\lambda_{min}$ - 7.53, 14.02 - $\lambda_{max}$] \\
SL2 &  [$\lambda_{min}$ - 5.23, 7.49 - $\lambda_{max}$] \\
SL3 &  [$\lambda_{min}$ - 7.4,  8.5 - $\lambda_{max}$] \\
LL1 & [$\lambda_{min}$ - 20.5, 37.4 - $\lambda_{max}$] \\
LL2 &  [$\lambda_{min}$ - 14.0, 20.52 - $\lambda_{max}$] \\
LL3 &  [$\lambda_{min}$ - 19.8, 21.5 - $\lambda_{max}$] \\
\hline
\end{tabular}
}
\label{t:IRSchar}
\end{center}
\end{table}

\subsection{Rescaling the spectra to the photometry}

\begin{figure}[h!tbp]
\begin{center}
\includegraphics[width=8.5cm]{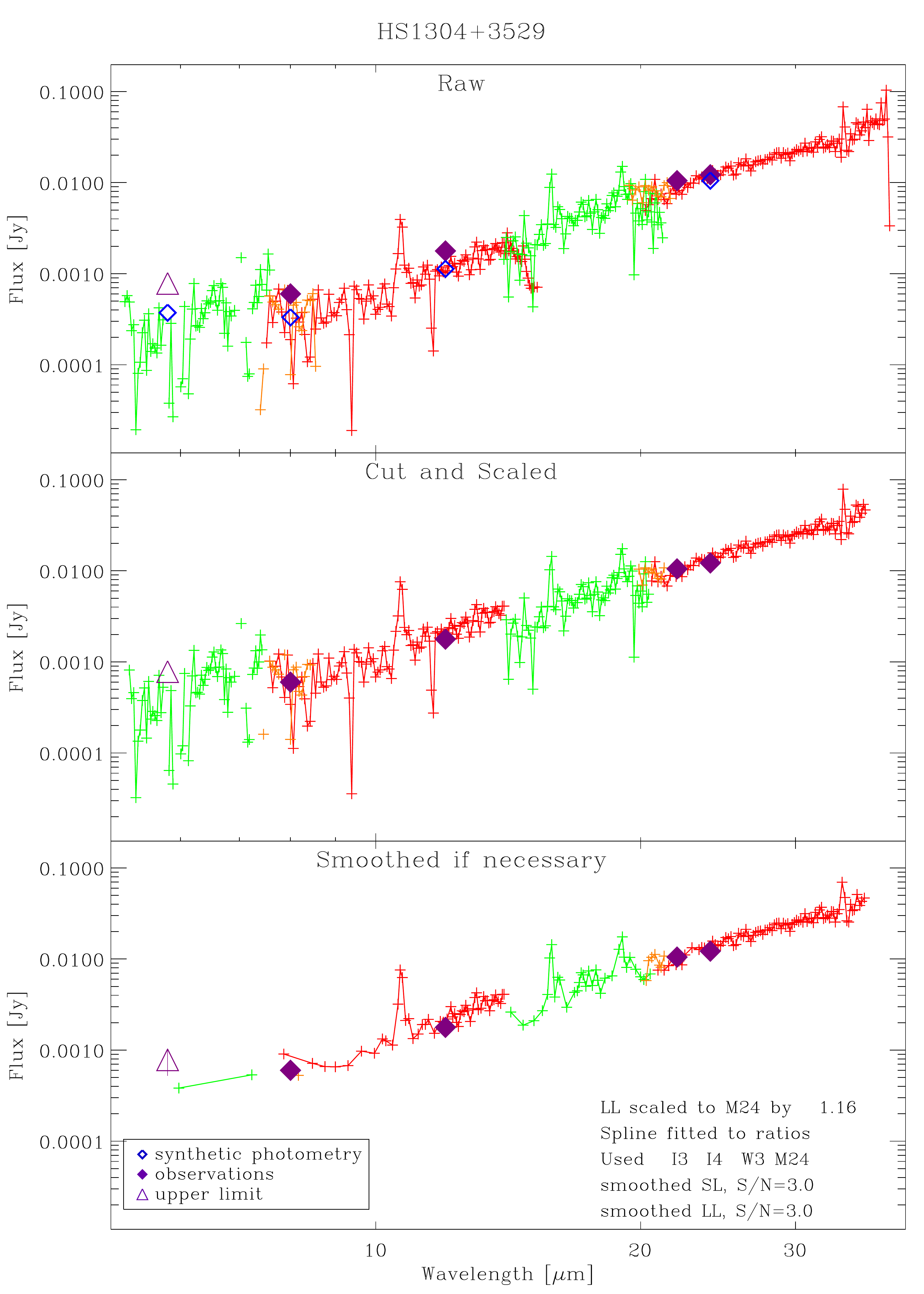}
\caption{Example of IRS treatment for HS~1304+3529: the raw spectrum is presented in the {\it top} panel. The photometry points are overlaid in filled purple diamonds as well as synthetic photometry as blue open diamonds. The galaxy is not detected at 5.8 \mic\ as indicated by the open purple triangle. Green, orange and red parts of the spectrum represents the SL2 and LL2, SL3 and LL3, SL1 and LL1 spectra, respectively. The {\it middle} panel shows the spectra after cutting-off the discrepant wavelengths and rescaling to match the photometry. The spline used to rescale SL is shown in Fig. \ref{f:IRSspline}. The {\it bottom} panel shows the spectra smoothed to a S/N of 3.}
\label{f:IRSex}
\end{center}
\end{figure}

\begin{figure}[h!tbp]
\begin{center}
\includegraphics[width=8.0cm]{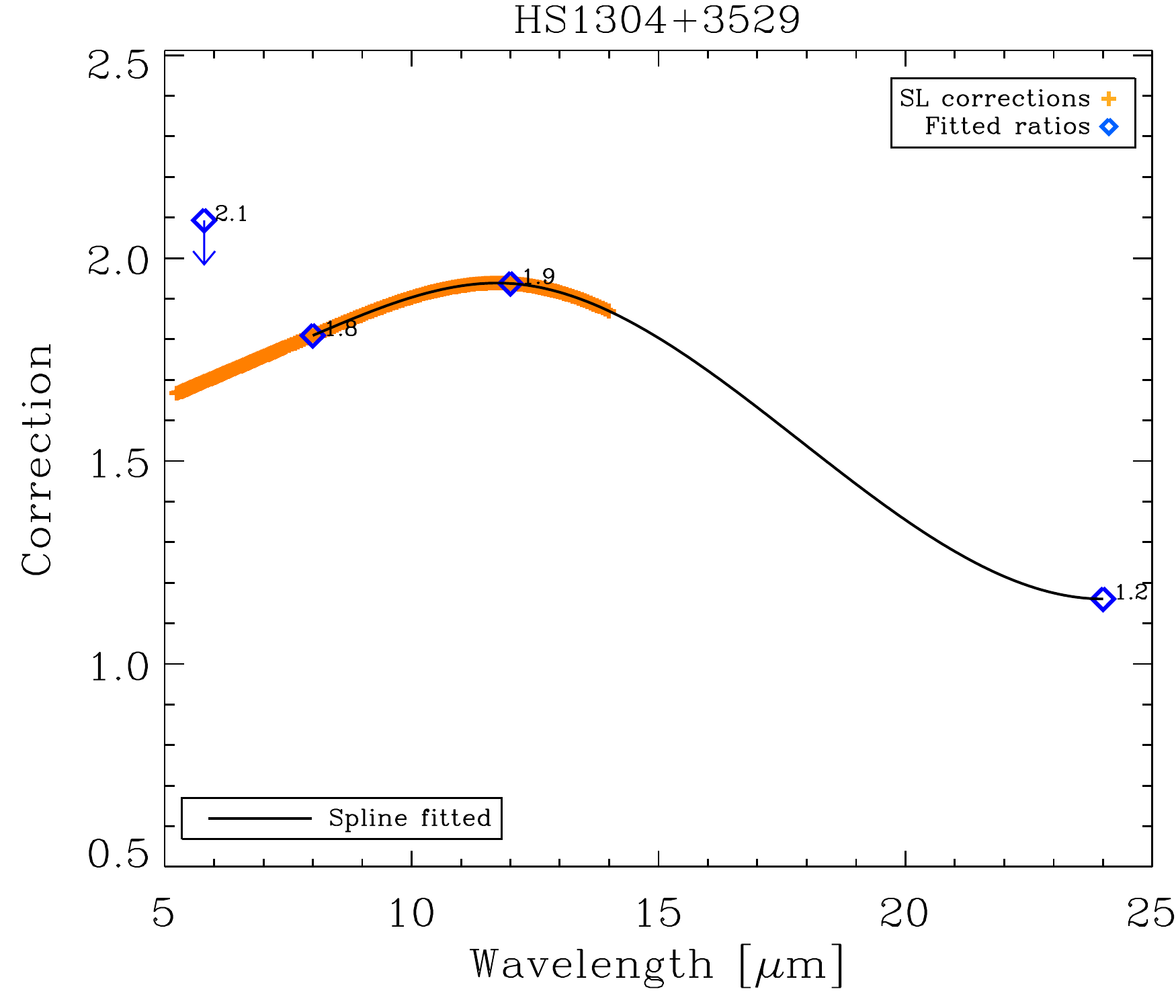}
\caption{Example of a spline used to rescale SL in the case of HS~1304+3529. The diamonds represent the ratio between the observed and the synthetic photometry. The spline has been fitted to the blue diamonds, except at 5.8 \mic\ where the galaxy is not detected. The black curve is the final adopted spline, sp(x$_0$), where {\it x$_0$} = 1.9. The orange crosses are the corrections applied to the SL part of the spectrum.}
\label{f:IRSspline}
\end{center}
\end{figure}

We describe now in more details the rescaling step of the IRS spectrum, and how we derive the correction factor for the SL and LL parts of the IRS spectrum.

The correction factor for LL, {\it C$_{LL}$} can be written: 

\begin{equation}
C_{LL} = F_{24}/F_{IRS}(M_{24})
\label{eq:CLL}
\end{equation}

\noindent where {\it F$_{24}$} is the observed 24 \mic\ MIPS flux density (from \cite{Bendo2012}) and {\it F$_{IRS}$(M$_{24}$)} is the synthetic photometry for the IRS LL, spectrum at 24 \mic. Eq. \ref{eq:CLL} is adapted for \wise\ 22 \mic\ when MIPS 24 \mic\ is not available.

Several constraints are available for the SL module, and used simultaneously to derive a correction that smoothly depends on wavelength, {\it C$_{SL}$($\lambda$)}. We fit a spline function, using the IDL {\sc spline\_p} procedure, to {\it C$_{SL}$(I3)}, {\it C$_{SL}$(I4)}, {\it C$_{SL}$(W3)} and {\it C$_{LL}$}. We use {\it C$_{LL}$} here to have a better constraint at the end of the spline. We impose a derivative of 0 at 24 \mic\ in order to have a smooth function. We also assume that the correction factor {\it C$_{SL}$($\lambda$)} should always be $\geq$~1.0, as the photometry measurement always encompasses more flux than the IRS slit.

{\it C$_{SL}$(I3)}, {\it C$_{SL}$(I4)}, {\it C$_{SL}$(W3)} are given by: 
 
\begin{equation}
\left  \{
 \begin{array}{rcl}
C_{SL}(I3) & = & F_{5.8}/F_{IRS}(I_{5.8}) \\
C_{SL}(I4) & = & F_{8.0}/F_{IRS}(I_{8.0}) \\
C_{SL}(W3) & = & F_{12}/F_{IRS, SL}(W_{12}) \end{array} 
 \right .
 \label{eq:CSL}
\end{equation}

\noindent where {\it F$_{5.8}$}, {\it F$_{8.0}$} and {\it F$_{12}$} are the observed 5.8 and 8.0 \mic\ IRAC and the 12 \mic\ WISE flux densities (from Tables \ref{t:irac} and \ref{t:wise}); and {\it F$_{IRS}$(I$_{5.8}$)}, {\it F$_{IRS}$(I$_{8.0}$)} and {\it F$_{IRS, SL}$(W$_{12}$)} are the synthetic photometry for the IRS SL spectrum at 5.8, 8.0 and 12 \mic.

Note that there is an overlap between the \wise\ 12 \mic\ filter and the LL wavelengths. As the LL spectrum has already been corrected, we must find the flux missing in the SL spectrum, i.e., {\it F$_{IRS, SL}$(W$_{12}$)}, to match {\it F$_{12}$}. But the integration over the \wise\ bandpass is not linear, i.e.,: {\it F$_{IRS, SL}$(W$_{12}$)} $\neq$ {\it F$_{IRS}$(W$_{12}$)} - {\it F$_{IRS, LL}$(W$_{12}$)}. Thus there is no simple way of deriving {\it C$_{SL}$(W3)}. Instead, we apply the following method:

\begin{enumerate}
\item We generate a grid of potential {\it C$_{SL}$(W3)}, \{x\}, from 0.01 to 10 ; 
\item find the spline going through {\it C$_{SL}$(I3)}, {\it C$_{SL}$(I4)}, {\it C$_{LL}$}, and each x$_i$, sp(x$_i$) ; 
 \item correct the SL spectrum with each sp(x$_i$) and 
\item compute the synthetic \wise\ 12 \mic\ photometry for each corrected {\it total} IRS spectrum, {\it F$_{IRS}$(W$_{12}$)(sp(x$_i$))}. 
\end{enumerate}

\noindent {\it C$_{SL}$(W3)} is the x$_0$ that gives {\it F$_{IRS}$(W$_{12}$)(sp(x$_0$))} = {\it F$_{12}$}. \\

The final IRS spectrum, {\it F$_{IRS, corr} (\lambda_{SL})$} and {\it F$_{IRS, corr} (\lambda_{LL})$}, is given by: 

\begin{equation}
\left  \{
 \begin{array}{rcl}
F_{IRS, corr} (\lambda_{SL}) & = & F_{IRS} (\lambda_{SL}) \times C_{SL}(\lambda_{SL}) \\
F_{IRS, corr} (\lambda_{LL}) & = & F_{IRS} (\lambda_{LL}) \times C_{LL} \\ \end{array} 
 \right .
 \label{eq:IRScorr}
\end{equation}

\begin{figure*}[h!tbp]
\begin{center}
\includegraphics[width=16cm]{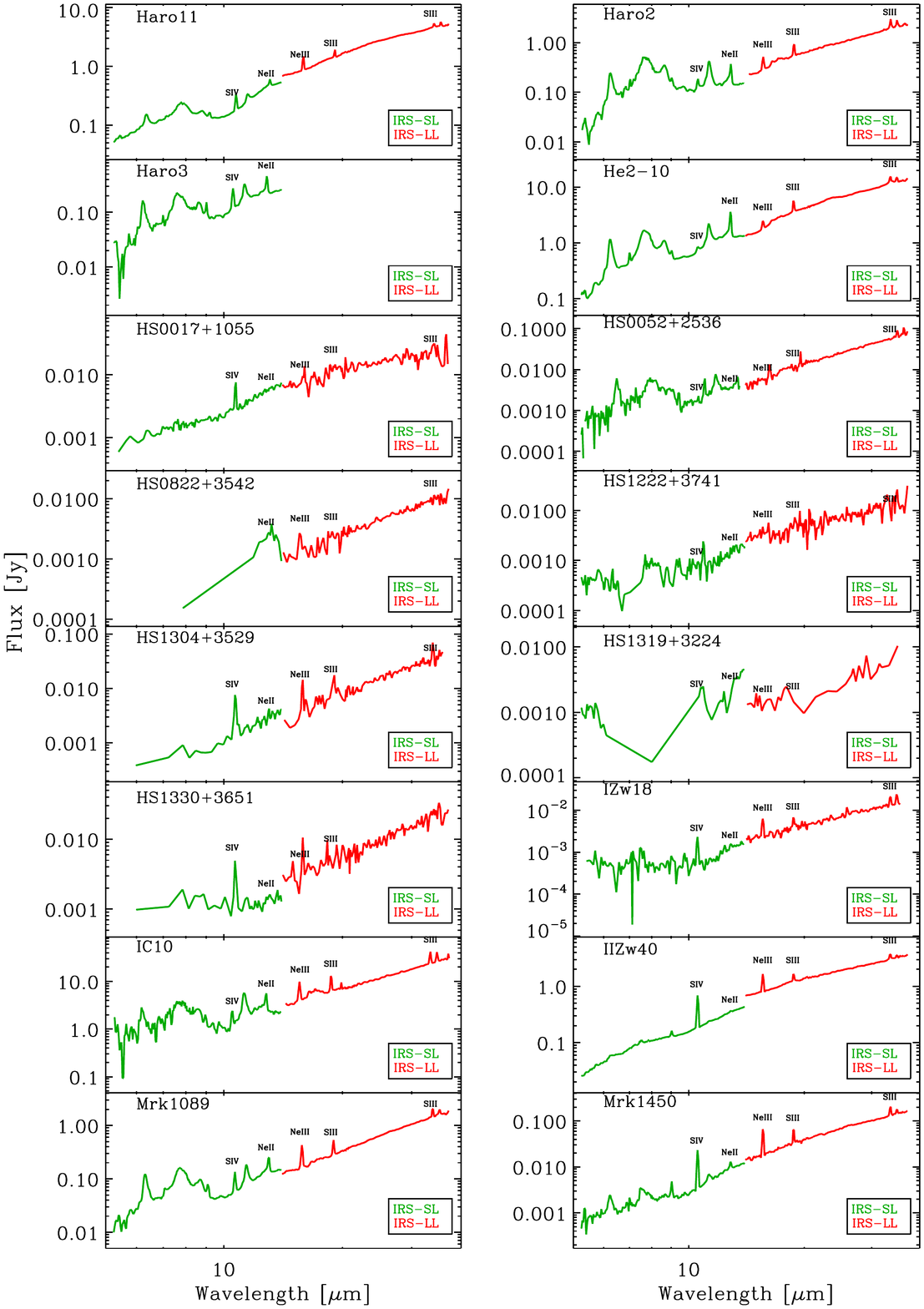}\caption{IRS spectra for the DGS galaxies. The SL module is shown in green and the LL module is shown in red. The positions of the [S{\sc IV}] 10.5 \mic, [Ne{\sc II}] 12.8 \mic, [Ne{\sc III}] 15.6 \mic, [S{\sc III}] 18.7 \mic\ and 33.5 \mic\ spectral lines are indicated. Line intensities have been extracted by \cite{Cormier2015}.
}
\label{f:IRSall}
\end{center}
\end{figure*}

The method presented here is adapted depending on the number of constraints for each galaxy. Upper limits are not considered for the correction of IRS spectrum. An example of this treatment of IRS spectrum is shown in Fig. \ref{f:IRSex} along with the spline used to correct the SL spectrum in Fig. \ref{f:IRSspline}. The final IRS spectra for the DGS galaxies are shown in Fig. \ref{f:IRSall}. 

In the case of NGC~1140 and NGC~1569, the IRS SL slit only covers a small part of these extended galaxies. 
We use the ISOCAM spectrum \citep[from][]{Galliano2003, Galliano2005} for comparison after applying the same rescaling step. The two spectra are consistent with each other and with the general NIR - MIR shape of the SED for both galaxies. Thus we consider that these IRS spectra are reliable and use them to constrain the SED model.

 \addtocounter{figure}{-1}
\begin{figure*}[h!tbp]
\begin{center}
\includegraphics[width=16cm]{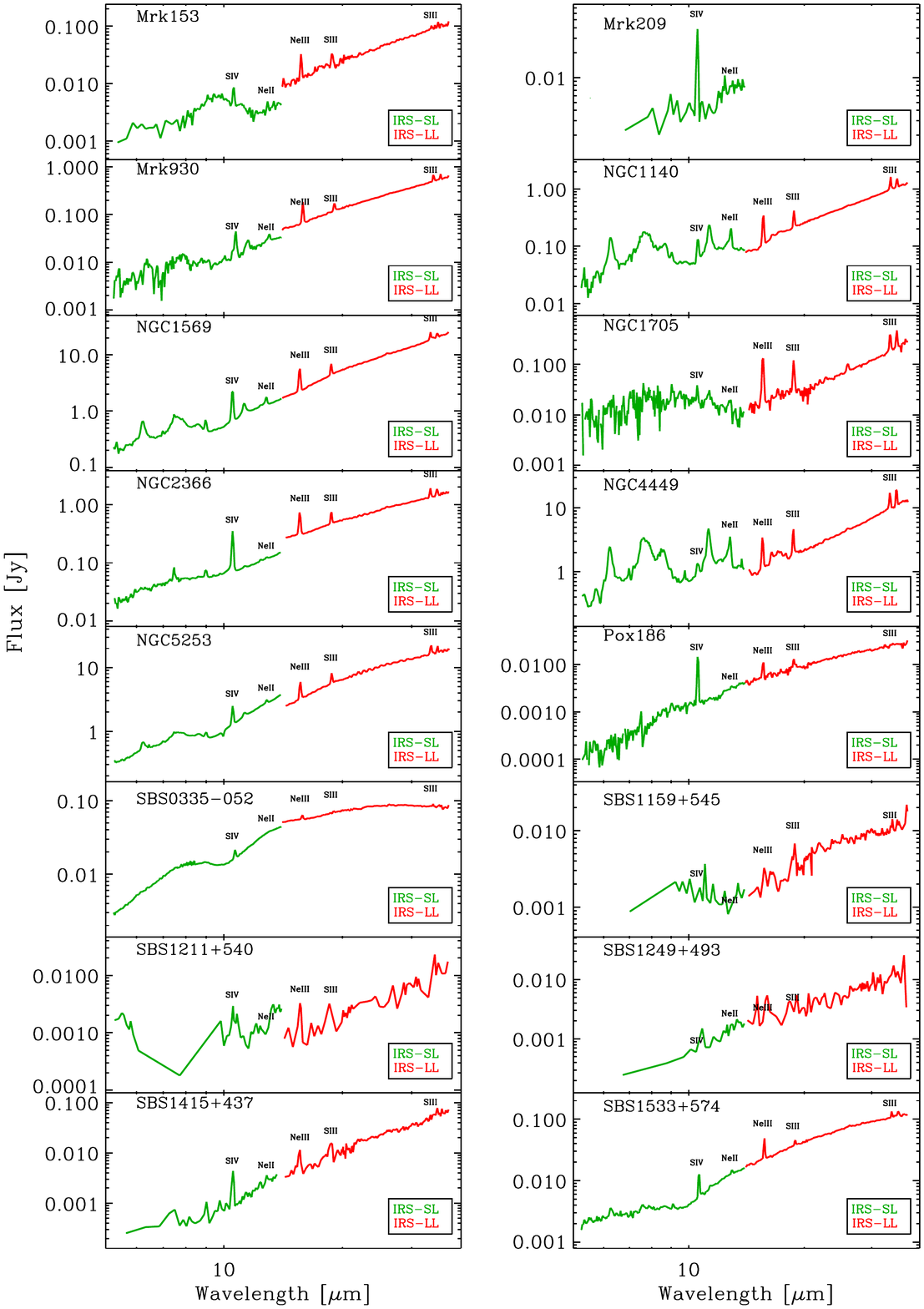}
\caption{IRS spectra for the DGS (continued).}
\end{center}
\end{figure*}

 \addtocounter{figure}{-1}
\begin{figure*}[h!tbp]
\begin{center}
\includegraphics[width=16cm]{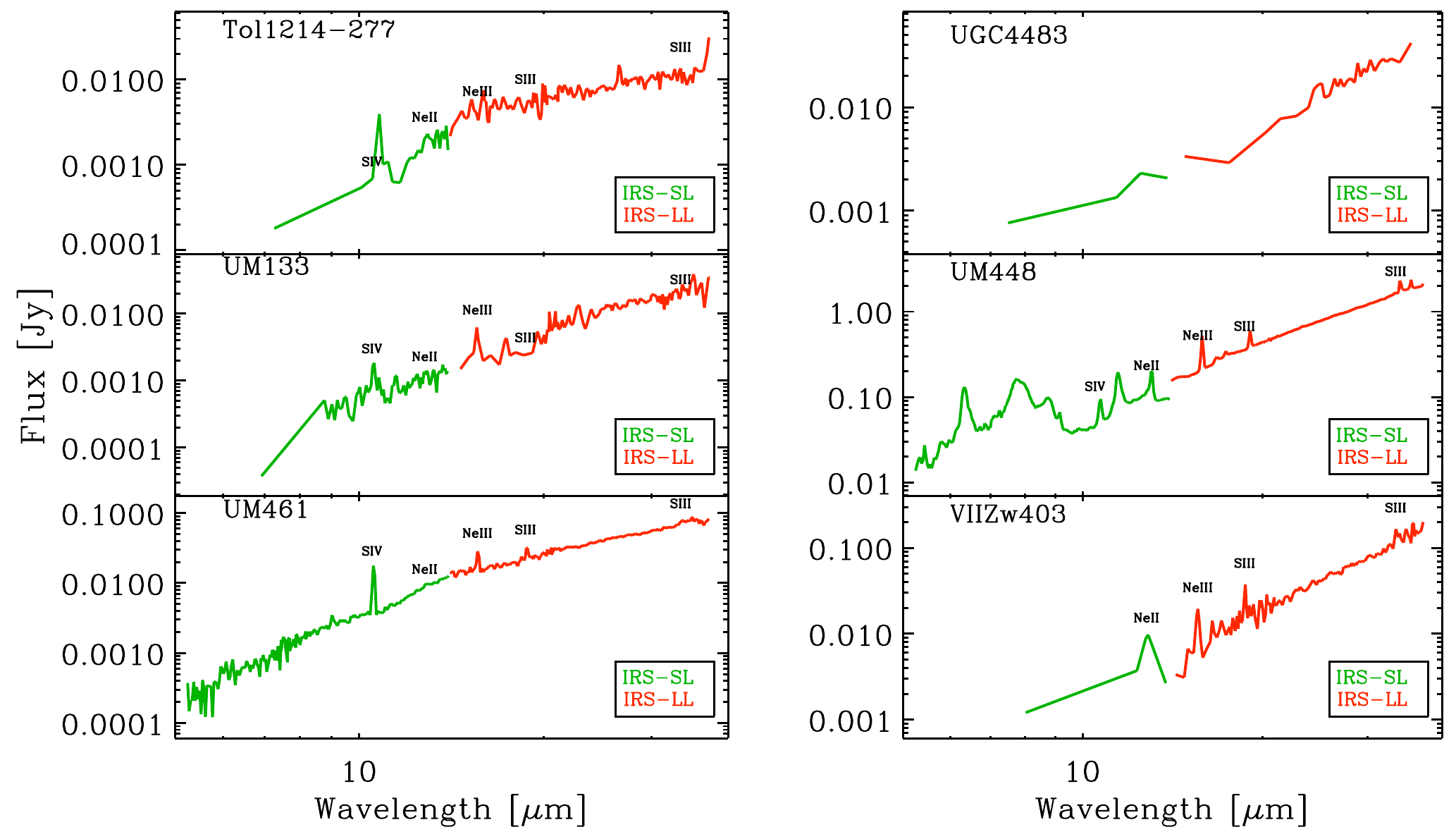}
\caption{IRS spectra for the DGS (continued).}
\end{center}
\end{figure*}

\section{Specific SED modelling}\label{ap:default}

In this Appendix, we detail the specific SED modelling for several DGS sources, and show the SEDs for the galaxies with and without any additional features (addition of a MIR modBB, including extinction, etc.) to show how the fit is improved. 

\begin{figure*}[h!bp]
\begin{center}
\includegraphics[width=17cm]{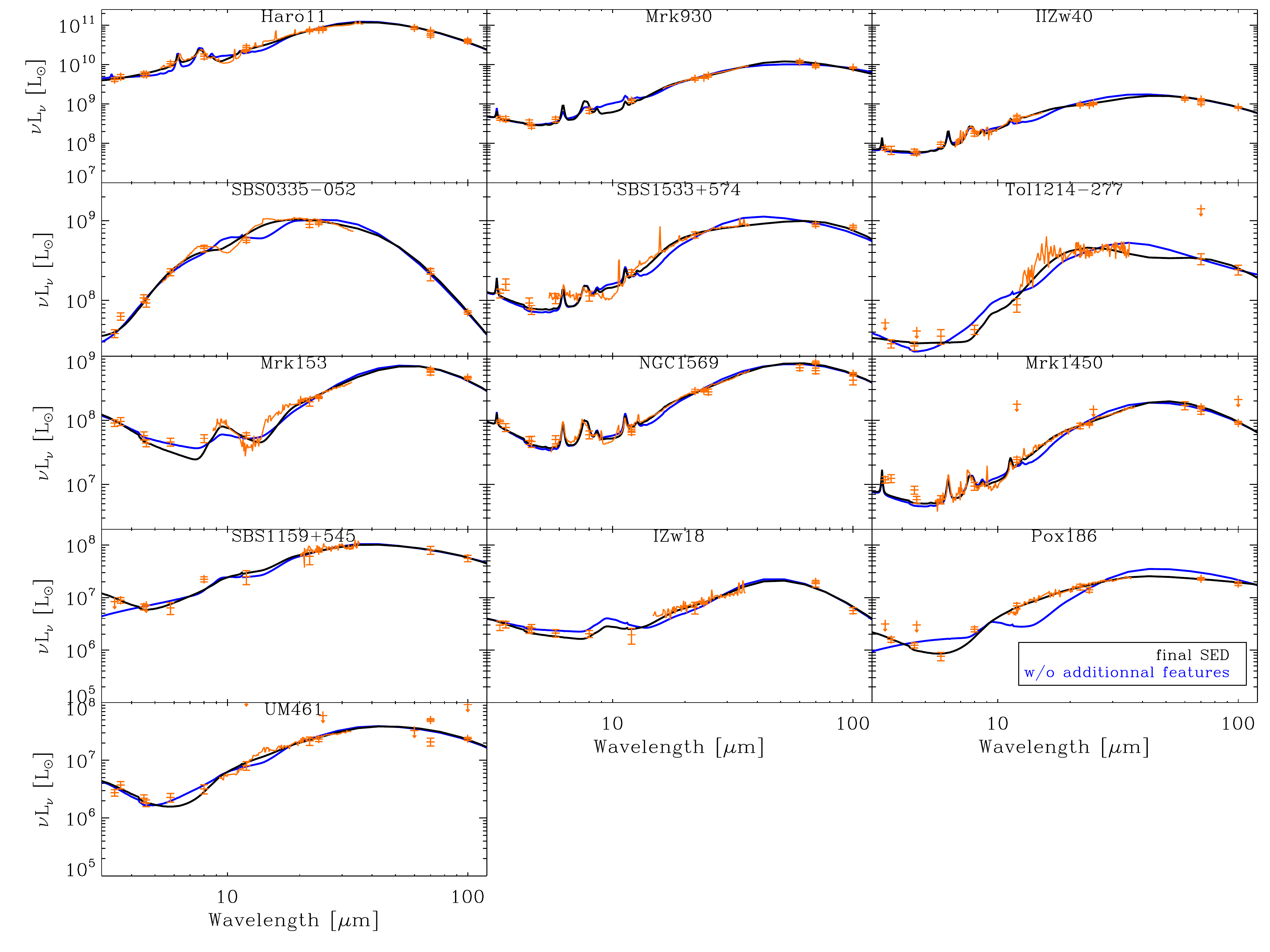}
\caption{Final SEDs in black compared to the SED without the additional features in blue (MIR modBB, extinction, see text for source per source details). Note how the final SEDs better reproduce the observed IR emission for these galaxies.}
\label{f:Noteseds}
\end{center}
\end{figure*}

\noindent {\it - Haro 11 - } As this galaxy has evident silicate absorption features in the IRS spectrum around 10 \mic, we add extinction in our modelling. The output luminosity is modified by: 

\begin{equation}
L_\nu^{tot,dust}(\lambda, A_V) =  L_\nu^{tot,dust}(\lambda,A_V=0) \times e^{-\tau_0(\lambda) A_V / 1.086}
\label{eq:totLAv}
\end{equation}

\noindent where {\it $\tau_0$} is the optical depth, determined from the optical properties and the size distribution of our dust composition. The extinction parameter {\it A$_V$} is left varying between 1 and 20, with a best fit value of {\it A$_V$}~=~5. The quality of the fit is clearly improved in the NIR-MIR range after introducing {\it A$_V$} (see Fig. \ref{f:Noteseds}). Because the stellar continuum is also affected by extinction, the stellar template used for the fit does not provide a reliable fit of the 2MASS NIR data anymore. Instead, we replace the stellar component by a blackbody and impose a temperature {\it T} $\geq$ 1000 K. We obtain a best fit temperature {\it T}=~30,000~K, corresponding to a maximum of the blackbody emission around 0.1 \mic\ in the UV - a sign of a young stellar population. This is consistent with the results of \cite{Adamo2010} who found a maximum in the cluster formation history $\sim$ 3.5 Myr ago, and with the UV part of the SED presented in \cite{Cormier2012}. \\ 

\noindent {\it - HS 0822+3542 -} The observed PACS upper limits at 70 and 100 \mic\ are not consistent with the rest of the MIR to FIR photometry (see Fig. \ref{f:SED_DGSall}). We recommend using the synthetic photometry provided by the model as upper limits for the 70 and 100 \mic\ PACS wavelengths, i.e., {\it F$_{70}$}~$\leq$~41 mJy and {\it F$_{100}$} $\leq$ 48 mJy. These values are reported in Table \ref{t:herschel}. \\ 

\noindent {\it - HS 2352+2733 -} There are only 5 detected points for this galaxy, and thus not enough constraints to fit a full SED. However, we can get a rough estimate of the dust mass by fitting a modBB with a fixed $\beta$=2.0, through MIPS 24 \mic, PACS 70 \mic\ and PACS 100 \mic\ where the galaxy is detected. We obtain a temperature of T= 52 K and M$_{BB}$ = 1.01 $\times$ 10$^4$ \msun. This mass can be seen as a lower limit to the real dust mass (see Sect. \ref{ssec:compBB}).  
This value is reported in Table \ref{t:Dustparam} and marked with a $^a$. \\

\noindent {\it - II Zw 40 -} This galaxy has an IRS spectrum with very good S/N. However, when compared to other MIR measurements, it seems that the \spitz\ IRS-LL spectrum is not consistent with the rest of the MIR - FIR photometry, meaning that for this galaxy, a wavelength dependent correction may also be needed for the IRS LL spectrum.
As an alternative, we use the ISOCAM spectrum for this galaxy \citep[from][]{Galliano2005}, which covers a 5.6 to 16.3 \mic\ wavelength range. The ISOCAM spectrum was matched to the broad band photometry by applying the same rescaling process as for the IRS SL spectrum. 
To properly match the end of the ISOCAM spectrum, we add an extra modBB ($\beta$ fixed to 2.0 and best fit temperature {\it T} = 113 K). \\

\noindent {\it - Mrk 153 -} This galaxy has a prominent silicate emission feature in its IRS spectrum around 10 \mic. 
This emission feature originates from hot small silicate grains, possibly in the accretion disk around an AGN. We allow the the silicate-to-(silicate+graphite) grain mass fraction to vary in our fit and get a ratio of 0.94 (i.e., 1.3 times the Galactic value). \\

\noindent {\it - SBS 0335-052 -} This galaxy has a very surprising SED with a IR peak around 15 - 30 \mic, and an IRS spectrum showing silicate absorption features superimposed on a featureless continuum \citep[see also][]{Thuan1999a, Dale2001, Houck2004, Galliano2008}. The silicate absorption feature around 10 \mic\ observed in the IRS spectrum indicates that extinction is present in this galaxy. We add extinction in our model as for Haro 11, and get {\it A$_V$} = 1.1. The peculiar shape of the IRS spectrum requires two MIR modified blackbodies to obtain a satisfactory fit ($\beta$ fixed to 2.0 and best fit temperatures {\it T$_1$} = 273 K and {\it T$_2$} = 123 K, see Fig. \ref{f:Noteseds}). \\

\noindent {\it - Tol 1214-277 -} The MIPS 70 \mic\ point is not consistent with the rest of the photometry: IRS and MIPS 24 \mic\ on one side and PACS 70 and 100 \mic\ on the other side (see Fig. \ref{f:SED_DGSall}). It is another confirmation that this point is discrepant \citep{RemyRuyer2013}, and thus we do not consider it in the modelling.

\end{document}

%% file: SFRparam_DGS_KF_fin.tex
\begin{table*}[h!tbp]                                                                               
\begin{center}                                                                                      
\caption{DGS and KINGFISH stellar mass and star-formation parameters.}                              
\label{t:sfr}                                                                                       
 \begin{tabular}{lccc | cc}                                                                         
\hline                                                                                              
\hline                                                                                              
Name &             Log[\mstar] &         Log[SFR] &         Log[\ssfr] &Log[L$_{H\alpha}$] &              Ref \\
 &                 Log[\msun] &         Log[\sfr] &     Log[yr$^{-1}$] &       Log[erg/s] &                   \\
 &                         (1) &               (2) &               (3) &              (4)  &              (5) \\
\hline                                                                                              
{\bf DGS} & & & &  \\                                                                               
Haro11      & $       10.24    \pm  0.21               $ & $        1.40    \pm  0.33               $ & $       -8.84    \pm  0.39               $ & $         42.45       $ & $        (1)       $  \\
Haro2       & $        9.56    \pm  0.21               $ & $       -0.09    \pm  0.20               $ & $       -9.66    \pm  0.29               $ & $         40.96       $ & $        (2)       $  \\
Haro3       & $        9.49    \pm  0.21               $ & $       -0.20    \pm  0.18               $ & $       -9.69    \pm  0.28               $ & $         40.83       $ & $        (3)       $  \\
He2-10      & $        9.50    \pm  0.21               $ & $       -0.27    \pm  0.14               $ & $       -9.77    \pm  0.25               $ & $         40.70       $ & $        (4)       $  \\
HS0017+1055 & $        7.84    \pm  0.16               $ & $         -                   $ & $         -                   $ & $           -       $ & $        -       $  \\
HS0052+2536 & $       10.31    \pm  0.30               $ & $        0.47    \pm  0.07               $ & $       -9.84    \pm  0.31               $ & $           -       $ & $        -       $  \\
HS0822+3542 & $        6.51    \pm  0.17               $ & $       -1.96    \pm  0.26               $ & $       -8.47    \pm  0.31               $ & $         39.03   $$^a$ & $        (5)       $  \\
HS1222+3741 & $        9.38    \pm  0.16               $ & $        0.29    \pm  0.16               $ & $       -9.08    \pm  0.22               $ & $         41.54       $ & $        (6)       $  \\
HS1236+3937 & $  \leq  9.35                   $ & $         -                   $ & $         -                   $ & $         40.37       $ & $        (6)       $  \\
HS1304+3529 & $        8.58    \pm  0.31               $ & $        0.04    \pm  0.01               $ & $       -8.54    \pm  0.31               $ & $         41.27       $ & $        (6)       $  \\
HS1319+3224 & $        8.05    \pm  0.24               $ & $       -1.35    \pm  0.13               $ & $       -9.41    \pm  0.27               $ & $           -       $ & $        -       $  \\
HS1330+3651 & $        8.95    \pm  0.28               $ & $       -0.21    \pm  0.03               $ & $       -9.16    \pm  0.28               $ & $         41.01       $ & $        (6)       $  \\
HS1442+4250 & $        7.92    \pm  0.13               $ & $       -1.67    \pm  0.06               $ & $       -9.58    \pm  0.14               $ & $         39.56       $ & $        (5)       $  \\
HS2352+2733 & $        8.70    \pm  0.16               $ & $         -                   $ & $         -                   $ & $           -       $ & $        -       $  \\
IZw18       & $        7.34    \pm  0.21               $ & $       -1.12    \pm  0.09               $ & $       -8.46    \pm  0.23               $ & $         40.14       $ & $        (5)       $  \\
IC10        & $        8.44    \pm  0.02               $ & $       -1.57    \pm  0.01               $ & $      -10.01    \pm  0.02               $ & $         39.54       $ & $        (4)       $  \\
IIZw40      & $        8.60    \pm  0.33               $ & $        0.04    \pm  0.09               $ & $       -8.56    \pm  0.34               $ & $         41.24       $ & $        (4)       $  \\
Mrk1089     & $       10.28    \pm  0.21               $ & $        0.65    \pm  0.16               $ & $       -9.63    \pm  0.26               $ & $         41.69       $ & $        (7)       $  \\
Mrk1450     & $        8.09    \pm  0.21               $ & $       -0.92    \pm  0.05               $ & $       -9.01    \pm  0.22               $ & $         40.29       $ & $        (5)       $  \\
Mrk153      & $        9.06    \pm  0.21               $ & $       -1.04    \pm  0.14               $ & $      -10.10    \pm  0.25               $ & $         39.86       $ & $        (3)       $  \\
Mrk209      & $        7.46    \pm  0.21               $ & $       -1.55    \pm  0.13               $ & $       -9.01    \pm  0.25               $ & $         39.70       $ & $        (4)       $  \\
Mrk930      & $        9.44    \pm  0.21               $ & $        0.62    \pm  0.15               $ & $       -8.82    \pm  0.26               $ & $         41.77       $ & $        (3)       $  \\
NGC1140     & $        9.64    \pm  0.21               $ & $       -0.11    \pm  0.04               $ & $       -9.76    \pm  0.22               $ & $         41.04       $ & $        (3)       $  \\
NGC1569     & $        8.93    \pm  0.21               $ & $       -0.21    \pm  0.04               $ & $       -9.13    \pm  0.21               $ & $         41.02       $ & $        (4)       $  \\
NGC1705     & $        8.34    \pm  0.34               $ & $       -1.26    \pm  0.02               $ & $       -9.60    \pm  0.34               $ & $         39.98       $ & $        (4)       $  \\
NGC2366     & $        8.29    \pm  0.31               $ & $       -1.12    \pm  0.07               $ & $       -9.41    \pm  0.32               $ & $         40.10       $ & $        (4)       $  \\
NGC4214     & $        9.03    \pm  0.21               $ & $       -0.96    \pm  0.04               $ & $       -9.99    \pm  0.21               $ & $         40.18       $ & $        (4)       $  \\
NGC4449     & $        9.43    \pm  0.21               $ & $       -0.41    \pm  0.07               $ & $       -9.84    \pm  0.22               $ & $         40.70       $ & $        (4)       $  \\
NGC4861     & $        8.49    \pm  0.21               $ & $       -0.66    \pm  0.13               $ & $       -9.15    \pm  0.25               $ & $         40.58       $ & $        (5)       $  \\
NGC5253     & $        8.77    \pm  0.21               $ & $       -0.57    \pm  0.18               $ & $       -9.34    \pm  0.28               $ & $         40.54       $ & $        (4)       $  \\
NGC625      & $        8.73    \pm  0.21               $ & $       -1.30    \pm  0.06               $ & $      -10.03    \pm  0.22               $ & $         39.83       $ & $        (4)       $  \\
NGC6822     & $        8.19    \pm  0.23               $ & $       -1.89    \pm  0.25               $ & $      -10.08    \pm  0.34               $ & $         39.13   $$^a$ & $        (4)       $  \\
Pox186      & $        7.06    \pm  0.17               $ & $       -1.46    \pm  0.03               $ & $       -8.52    \pm  0.17               $ & $         39.77       $ & $        (5)       $  \\
SBS0335-052 & $        8.00    \pm  0.15               $ & $       -0.13    \pm  0.16               $ & $       -8.13    \pm  0.22               $ & $         41.08       $ & $        (8)       $  \\
SBS1159+545 & $        7.84    \pm  0.18               $ & $       -0.90    \pm  0.06               $ & $       -8.75    \pm  0.19               $ & $         40.33       $ & $        (9)       $  \\
SBS1211+540 & $        7.60    \pm  0.21               $ & $       -1.73    \pm  0.29               $ & $       -9.33    \pm  0.36               $ & $         39.39   $$^a$ & $        (9)       $  \\
SBS1249+493 & $        8.42    \pm  0.39               $ & $       -0.40    \pm  0.11               $ & $       -8.82    \pm  0.40               $ & $         40.82       $ & $       (10)       $  \\
SBS1415+437 & $        7.80    \pm  0.26               $ & $       -1.20    \pm  0.11               $ & $       -9.00    \pm  0.29               $ & $         40.04       $ & $        (5)       $  \\
SBS1533+574 & $        9.33    \pm  0.21               $ & $       -0.17    \pm  0.13               $ & $       -9.50    \pm  0.25               $ & $         41.02       $ & $        (5)       $  \\
Tol0618-402 & $       10.43    \pm  0.21               $ & $       -0.57    \pm  0.11               $ & $      -11.00    \pm  0.24               $ & $         40.63       $ & $       (11)       $  \\
Tol1214-277 & $        8.17    \pm  0.18               $ & $       -0.10    \pm  0.13               $ & $       -8.27    \pm  0.22               $ & $         41.14       $ & $       (11)       $  \\
UGC4483     & $        6.89    \pm  0.22               $ & $       -2.21    \pm  0.18               $ & $       -9.11    \pm  0.28               $ & $         38.61   $$^a$ & $        (4)       $  \\
UGCA20      & $  \leq  7.14                   $ & $       -1.79    \pm  0.23               $ & $  \geq -8.93                   $ & $         39.30   $$^a$ & $        (4)       $  \\
UM133       & $  \leq  7.99                   $ & $       -1.11    \pm  0.06               $ & $  \geq -9.09                   $ & $         40.13       $ & $        (5)       $  \\
UM311       & $        9.78    \pm  0.13               $ & $        0.38    \pm  0.18               $ & $       -9.40    \pm  0.22               $ & $         41.60       $ & $       (12)       $  \\
UM448       & $       10.62    \pm  0.21               $ & $        1.03    \pm  0.10               $ & $       -9.59    \pm  0.23               $ & $         42.07       $ & $        (3)       $  \\
UM461       & $        7.66    \pm  0.22               $ & $       -1.42    \pm  0.15               $ & $       -9.08    \pm  0.26               $ & $         39.80       $ & $        (3)       $  \\
VIIZw403    & $        7.21    \pm  0.21               $ & $       -1.89    \pm  0.25               $ & $       -9.10    \pm  0.33               $ & $         39.14   $$^a$ & $        (4)       $  \\
& & & & \\                                                                                          
\hline                                                                                              
{\bf KINGFISH} & & & & \\                                                                           
NGC0337     & $       10.11    \pm  0.17               $ & $        0.18    \pm  0.05               $ & $       -9.93    \pm  0.18               $ & $         41.25       $ & $        (2)       $  \\
NGC0584     & $       10.87    \pm  0.20               $ & $       -1.15    \pm  0.10               $ & $      -12.02    \pm  0.22               $ & $           -       $ & $        -       $  \\
NGC0628     & $       10.29    \pm  0.21               $ & $       -0.14    \pm  0.08               $ & $      -10.43    \pm  0.22               $ & $         40.87       $ & $        (2)       $  \\
NGC0855     & $        9.20    \pm  0.21               $ & $       -1.27    \pm  0.05               $ & $      -10.47    \pm  0.21               $ & $         39.81       $ & $        (2)       $  \\
NGC0925     & $        9.97    \pm  0.20               $ & $       -0.24    \pm  0.06               $ & $      -10.21    \pm  0.21               $ & $         40.89       $ & $        (2)       $  \\
NGC1097     & $       11.00    \pm  0.20               $ & $        0.48    \pm  0.09               $ & $      -10.53    \pm  0.22               $ & $         41.23       $ & $        (2)       $  \\
NGC1266     & $       10.18    \pm  0.21               $ & $        0.17    \pm  0.10               $ & $      -10.01    \pm  0.24               $ & $         40.19       $ & $        (3)       $  \\
NGC1291     & $       11.02    \pm  0.20               $ & $       -0.47    \pm  0.06               $ & $      -11.49    \pm  0.21               $ & $         40.64       $ & $        (4)       $  \\
NGC1316     & $       11.68    \pm  0.20               $ & $       -0.40    \pm  0.14               $ & $      -12.08    \pm  0.24               $ & $         40.29       $ & $       (13)       $  \\
NGC1377     & $        9.47    \pm  0.21               $ & $         -                   $ & $         -                   $ & $           -       $ & $        (3)       $  \\
NGC1404     & $       11.15    \pm  0.20               $ & $       -0.70    \pm  0.20               $ & $      -11.85    \pm  0.29               $ & $         40.59       $ & $       (13)       $  \\
IC0342      & $       10.91    \pm  0.04               $ & $        0.03    \pm  0.10               $ & $      -10.88    \pm  0.11               $ & $         40.73       $ & $        (4)       $  \\
NGC1482     & $       10.55    \pm  0.21               $ & $        0.46    \pm  0.15               $ & $      -10.10    \pm  0.26               $ & $         40.80       $ & $        (2)       $  \\
\hline                                                                                              
\end{tabular}                                                                                       
\end{center}                                                                                        
\end{table*}                                                                                        
\addtocounter{table}{-1}                                                                            
\begin{table*}[h!tbp]                                                                               
\begin{center}                                                                                      
\caption{(continued) DGS and KINGFISH stellar mass and star-formation parameters.}                  
 \begin{tabular}{lccc | cc}                                                                         
\hline                                                                                              
\hline                                                                                              
Name &             Log[\mstar] &         Log[SFR] &         Log[\ssfr] &Log[L$_{H\alpha}$] &              Ref \\
 &                 Log[\msun] &         Log[\sfr] &     Log[yr$^{-1}$] &       Log[erg/s] &                   \\
 &                         (1) &               (2) &               (3) &              (4)  &              (5) \\
\hline                                                                                              
NGC1512     & $       10.36    \pm  0.19               $ & $       -0.52    \pm  0.15               $ & $      -10.89    \pm  0.24               $ & $         40.42       $ & $        (2)       $  \\
NGC2146     & $       10.94    \pm  0.04               $ & $        0.90    \pm  0.14               $ & $      -10.05    \pm  0.14               $ & $         41.37       $ & $        (3)       $  \\
HoII        & $        8.23    \pm  0.21               $ & $       -1.51    \pm  0.15               $ & $       -9.74    \pm  0.25               $ & $         39.73       $ & $        (4)       $  \\
DDO053      & $        7.24    \pm  0.32               $ & $       -2.11    \pm  0.08               $ & $       -9.35    \pm  0.33               $ & $         38.78   $$^a$ & $        (2)       $  \\
NGC2798     & $       10.36    \pm  0.25               $ & $        0.39    \pm  0.20               $ & $       -9.98    \pm  0.32               $ & $         40.95       $ & $        (2)       $  \\
NGC2841     & $       11.08    \pm  0.20               $ & $       -0.10    \pm  0.08               $ & $      -11.18    \pm  0.21               $ & $         40.81       $ & $        (2)       $  \\
NGC2915     & $        8.49    \pm  0.22               $ & $       -1.78    \pm  0.17               $ & $      -10.28    \pm  0.27               $ & $         39.43       $ & $        (2)       $  \\
HoI         & $        7.84    \pm  0.14               $ & $       -2.08    \pm  0.06               $ & $       -9.92    \pm  0.16               $ & $         38.83   $$^a$ & $        (4)       $  \\
NGC2976     & $        9.32    \pm  0.21               $ & $       -1.11    \pm  0.04               $ & $      -10.43    \pm  0.21               $ & $         39.89       $ & $        (2)       $  \\
NGC3049     & $        9.71    \pm  0.19               $ & $       -0.41    \pm  0.06               $ & $      -10.12    \pm  0.20               $ & $         40.58       $ & $        (2)       $  \\
NGC3077     & $        9.46    \pm  0.20               $ & $       -1.06    \pm  0.02               $ & $      -10.52    \pm  0.20               $ & $         39.98       $ & $        (4)       $  \\
M81dwB      & $        7.24    \pm  0.32               $ & $       -2.36    \pm  0.34               $ & $       -9.61    \pm  0.47               $ & $         38.37   $$^a$ & $        (4)       $  \\
NGC3190     & $       10.74    \pm  0.20               $ & $       -0.47    \pm  0.18               $ & $      -11.21    \pm  0.27               $ & $         39.59       $ & $        (2)       $  \\
NGC3184     & $       10.49    \pm  0.21               $ & $       -0.07    \pm  0.09               $ & $      -10.56    \pm  0.23               $ & $         40.93       $ & $        (2)       $  \\
NGC3198     & $       10.35    \pm  0.21               $ & $       -0.12    \pm  0.10               $ & $      -10.47    \pm  0.23               $ & $         40.87       $ & $        (2)       $  \\
IC2574      & $        8.99    \pm  0.20               $ & $       -1.18    \pm  0.08               $ & $      -10.17    \pm  0.22               $ & $         40.01       $ & $        (2)       $  \\
NGC3265     & $        9.55    \pm  0.21               $ & $       -0.65    \pm  0.09               $ & $      -10.21    \pm  0.23               $ & $         40.22       $ & $        (2)       $  \\
NGC3351     & $       10.47    \pm  0.20               $ & $       -0.27    \pm  0.08               $ & $      -10.74    \pm  0.22               $ & $         40.58       $ & $        (2)       $  \\
NGC3521     & $       10.99    \pm  0.20               $ & $        0.37    \pm  0.05               $ & $      -10.62    \pm  0.21               $ & $         41.17       $ & $        (2)       $  \\
NGC3621     & $       10.20    \pm  0.20               $ & $        0.01    \pm  0.05               $ & $      -10.19    \pm  0.20               $ & $         41.11       $ & $        (2)       $  \\
NGC3627     & $       10.79    \pm  0.20               $ & $        0.29    \pm  0.04               $ & $      -10.50    \pm  0.20               $ & $         41.11       $ & $        (2)       $  \\
NGC3773     & $        9.14    \pm  0.20               $ & $       -0.93    \pm  0.02               $ & $      -10.08    \pm  0.20               $ & $         40.21       $ & $        (2)       $  \\
NGC3938     & $       10.60    \pm  0.19               $ & $        0.21    \pm  0.02               $ & $      -10.39    \pm  0.19               $ & $         41.20       $ & $        (2)       $  \\
NGC4236     & $        9.08    \pm  0.19               $ & $       -0.87    \pm  0.06               $ & $       -9.95    \pm  0.20               $ & $         40.33       $ & $        (4)       $  \\
NGC4254     & $       10.73    \pm  0.20               $ & $        0.49    \pm  0.04               $ & $      -10.25    \pm  0.21               $ & $         41.38       $ & $        (2)       $  \\
NGC4321     & $       10.86    \pm  0.20               $ & $        0.35    \pm  0.04               $ & $      -10.51    \pm  0.21               $ & $         41.20       $ & $        (2)       $  \\
NGC4536     & $       10.44    \pm  0.19               $ & $        0.17    \pm  0.11               $ & $      -10.26    \pm  0.22               $ & $         40.88       $ & $        (2)       $  \\
NGC4559     & $        9.82    \pm  0.21               $ & $       -0.40    \pm  0.05               $ & $      -10.23    \pm  0.21               $ & $         40.71       $ & $        (2)       $  \\
NGC4569     & $       10.51    \pm  0.19               $ & $       -0.45    \pm  0.08               $ & $      -10.96    \pm  0.21               $ & $         40.29       $ & $        (2)       $  \\
NGC4579     & $       11.03    \pm  0.20               $ & $       -0.05    \pm  0.06               $ & $      -11.09    \pm  0.21               $ & $         40.86       $ & $        (2)       $  \\
NGC4594     & $       11.19    \pm  0.20               $ & $       -0.75    \pm  0.17               $ & $      -11.95    \pm  0.26               $ & $         39.70       $ & $        (4)       $  \\
NGC4625     & $        9.27    \pm  0.19               $ & $       -1.23    \pm  0.15               $ & $      -10.51    \pm  0.24               $ & $         39.81       $ & $        (2)       $  \\
NGC4631     & $       10.44    \pm  0.20               $ & $        0.27    \pm  0.04               $ & $      -10.17    \pm  0.20               $ & $         41.19       $ & $        (2)       $  \\
NGC4725     & $       10.85    \pm  0.20               $ & $       -0.09    \pm  0.20               $ & $      -10.94    \pm  0.28               $ & $         40.98       $ & $       (13)       $  \\
NGC4736     & $       10.51    \pm  0.20               $ & $       -0.36    \pm  0.10               $ & $      -10.87    \pm  0.22               $ & $         40.48       $ & $        (2)       $  \\
DDO154      & $        7.35    \pm  0.41               $ & $       -2.24    \pm  0.20               $ & $       -9.59    \pm  0.46               $ & $         38.57   $$^a$ & $        (2)       $  \\
NGC4826     & $       10.48    \pm  0.20               $ & $       -0.57    \pm  0.10               $ & $      -11.05    \pm  0.22               $ & $         40.18       $ & $        (2)       $  \\
DDO165      & $        8.01    \pm  0.20               $ & $       -2.30    \pm  0.26               $ & $      -10.31    \pm  0.33               $ & $         38.48   $$^a$ & $        (2)       $  \\
NGC5055     & $       10.77    \pm  0.20               $ & $        0.11    \pm  0.06               $ & $      -10.66    \pm  0.21               $ & $         40.91       $ & $        (2)       $  \\
NGC5398     & $        8.99    \pm  0.02               $ & $       -1.18    \pm  0.06               $ & $      -10.18    \pm  0.07               $ & $         39.94       $ & $       (13)       $  \\
NGC5408     & $        8.60    \pm  0.20               $ & $       -1.06    \pm  0.02               $ & $       -9.67    \pm  0.20               $ & $         40.15       $ & $        (2)       $  \\
NGC5457     & $       10.67    \pm  0.20               $ & $        0.32    \pm  0.12               $ & $      -10.35    \pm  0.23               $ & $         41.33       $ & $        (4)       $  \\
NGC5474     & $        9.17    \pm  0.17               $ & $       -1.11    \pm  0.25               $ & $      -10.28    \pm  0.30               $ & $         40.01       $ & $        (2)       $  \\
NGC5713     & $       10.50    \pm  0.22               $ & $        0.32    \pm  0.10               $ & $      -10.18    \pm  0.24               $ & $         40.96       $ & $        (2)       $  \\
NGC5866     & $       10.80    \pm  0.20               $ & $       -0.38    \pm  0.13               $ & $      -11.18    \pm  0.24               $ & $         40.53       $ & $       (14)       $  \\
NGC6946     & $       10.77    \pm  0.20               $ & $        0.53    \pm  0.04               $ & $      -10.24    \pm  0.21               $ & $         41.53       $ & $        (2)       $  \\
NGC7331     & $       11.15    \pm  0.20               $ & $        0.49    \pm  0.07               $ & $      -10.66    \pm  0.21               $ & $         41.22       $ & $        (2)       $  \\
NGC7793     & $        9.72    \pm  0.19               $ & $       -0.57    \pm  0.05               $ & $      -10.28    \pm  0.20               $ & $         40.56       $ & $        (2)       $  \\
\hline                                                                                              
\end{tabular}                                                                                       
\end{center}                                                                                        
{\footnotesize                                                                                      
\noindent                                                                                           
{\bf Notes:}                                                                                        
(1) Log of stellar masses in Log[\msun], estimated with the formula from \cite{Eskew2012};                                                                                                                                                                                                                  
(2) Log of star-formation rates from H$\alpha$+\ltir, in Log[\sfr] (see Section \ref{sssec:sfr});                                                                                                       
(3) Log of specific star-formation rates defined by SFR/\mstar, in Log[yr$^{-1}$];                                                                                                                      
(4) Log of H$\alpha$ luminosities corrected for underlying stellar absorption, N{\sc II} line contamination and foreground Galactic extinction, in Log[erg/s];                                                                                                                                              
(5) References for L$_{H_\alpha}$.                                                                  
                                                                                                    
\noindent                                          
$^a$ SFR estimated from \cite{Lee2009} (see Section \ref{sssec:sfr}). \\
                                                 
{\bf References for H$\alpha$ luminosities} :                                                       
(1) \cite{Schmitt2006} ; 
(2) \cite{Kennicutt2009} ; 
(3) \cite{Moustakas2006a} ; 
(4) \cite{Kennicutt2008} ; 
(5) \cite{DePaz2003} ; 
(6) \cite{Popescu2000} ; 
(7) \cite{IglesiasParamo1997} ; 
(8) \cite{Pustilnik2004} ; 
(9) \cite{Izotov1994} ; 
(10) \cite{Izotov1998} ; 
(11) \cite{Terlevich1991} ; 
(12) \cite{Moles1994} ; 
(13) \cite{Skibba2011} ; 
(14) \cite{Plana1998} \\
}                                                                                                   
\end{table*}

%% file: Herschel_fluxes_fin.tex
\begin{table*}[h!tbp]
\begin{center}
\caption{Updated {\it Herschel} flux densities for the DGS sample.}
\label{t:herschel}
{\small
 \begin{tabular}{lcccc | ccc}
\hline
\hline
& \multicolumn{4}{c}{PACS} &  \multicolumn{3}{c}{SPIRE} \\
Name &              F$_{70}$ &       F$_{100}$ &       F$_{160}$ &             Map &       F$_{250}$ &       F$_{350}$ &      F$_{500}$ \\
 &                      [Jy] &            [Jy] &            [Jy] &      method$^a$ &            [Jy] &            [Jy] &           [Jy] \\
\hline
Haro11       &        6.34 $\pm$ 0.32     &        5.04 $\pm$ 0.25     &        2.15 $\pm$ 0.11     &                S &       0.59 $\pm$ 0.05     &        0.22 $\pm$ 0.02     &        0.080 $\pm$ 0.009     \\
Haro2        &        5.09 $\pm$ 0.25     &        5.47 $\pm$ 0.27     &        3.66 $\pm$ 0.18     &                S &       1.23 $\pm$ 0.10     &        0.52 $\pm$ 0.04     &        0.17 $\pm$ 0.02     \\
Haro3        &        5.59 $\pm$ 0.28     &        6.09 $\pm$ 0.30     &        4.73 $\pm$ 0.24     &                S &       1.72 $\pm$ 0.14     &        0.76 $\pm$ 0.06     &        0.25 $\pm$ 0.02     \\
He2-10       &        26.9 $\pm$  1.3     &        26.4 $\pm$  1.3     &        17.7 $\pm$  0.9     &                S &       6.47 $\pm$ 0.52     &        2.56 $\pm$ 0.21     &        0.90 $\pm$ 0.08     \\
HS0017+1055  &        0.046 $\pm$ 0.005     &        0.033 $\pm$ 0.004     &        0.019 $\pm$ 0.004     &                P &$\leq$ 0.030     & $\leq$ 0.050     & $\leq$ 0.045     \\
HS0052+2536  &        0.22 $\pm$ 0.01     &        0.21 $\pm$ 0.01     &        0.139 $\pm$ 0.008     &                P &       0.056 $\pm$ 0.007     &        0.030 $\pm$ 0.010     &        0.021 $\pm$ 0.008     \\
HS0822+3542  & $\leq$ 0.041$^b$ & $\leq$ 0.048$^b$ &        0.034 $\pm$ 0.003     &                P &-     & -     & -     \\
HS1222+3741  &        0.025 $\pm$ 0.004     & $\leq$ 0.036     & $\leq$ 0.022     &                P &-     & -     & -     \\
HS1236+3937  & $\leq$ 0.029     & $\leq$ 0.035     & $\leq$ 0.028     &                P &$\leq$ 0.030     & $\leq$ 0.050     & $\leq$ 0.045     \\
HS1304+3529  &        0.121 $\pm$ 0.007     &        0.150 $\pm$ 0.009     &        0.069 $\pm$ 0.005     &                P &       0.029 $\pm$ 0.005     & $\leq$ 0.050     & $\leq$ 0.045     \\
HS1319+3224  &        0.012 $\pm$ 0.003     &        0.013 $\pm$ 0.002     & $\leq$ 0.015     &                P &-     & -     & -     \\
HS1330+3651  &        0.093 $\pm$ 0.006     &        0.112 $\pm$ 0.007     &        0.091 $\pm$ 0.005     &                P &-     & -     & -     \\
HS1442+4250  &        0.09 $\pm$ 0.01     & $\leq$ 0.054     & $\leq$ 0.047     &                P &$\leq$ 0.030     & $\leq$ 0.050     & $\leq$ 0.045     \\
HS2352+2733  &        0.039 $\pm$ 0.003     &        0.016 $\pm$ 0.002     & $\leq$ 0.016     &                P &$\leq$ 0.030     & $\leq$ 0.050     & $\leq$ 0.045     \\
IZw18        &        0.045 $\pm$ 0.003     &        0.018 $\pm$ 0.002     & $\leq$ 0.011     &                P &$\leq$ 0.030     & $\leq$ 0.050     & $\leq$ 0.045     \\
IC10         &        145. $\pm$   7.     &        211. $\pm$  11.     &        209. $\pm$  10.     &                S &       99.5 $\pm$ 30.9     &        48.1 $\pm$ 14.9     &        19.3 $\pm$  6.0     \\
IIZw40       &        6.66 $\pm$ 0.33     &        6.01 $\pm$ 0.30     &        3.27 $\pm$ 0.16     &                S &       1.27 $\pm$ 0.11     &        0.53 $\pm$ 0.05     &        0.15 $\pm$ 0.01     \\
Mrk1089      &        4.86 $\pm$ 0.24     &        5.17 $\pm$ 0.26     &        4.41 $\pm$ 0.22     &                S &       1.72 $\pm$ 0.14     &        0.79 $\pm$ 0.07     &        0.29 $\pm$ 0.03     \\
Mrk1450      &        0.30 $\pm$ 0.02     &        0.25 $\pm$ 0.01     &        0.127 $\pm$ 0.007     &                P &       0.046 $\pm$ 0.006     & $\leq$ 0.050     & $\leq$ 0.045     \\
Mrk153       &        0.28 $\pm$ 0.01     &        0.30 $\pm$ 0.02     &        0.137 $\pm$ 0.009     &                P &       0.045 $\pm$ 0.007     & $\leq$ 0.050     & $\leq$ 0.045     \\
Mrk209       &        0.36 $\pm$ 0.02     &        0.37 $\pm$ 0.02     &        0.17 $\pm$ 0.01     &                S &       0.059 $\pm$ 0.007     &        0.033 $\pm$ 0.009     & $\leq$ 0.045     \\
Mrk930       &        1.20 $\pm$ 0.06     &        1.48 $\pm$ 0.07     &        0.96 $\pm$ 0.05     &                S &       0.39 $\pm$ 0.03     &        0.19 $\pm$ 0.02     &        0.08 $\pm$ 0.01     \\
NGC1140      &        4.24 $\pm$ 0.21     &        4.44 $\pm$ 0.22     &        4.47 $\pm$ 0.22     &                S &       1.89 $\pm$ 0.16     &        0.92 $\pm$ 0.08     &        0.33 $\pm$ 0.04     \\
NGC1569      &        61.7 $\pm$  3.1     &        58.9 $\pm$  2.9     &        37.4 $\pm$  1.9     &                S &       12.1 $\pm$  1.0     &        5.03 $\pm$ 0.41     &        1.85 $\pm$ 0.16     \\
NGC1705      &        1.10 $\pm$ 0.06     &        1.37 $\pm$ 0.07     &        1.14 $\pm$ 0.06     &            S$^c$ &       0.59 $\pm$ 0.05     &        0.30 $\pm$ 0.03     &        0.12 $\pm$ 0.02     \\
NGC2366      &        5.82 $\pm$ 0.29     &        7.05 $\pm$ 0.35     &        4.30 $\pm$ 0.22     &                S &       2.00 $\pm$ 0.17     &        1.07 $\pm$ 0.09     &        0.45 $\pm$ 0.04     \\
NGC4214      &        26.8 $\pm$  1.3     &        35.0 $\pm$  1.8     &        33.0 $\pm$  1.6     &                S &       18.6 $\pm$  1.5     &        9.96 $\pm$ 0.81     &        4.53 $\pm$ 0.37     \\
NGC4449      &        50.4 $\pm$  2.5     &        79.6 $\pm$  4.0     &        75.6 $\pm$  3.8     &                S &       32.2 $\pm$  2.6     &        14.8 $\pm$  1.2     &        5.90 $\pm$ 0.48     \\
NGC4861      &        2.31 $\pm$ 0.12     &        2.27 $\pm$ 0.11     &        2.36 $\pm$ 0.12     &                S &       1.06 $\pm$ 0.09     &        0.55 $\pm$ 0.05     &        0.23 $\pm$ 0.03     \\
NGC5253      &        33.7 $\pm$  1.7     &        33.5 $\pm$  1.7     &        22.5 $\pm$  1.1     &                S &       7.78 $\pm$ 0.63     &        3.59 $\pm$ 0.29     &        1.37 $\pm$ 0.11     \\
NGC625       &        6.14 $\pm$ 0.31     &        9.98 $\pm$ 0.50     &        8.32 $\pm$ 0.42     &                S &       4.34 $\pm$ 0.35     &        2.18 $\pm$ 0.18     &        0.95 $\pm$ 0.08     \\
NGC6822      &        56.2 $\pm$  2.9     &        72.5 $\pm$  3.7     &        79.0 $\pm$  4.0     &                S &       49.7 $\pm$ 15.5     &        30.8 $\pm$  9.6     &        16.1 $\pm$  5.0     \\
Pox186       &        0.051 $\pm$ 0.003     &        0.059 $\pm$ 0.005     &        0.067 $\pm$ 0.001     &                P &       0.042 $\pm$ 0.006     & $\leq$ 0.050     & $\leq$ 0.045     \\
SBS0335-052  &        0.056 $\pm$ 0.004     &        0.024 $\pm$ 0.001     &        0.007 $\pm$ 0.001     &                P &$\leq$ 0.030     & $\leq$ 0.050     & $\leq$ 0.045     \\
SBS1159+545  &        0.019 $\pm$ 0.003     &        0.019 $\pm$ 0.003     & $\leq$ 0.018     &                P &$\leq$ 0.030     & $\leq$ 0.050     & $\leq$ 0.045     \\
SBS1211+540  &        0.034 $\pm$ 0.003     &        0.018 $\pm$ 0.002     &        0.013 $\pm$ 0.002     &                P &$\leq$ 0.030     & $\leq$ 0.050     & $\leq$ 0.045     \\
SBS1249+493  &        0.032 $\pm$ 0.005     & $\leq$ 0.034     & $\leq$ 0.042     &                P &$\leq$ 0.030     & $\leq$ 0.050     & $\leq$ 0.045     \\
SBS1415+437  &        0.18 $\pm$ 0.01     &        0.16 $\pm$ 0.01     &        0.065 $\pm$ 0.007     &                P &-     & -     & -     \\
SBS1533+574  &        0.23 $\pm$ 0.01     &        0.30 $\pm$ 0.02     &        0.174 $\pm$ 0.010     &                S &$\leq$ 0.145     & $\leq$ 0.050     & $\leq$ 0.045     \\
Tol0618-402  & $\leq$ 0.014     & $\leq$ 0.011     & $\leq$ 0.013     &                P &-     & -     & -     \\
Tol1214-277  &        0.017 $\pm$ 0.003     &        0.018 $\pm$ 0.002     & $\leq$ 0.018     &                P &$\leq$ 0.030     & $\leq$ 0.050     & $\leq$ 0.045     \\
UGC4483      &        0.11 $\pm$ 0.02     &        0.27 $\pm$ 0.02     &        0.091 $\pm$ 0.010     &            S$^c$ &       0.015 $\pm$ 0.005     & $\leq$ 0.050     & $\leq$ 0.045     \\
UGCA20       & $\leq$ 0.052     & $\leq$ 0.057     & $\leq$ 0.048     &                P &-     & -     & -     \\
UM133        &        0.12 $\pm$ 0.02     &        0.13 $\pm$ 0.02     &        0.078 $\pm$ 0.009     &            P$^d$ &       0.025 $\pm$ 0.006     & $\leq$ 0.050     & $\leq$ 0.045     \\
UM311        &        3.13 $\pm$ 0.16     &        5.52 $\pm$ 0.28     &        6.12 $\pm$ 0.31     &                S &       3.79 $\pm$ 0.31     &        1.91 $\pm$ 0.16     &        0.81 $\pm$ 0.07     \\
UM448        &        5.23 $\pm$ 0.26     & -     &        3.32 $\pm$ 0.17     &                S &       0.95 $\pm$ 0.08     &        0.36 $\pm$ 0.03     &        0.11 $\pm$ 0.01     \\
UM461        &        0.21 $\pm$ 0.01     &        0.145 $\pm$ 0.009     &        0.113 $\pm$ 0.007     &                P &       0.025 $\pm$ 0.006     &        0.024 $\pm$ 0.010     & $\leq$ 0.045     \\
VIIZw403     &        0.47 $\pm$ 0.03     &        0.56 $\pm$ 0.03     &        0.34 $\pm$ 0.02     &                P &       0.14 $\pm$ 0.01     &        0.062 $\pm$ 0.008     &        0.027 $\pm$ 0.008     \\
\hline
\end{tabular}
}
\end{center}
{\footnotesize                                                                                      
{\bf Notes : } \\                                                                                   
\noindent                                                                                           
$^a$ P = {\sc PhotProject}, S = {\sc Scanamorphos}. \\                                                                                                                                                  
$^b$ These upper limits come from the SED modelling (see Appendix \ref{ap:default}). \\                                                                                                               
$^c$ For these galaxies, the PACS map-making method was changed from {\sc PhotProject} to {\sc Scanamorphos} since \cite{RemyRuyer2013}. \\                                                             
$^d$ For these galaxies, the PACS map-making method was changed from {\sc Scanamorphos} to {\sc PhotProject} since \cite{RemyRuyer2013}. \\                                                             
}                                                                                                   
 \end{table*}

%% file: IRAC_fluxes_fin.tex
\begin{table*}[h!tbp]
\begin{center}
\caption{{\it Spitzer} IRAC flux densities for the DGS sample.}
\label{t:irac}
{\small
 \begin{tabular}{lcccccc}
\hline
\hline
Source &           F$_{3.6}$ &       F$_{4.5}$ &       F$_{5.8}$ &       F$_{8.0}$ & Aperture radius &    Aper corr ? \\
 &                     [mJy] &           [mJy] &           [mJy] &           [mJy] &       [\arcsec] &                \\
\hline
Haro11       &            22.4    $\pm$      3.2        &            32.3    $\pm$      4.6        &            77.1    $\pm$     10.9        &            164.    $\pm$      23.        &                     45$^a$ &                       yes        \\
Haro2        &            24.7    $\pm$      3.5        &            18.2    $\pm$      2.6        &            49.3    $\pm$      7.0        &            257.    $\pm$      36.        &                     50$^a$ &                       yes        \\
Haro3        &            28.2    $\pm$      4.0        &            21.5    $\pm$      3.1        &            43.9    $\pm$      6.3        &            116.    $\pm$      16.        &                     60$^a$ &                       yes        \\
He2-10       &            134.    $\pm$      19.        &            99.2    $\pm$     14.1        &            296.    $\pm$      42.        &            836.    $\pm$     118.        &                    108$^a$ &                       yes        \\
HS0017+1055  &        0.17    $\pm$ 0.02        &        0.31    $\pm$ 0.03        &        0.76    $\pm$ 0.08        &            1.53    $\pm$     0.16        &                     12$^b$ &                                  \\
HS0052+2536  &        0.7    $\pm$ 0.1        &        0.32    $\pm$ 0.09        &        0.7    $\pm$ 0.2        &            3.17    $\pm$     0.46        &                     17$^c$ &                       yes        \\
HS0822+3542  &        0.12    $\pm$ 0.01        &        0.10    $\pm$ 0.01        &        0.10    $\pm$ 0.02        &        0.09    $\pm$ 0.02        &                      5$^b$ &                       yes        \\
HS1222+3741  &        0.25    $\pm$ 0.03        &        0.19    $\pm$ 0.02        &        0.27    $\pm$ 0.05        &        0.57    $\pm$ 0.07        &                     14$^a$ &                                  \\
HS1236+3937  & $\leq$  0.56            & $\leq$  0.38            & $\leq$  0.66            & $\leq$  0.32            &                     15$^a$ &                                  \\
HS1304+3529  &        0.38    $\pm$ 0.08        &        0.41    $\pm$ 0.09        & $\leq$  0.78            &        0.6    $\pm$ 0.1        &                     18$^a$ &                                  \\
HS1319+3224  &        0.061    $\pm$ 0.010        &        0.051    $\pm$ 0.008        &        0.10    $\pm$ 0.02        &        0.12    $\pm$ 0.02        &                      8$^a$ &                                  \\
HS1330+3651  &        0.56    $\pm$ 0.09        &        0.5    $\pm$ 0.1        & $\leq$  0.80            &        1.0    $\pm$ 0.2        &                     20$^a$ &                                  \\
HS1442+4250  & -            &            1.24    $\pm$     0.21        & -            &            1.59    $\pm$     0.25        &                     51$^a$ &                       yes        \\
HS2352+2733  & $\leq$  0.04            & $\leq$  0.04            & $\leq$  0.13            & $\leq$  0.11            &                     15$^a$ &                                  \\
IZw18        &        0.36    $\pm$ 0.05        &        0.35    $\pm$ 0.05        &        0.39    $\pm$ 0.07        &        0.53    $\pm$ 0.08        &                     12$^b$ &                       yes        \\
IC10$^f$         & -            & -            & -            & -            &                    - &                           -        \\
IIZw40       &            18.1    $\pm$      4.3        &            19.9    $\pm$      3.5        &            39.8    $\pm$      5.7        &            105.    $\pm$      15.        &                     45 $\times$ 33$^c$ &                       yes        \\
Mrk1089      &            20.4    $\pm$      2.9        &            15.5    $\pm$      2.2        &            31.4    $\pm$      4.4        &            82.2    $\pm$     11.6        &                     75$^a$ &                       yes        \\
Mrk1450      &            1.22    $\pm$     0.17        &            1.01    $\pm$     0.14        &        0.9    $\pm$ 0.1        &            2.06    $\pm$     0.29        &                     20$^a$ &                       yes        \\
Mrk153       &            2.29    $\pm$     0.33        &            1.70    $\pm$     0.24        &            1.76    $\pm$     0.25        &            2.75    $\pm$     0.39        &                     25$^b$ &                       yes        \\
Mrk209       &            2.90    $\pm$     0.41        &            2.21    $\pm$     0.32        &            1.78    $\pm$     0.27        &            2.20    $\pm$     0.32        &                     39$^d$ &                       yes        \\
Mrk930       &            2.66    $\pm$     0.38        &            2.73    $\pm$     0.39        &            4.17    $\pm$     0.59        &             10.    $\pm$       1.        &                     13$^b$ &                       yes        \\
NGC1140      &            32.8    $\pm$      4.8        &            23.3    $\pm$      3.4        &            41.0    $\pm$      5.8        &            93.9    $\pm$     13.3        &                    118$^a$ &                       yes        \\
NGC1569      &            312.    $\pm$      45.        &            244.    $\pm$      35.        &            324.    $\pm$      46.        &            533.    $\pm$      75.        &                    150$^a$ &                       yes        \\
NGC1705      &            25.5    $\pm$      5.9        &            18.3    $\pm$      4.2        &            15.0    $\pm$      2.8        &            17.5    $\pm$      2.6        &                     72$^a$ &                       yes        \\
NGC2366      &            61.5    $\pm$     12.4        &            45.7    $\pm$     10.4        &            35.8    $\pm$      7.6        &            52.8    $\pm$      8.1        &                    200$^d$ &                       yes        \\
NGC4214      &            334.    $\pm$      47.        &            253.    $\pm$      36.        &            280.    $\pm$      40.        &            748.    $\pm$     106.        &                   205 $\times$ 160$^c$ &                       yes        \\
NGC4449      &            472.    $\pm$      67.        &            342.    $\pm$      48.        &            709.    $\pm$     100.        &           1622.    $\pm$     229.        &                190 $\times$ 170$^c$ &                       yes        \\
NGC4861      &            18.3    $\pm$      2.6        &            13.8    $\pm$      2.0        &            16.8    $\pm$      2.5        &            17.7    $\pm$      2.5        &                    100 $\times$ 50$^c$ &                       yes        \\
NGC5253      &            235.    $\pm$      34.        &            252.    $\pm$      36.        &            458.    $\pm$      65.        &            812.    $\pm$     115.        &                    120$^a$ &                       yes        \\
NGC625       &            110.    $\pm$      16.        &            79.7    $\pm$     11.4        &            71.3    $\pm$     10.1        &            147.    $\pm$      21.        &                    170$^a$ &                       yes        \\
NGC6822      &           1780.    $\pm$     272.        &           1240.    $\pm$     196.        &            951.    $\pm$     141.        &           1226.    $\pm$     175.        &                    440$^a$ &                       yes        \\
Pox186       &        0.18    $\pm$ 0.02        &        0.18    $\pm$ 0.02        &        0.14    $\pm$ 0.03        &        0.64    $\pm$ 0.07        &                      9$^b$ &                                  \\
SBS0335-052  &        0.78    $\pm$ 0.08        &            1.64    $\pm$     0.17        &            4.50    $\pm$     0.45        &            12.3    $\pm$      1.2        &                     10$^a$ &                                  \\
SBS1159+545  &        0.11    $\pm$ 0.01        &        0.10    $\pm$ 0.01        &        0.12    $\pm$ 0.03        &        0.59    $\pm$ 0.07        &                     12$^d$ &                                  \\
SBS1211+540  &        0.16    $\pm$ 0.02        &        0.08    $\pm$ 0.01        & $\leq$  0.12            &        0.30    $\pm$ 0.05        &                     15$^a$ &                                  \\
SBS1249+493  &        0.08    $\pm$ 0.02        &        0.07    $\pm$ 0.02        & $\leq$  0.20            &        0.26    $\pm$ 0.04        &                      5$^b$ &                                  \\
SBS1415+437  & -            &        0.8    $\pm$ 0.1        & -            &        0.50    $\pm$ 0.08        &                     34$^a$ &                       yes        \\
SBS1533+574  &            2.09    $\pm$     0.34        &            1.52    $\pm$     0.24        &            2.26    $\pm$     0.33        &            3.32    $\pm$     0.48        &                     30$^a$ &                       yes        \\
Tol0618-402  &            2.68    $\pm$     0.38        &            1.64    $\pm$     0.23        &            1.13    $\pm$     0.16        &        0.9    $\pm$ 0.1        &                     18$^a$ &                       yes        \\
Tol1214-277  &        0.076    $\pm$ 0.009        &        0.09    $\pm$ 0.01        &        0.15    $\pm$ 0.03        &        0.25    $\pm$ 0.03        &                      8$^b$ &                                  \\
UGC4483      &            1.69    $\pm$     0.25        &            1.02    $\pm$     0.15        & $\leq$  0.62 &        0.9    $\pm$ 0.1        &                   43 $\times$ 24$^c$ &                       yes        \\
UGCA20       & $\leq$  0.85            & $\leq$  0.60            & $\leq$  1.07            & $\leq$  0.46            &                     20$^a$ &                                  \\
UM133        &        0.39    $\pm$ 0.08        &        0.28    $\pm$ 0.06        & $\leq$  0.46            & $\leq$  0.27            &                      8$^a$ &                                  \\
UM311        & -            &            33.4    $\pm$      4.7        & -            &            114.    $\pm$      16.        &                    115$^e$ &                       yes        \\
UM448        &            17.7    $\pm$      2.5        &            13.2    $\pm$      1.9        &            34.1    $\pm$      4.8        &            83.0    $\pm$     11.7        &                     53 $\times$ 31$^c$ &                       yes        \\
UM461        &        0.8    $\pm$ 0.1        &        0.61    $\pm$ 0.09        &        0.8    $\pm$ 0.1        &            1.58    $\pm$     0.29        &                     17$^a$ &                       yes        \\
VIIZw403     &            2.86    $\pm$     0.41        &            2.23    $\pm$     0.32        &            1.69    $\pm$     0.25        &            2.56    $\pm$     0.37        &                     40$^a$ &                       yes        \\
\hline
\end{tabular}
}
\end{center}
{\footnotesize                                                                                      
{\bf Notes : } \\                                                                                   
\noindent                                                                                           
$^a$ The aperture is the same as the one used for \hersc. \\                                                                                                                                            
$^b$ The \hersc\ aperture has been shorten to avoid a contaminating source. \\                                                                                                                          
$^c$ The \hersc\ aperture has been adapted to match the peculiar morphology of the source in the NIR. \\                                                                          
$^d$ The \hersc\ aperture has been enlarged to encompass all of the NIR emission. \\                                                                                                                    
$^e$ The IRAC map is smaller than the \hersc\ aperture. The aperture thus had to be shorten. \\                                                                                                        
$^f$  The IRAC maps do not cover the full galaxy. Thus we do not report flux densities for this source. \\                                                       
}                                                                                                   
 \end{table*}

%% file: WISE_fluxes_fin.tex
\begin{table*}[h!tbp]
\begin{center}
\caption{WISE flux densities for the DGS sample.}
\label{t:wise}
{\small
 \begin{tabular}{lcccccc}
\hline
\hline
Source &           F$_{3.4}$ &       F$_{4.6}$ &        F$_{12}$ &        F$_{22}$ &                 Aperture radius \\
 &                     [mJy] &           [mJy] &           [mJy] &           [mJy] &                       [\arcsec] \\
\hline
Haro11       &            17.9    $\pm$      1.9        &            32.6    $\pm$      3.4        &            353.    $\pm$      19.        &           2054.    $\pm$     148.        &                   45$^a$, 90$^b$ \\
Haro2        &            24.5    $\pm$      2.6        &            16.5    $\pm$      1.7        &            152.    $\pm$       8.        &            760.    $\pm$      55.        &                   50$^a$, 65$^b$ \\
Haro3        &            25.8    $\pm$      2.7        &            17.7    $\pm$      1.9        &            162.    $\pm$       9.        &            766.    $\pm$      56.        &                           60$^a$ \\
He2-10       &            133.    $\pm$      17.        &            92.7    $\pm$     11.2        &            996.    $\pm$      53.        &           5018.    $\pm$     362.        &                           81$^b$ \\
HS0017+1055  &        0.16    $\pm$ 0.03        &        0.25    $\pm$ 0.04        &            3.83    $\pm$     0.49        &            11.7    $\pm$      2.0        &                   12$^a$, 25$^b$ \\
HS0052+2536  &        0.65    $\pm$ 0.08        &        0.52    $\pm$ 0.07        &            3.71    $\pm$     0.40        &            16.2    $\pm$      2.3        &                   17$^a$, 23$^a$ \\
HS0822+3542  &        0.106    $\pm$ 0.009$^c$ &        0.082    $\pm$ 0.008$^c$ & $\leq$  0.47$^c$ & $\leq$  3.85$^c$ &                                - \\
HS1222+3741  &        0.15    $\pm$ 0.03        &        0.15    $\pm$ 0.03        &        1.0    $\pm$ 0.2$^c$ &            5.94    $\pm$     0.73$^c$ &                        14$^a$, - \\
HS1236+3937  &        0.22    $\pm$ 0.05        &        0.16    $\pm$ 0.04        & $\leq$  0.65     & $\leq$  2.72     &                           15$^a$ \\
HS1304+3529  &        0.4    $\pm$ 0.1        &        0.43    $\pm$ 0.09        &            1.78    $\pm$     0.21$^c$ &            10.5    $\pm$      0.9$^c$ &                        18$^a$, - \\
HS1319+3224  &        0.06    $\pm$ 0.02        &        0.07    $\pm$ 0.02        & $\leq$  0.70$^c$ & $\leq$  1.93$^c$ &                        12$^b$, - \\
HS1330+3651  &        0.51    $\pm$ 0.07        &        0.36    $\pm$ 0.07        &            1.23    $\pm$     0.26        &              5.31    $\pm$       1.26       &                           20$^a$ \\
HS1442+4250  &            1.20    $\pm$     0.14        &        0.8    $\pm$ 0.1        &            1.54    $\pm$     0.33        & $\leq$  9.40     &                           32$^b$ \\
HS2352+2733  &        0.15    $\pm$ 0.04        &        0.13    $\pm$ 0.04        & $\leq$  1.03$^c$ & $\leq$  2.98$^c$ &                        15$^a$, - \\
IZw18        &        0.33    $\pm$ 0.07        &        0.39    $\pm$ 0.07        &        0.8    $\pm$ 0.3        &              4.64    $\pm$       1.19        &                           23$^b$ \\
IC10         &           1454.    $\pm$     154.        &            917.    $\pm$      98.        &           3521.    $\pm$     191.        &           9991.    $\pm$     728.        &    205 $\times$ 150$^b$, 230$^b$ \\
IIZw40       &            18.6    $\pm$      2.0        &            19.3    $\pm$      2.1        &            335.    $\pm$      18.        &           1540.    $\pm$     112.        &       45 $\times$ 33$^a$, 90$^b$ \\
Mrk1089      &            19.0    $\pm$      2.3        &            13.9    $\pm$      3.1        &            102.    $\pm$       6.        &            453.    $\pm$      34.        &                           75$^a$ \\
Mrk1450      &            1.10    $\pm$     0.12        &        0.73    $\pm$ 0.09        &            8.00    $\pm$     0.74        &            49.3    $\pm$      4.8        &                   20$^a$, 40$^b$ \\
Mrk153       &            2.03    $\pm$     0.23        &            1.34    $\pm$     0.17        &            4.61    $\pm$     0.41        &            28.1    $\pm$      3.9        &                           25$^a$ \\
Mrk209       &            2.61    $\pm$     0.39        &            1.70    $\pm$     0.31        &            7.30    $\pm$     0.85        &            46.4    $\pm$      5.4        &                           39$^a$ \\
Mrk930       &            2.61    $\pm$     0.30        &            2.26    $\pm$     0.27        &            25.8    $\pm$      2.6        &            170.    $\pm$      14.        &                   34$^b$, 60$^a$ \\
NGC1140      &            32.7    $\pm$      3.5        &            21.6    $\pm$      2.3        &            94.9    $\pm$      5.5        &            367.    $\pm$      28.        &                          118$^a$ \\
NGC1569      &            359.    $\pm$      24.        &            240.    $\pm$      47.        &           1021.    $\pm$     116.        &           6971.    $\pm$     503.        &                          120$^b$ \\
NGC1705      &            25.7    $\pm$      2.7        &            16.1    $\pm$      1.7        &            20.3    $\pm$      1.5        &            46.2    $\pm$      6.1        &       75 $\times$ 46$^b$, 72$^a$ \\
NGC2366      &            75.8    $\pm$      5.7        &            39.4    $\pm$      6.2        &            121.    $\pm$      12.        &            612.    $\pm$      65.        &    260 $\times$ 125$^b$, 150$^a$ \\
NGC4214      &            294.    $\pm$      31.        &            184.    $\pm$      19.        &            546.    $\pm$      29.        &           1830.    $\pm$     133.        &                          181$^a$ \\
NGC4449      &            483.    $\pm$      50.        &            307.    $\pm$      32.        &          1323.8    $\pm$     70.0        &           3119.    $\pm$     227.        &    190 $\times$ 170$^a$, 250$^a$ \\
NGC4861      &            18.1    $\pm$      1.1        &            10.1    $\pm$      1.9        &            52.8    $\pm$      5.7        &            330.    $\pm$      26.        &     120 $\times$ 50$^b$, 120$^a$ \\
NGC5253      &            212.    $\pm$      22.        &            255.    $\pm$      27.        &           2002.    $\pm$     105.        &          10345.    $\pm$     746.        &     120 $\times$ 80$^b$, 120$^a$ \\
NGC625       &            106.    $\pm$      11.        &            64.7    $\pm$      6.8        &            194.    $\pm$      11.        &            797.    $\pm$      59.        &                  200 $\times$ 75 \\
NGC6822$^d$  & - & - & - & -        &                                - \\
Pox186       & $\leq$  0.34     & $\leq$  0.44     &            2.67    $\pm$     0.33        &            11.2    $\pm$      1.4        &                           25$^b$ \\
SBS0335-052  &        0.45    $\pm$ 0.05        &            1.45    $\pm$     0.16        &            23.4    $\pm$      1.4        &            67.8    $\pm$      5.8        &                   14$^b$, 35$^b$ \\
SBS1159+545  & $\leq$  0.09     & $\leq$  0.11     &        1.0    $\pm$ 0.3        &              4.44    $\pm$       1.38        &                   12$^a$, 25$^b$ \\
SBS1211+540  &        0.133    $\pm$ 0.009        &        0.10    $\pm$ 0.03        & $\leq$  0.80     & $\leq$  3.60     &                           15$^a$ \\
SBS1249+493  & $\leq$  0.05$^c$ & $\leq$  0.07$^c$ & $\leq$  1.17     & $\leq$  3.76     &                          -, 20$^b$ \\
SBS1415+437  &        0.8    $\pm$ 0.1        &        0.53    $\pm$ 0.08        &            2.30    $\pm$     0.55        &            15.5    $\pm$      2.8        &       28 $\times$ 15$^b$, 34$^a$ \\
SBS1533+574  &            1.52    $\pm$     0.19        &            1.28    $\pm$     0.17        &            9.76    $\pm$     0.84        &            53.3    $\pm$      4.7        &                   30$^a$, 50$^b$ \\
Tol0618-402  &            2.33    $\pm$     0.33        &            1.32    $\pm$     0.21        & $\leq$  0.71     & $\leq$  1.73     &                           18$^a$ \\
Tol1214-277  & $\leq$  0.13     & $\leq$  0.14     &        0.8    $\pm$ 0.2$^c$ & $\leq$  6.89$^c$ &                        12$^a$, - \\
UGC4483      &            1.12    $\pm$     0.26        & $\leq$  1.01     & $\leq$  2.30     & $\leq$ 12.72     &                   43 $\times$ 24 \\
UGCA20       &        0.26    $\pm$ 0.08        & $\leq$  0.24     & $\leq$  0.74     & $\leq$  3.27     &                           20$^a$ \\
UM133        & $\leq$ 10.83     & $\leq$  6.74     & $\leq$  1.49     & $\leq$  6.70     &                           26$^a$ \\
UM311        &            48.5    $\pm$      3.0        &            29.1    $\pm$      3.5        &            120.    $\pm$       7.        &            292.    $\pm$      25.        &                          115$^a$ \\
UM448        &            15.4    $\pm$      1.6        &            11.7    $\pm$      1.3        &            103.    $\pm$       6.        &            552.    $\pm$      40.        &                           64$^a$ \\
UM461        &        0.58    $\pm$ 0.07        &        0.52    $\pm$ 0.07        &              5.98    $\pm$       1.02       &            29.4    $\pm$      4.9        &                   17$^a$, 35$^b$ \\
VIIZw403     &            2.59    $\pm$     0.29        &            1.89    $\pm$     0.23        &            4.54    $\pm$     0.61        &            26.3    $\pm$      3.6        &                           40$^a$ \\
\hline
\end{tabular}
}
\end{center}
{\footnotesize                                                                                      
{\bf Notes : } \\                                                                                   
\noindent                                                                                           
In the aperture column, a single value indicates that only one aperture has been used for the four wavelengths. Two values separated by a comma indicate that a different aperture has been used for WISE1 and WISE2 on one side (first value), and WISE3 and WISE4 on the other side (second value, see text for details). For several galaxies elliptical apertures were used to match the morphology of the source and are indicated by semi-major axis $\times$ semi-minor axis. 
A dash indicates that we take the profile fit photometry from the All WISE database. \\                                                                                                                                                                                                                                                                                                                                                   
$^a$ The aperture is the same as that used for IRAC or \hersc\ photometry. \\                                                                                                                       
$^b$ The aperture has been adapted to match the peculiar morphology of the source in the NIR/MIR or to avoid a contaminating source. \\                                                                 
$^c$ This galaxy is not resolved at these wavelengths and we thus use the profile fit photometry provided by the WISE database. \\                                                                                                                                                                          
$^d$ The WISE map does not cover the full galaxy. Thus we do not report flux densities for this source. \\                                                                                              
}                                                                                                   
 \end{table*}

%% file: Literature_fluxes_fin.tex
\begin{landscape}
\begin{table}[h!tbp]
\begin{center}
\caption{2MASS and IRAS flux densities from the literature for the DGS sample.}
\label{t:jhkiras}
{\scriptsize
 \begin{tabular}{lcccc | ccccc}
\hline
\hline
& \multicolumn{4}{c}{2MASS} &  \multicolumn{5}{c}{IRAS} \\
Source &                   J &               H &               K &             Ref &        F$_{12}$ &        F$_{25}$ &        F$_{60}$ &       F$_{100}$ &            Ref \\
 &                     [mJy] &           [mJy] &           [mJy] &                 &            [Jy] &            [Jy] &            [Jy] &            [Jy] &                \\
\hline
Haro11       &                     13.0 $\pm$         0.4 &                   13.0 $\pm$         0.6 &                   14.1 $\pm$         0.7 &                         1 &               0.42 $\pm$    0.05 &                   2.49 $\pm$        0.16 &                   6.48 $\pm$        0.57 &                   5.01 $\pm$        0.61 &                         6 \\       
Haro2        &                     43.6 $\pm$         1.4 &                   53.5 $\pm$         2.3 &                   44.9 $\pm$         2.3 &                         2 &               0.21 $\pm$    0.03 &               0.95 $\pm$    0.06 &                   4.68 $\pm$        0.41 &                   5.32 $\pm$        0.62 &                         6 \\       
Haro3        &                     44.7 $\pm$         1.4 &                   53.2 $\pm$         2.3 &                   37.9 $\pm$         1.9 &                         1 &               0.21 $\pm$    0.03 &               0.94 $\pm$    0.06 &                   4.95 $\pm$        0.51 &                   6.75 $\pm$        0.79 &                         6 \\       
He2-10       &                     178. $\pm$          6. &                   201. $\pm$          9. &                   167. $\pm$          8. &                         1 &                   1.18 $\pm$        0.14 &                   6.78 $\pm$        0.84 &                   23.4 $\pm$         3.2 &                   26.3 $\pm$         4.7 &                         7 \\       
HS0017+1055  &                 0.38 $\pm$    0.01 &               0.59 $\pm$    0.03 &               0.36 $\pm$    0.02 &                         3 &        - & - & - & - & - \\       
HS0052+2536  &                 0.54 $\pm$    0.02 &               0.55 $\pm$    0.02 &               0.40 $\pm$    0.02 &                         3 &               0.108 $\pm$    0.005$^b$ &           0.148 $\pm$    0.009$^b$ &           0.25 $\pm$    0.02$^b$ &           0.71 $\pm$    0.07$^b$ &                     6 \\       
HS0822+3542  &                 0.15 $\pm$    0.02 &               0.34 $\pm$    0.04 &               0.16 $\pm$    0.03 &                         4 &        - & - & - & - & - \\       
HS1222+3741  &                 0.90 $\pm$    0.03 &                   1.18 $\pm$        0.05 &               0.83 $\pm$    0.04 &                         3 &        - & - & - & - & - \\       
HS1236+3937  &          - & - & - & -  &        - & - & - & - & - \\       
HS1304+3529  &                 0.315 $\pm$    0.010 &               0.27 $\pm$    0.01 &        $\leq$  0.38  &                         3 &        - & - & - & - & - \\       
HS1319+3224  &          - & - & - & -  &        - & - & - & - & - \\       
HS1330+3651  &          - & - & - & -  &        - & - & - & - & - \\       
HS1442+4250  &          - & - & - & -  &        - & - & - & - & - \\       
HS2352+2733  &          - & - & - & -  &        - & - & - & - & - \\       
IZw18        &                 0.75 $\pm$    0.08 &               0.64 $\pm$    0.07 &               0.52 $\pm$    0.06 &                         3 &        - & - & - & - & - \\       
IC10         &                   2051.6 $\pm$        63.9 &                  2972. $\pm$        129. &                  2632. $\pm$        133. &                         2 &        - & - & - & - & - \\       
IIZw40       &                     15.1 $\pm$         1.5 &                   21.0 $\pm$         2.2 &                   20.3 $\pm$         2.1 &                         4 &               0.41 $\pm$    0.04 &                   1.88 $\pm$        0.12 &                   6.02 $\pm$        0.57 &        $\leq$  1.97  &                         6 \\       
Mrk1089      &                     12.2 $\pm$         0.4 &                   13.9 $\pm$         0.6 &                   12.2 $\pm$         0.6 &                         1 &               0.25 $\pm$    0.01$^b$ &           0.70 $\pm$    0.04 &                   4.06 $\pm$        0.26 &                   5.64 $\pm$        0.56 &                         8 \\       
Mrk1450      &                     1.05 $\pm$        0.03 &               0.90 $\pm$    0.04 &               0.66 $\pm$    0.03 &                         3 &        $\leq$  0.06  &        $\leq$  0.10  &               0.28 $\pm$    0.04 &        $\leq$  0.57  &                         6 \\       
Mrk153       &                     4.64 $\pm$        0.56 &                   4.27 $\pm$        0.59 &                   3.63 $\pm$        0.51 &                         4 &        $\leq$  0.08  &        $\leq$  0.09  &               0.28 $\pm$    0.04 &        $\leq$  0.05  &                         6 \\       
Mrk209       &                       7. $\pm$          1. &                     8. $\pm$          2. &                     7. $\pm$          2. &                         5 &        - & - & - & - & - \\       
Mrk930       &                     3.52 $\pm$        0.36 &                   4.64 $\pm$        0.48 &                   3.53 $\pm$        0.36 &                         4 &        $\leq$  0.08  &               0.23 $\pm$    0.02 &                   1.25 $\pm$        0.12 &        $\leq$  2.15  &                         6 \\       
NGC1140      &                     48.9 $\pm$         1.5 &                   57.2 $\pm$         2.5 &                   41.8 $\pm$         2.1 &                         2 &               0.10 $\pm$    0.03 &               0.49 $\pm$    0.06 &                   3.34 $\pm$        0.38 &                   4.92 $\pm$        0.81 &                         7 \\       
NGC1569      &                     473. $\pm$         49. &                   547. $\pm$         57. &                   479. $\pm$         50. &                         2 &               0.87 $\pm$    0.08 &                   7.73 $\pm$        0.80 &                   44.0 $\pm$         4.5 &                   47.1 $\pm$         7.5 &                         7 \\       
NGC1705      &                     52.1 $\pm$         1.6 &                   50.9 $\pm$         2.2 &                   41.1 $\pm$         2.1 &                         2 &        $\leq$  0.05  &        $\leq$  0.11  &        $\leq$  0.87  &        $\leq$  1.82  &                         6 \\       
NGC2366      &                     145. $\pm$         10. &                   147. $\pm$         13. &                   110. $\pm$         14. &                         5 &        $\leq$  0.12  &               0.70 $\pm$    0.08 &                   3.51 $\pm$        0.29$^b$ &               4.67 $\pm$        0.54$^b$ &                     6 \\       
NGC4214      &                     521. $\pm$         16. &                   614. $\pm$         27. &                   458. $\pm$         23. &                         2 &               0.65 $\pm$    0.07 &                   2.58 $\pm$        0.25 &                   17.9 $\pm$         1.7 &                   29.0 $\pm$         4.3 &                         7 \\       
NGC4449      &                     916. $\pm$         29. &                 1068.3 $\pm$        46.3 &                   839. $\pm$         42. &                         2 &                   2.14 $\pm$        0.25$^b$ &               5.15 $\pm$        0.73 &                   36.6 $\pm$         4.6 &                   73.0 $\pm$        13.5 &                         7 \\       
NGC4861$^{a}$ &               16.9 $\pm$         0.5 &                   14.0 $\pm$         0.6 &                   13.1 $\pm$         0.7 &                         2 &               0.25 $\pm$    0.03$^b$ &           0.41 $\pm$    0.09$^b$ &               1.82 $\pm$        0.40$^b$ &               2.39 $\pm$        0.60$^b$ &                     7 \\       
NGC5253      &                     380. $\pm$         12. &                   414. $\pm$         18. &                   334. $\pm$         17. &                         2 &                   2.81 $\pm$        0.24$^b$ &               12.3 $\pm$         1.3 &                   29.0 $\pm$         3.0 &                   29.1 $\pm$         4.7 &                         7 \\       
NGC625       &                     218. $\pm$          7. &                   236. $\pm$         10. &                   184. $\pm$          9. &                         2 &               0.20 $\pm$    0.04 &                   1.30 $\pm$        0.09 &                   5.73 $\pm$        0.37 &                   8.63 $\pm$        0.87 &                         9 \\       
NGC6822$^{a}$ &             1619.2 $\pm$        50.4 &                 1753.5 $\pm$        76.0 &                 1362.5 $\pm$        69.0 &                         2 &               0.25 $\pm$    0.04$^b$ &               2.46 $\pm$        0.40$^b$ &               47.6 $\pm$         7.8$^b$ &               95.4 $\pm$        17.2$^b$ &                    10 \\       
Pox186       &          - & - & - & -  &        - & - & - & - & - \\       
SBS0335-052  &                 0.303 $\pm$    0.009 &               0.28 $\pm$    0.01 &        $\leq$  0.38  &                         3 &        - & - & - & - & - \\       
SBS1159+545  &          - & - & - & -  &        - & - & - & - & - \\       
SBS1211+540  &          - & - & - & -  &        - & - & - & - & - \\       
SBS1249+493  &          - & - & - & -  &        - & - & - & - & - \\       
SBS1415+437  &          - & - & - & -  &        - & - & - & - & - \\       
SBS1533+574  &                     1.95 $\pm$        0.06 &                   2.14 $\pm$        0.09 &                   1.54 $\pm$        0.08 &                         3 &        $\leq$  0.06  &        $\leq$  0.07  &               0.26 $\pm$    0.03$^b$ &           0.4 $\pm$    0.1$^b$ &                     6 \\       
Tol0618-402  &                     4.32 $\pm$        0.13 &                   5.84 $\pm$        0.25 &                   4.04 $\pm$        0.20 &                         4 &        - & - & - & - & - \\       
Tol1214-277  &          - & - & - & -  &        - & - & - & - & - \\       
UGC4483      &                     2.20 $\pm$        0.23 &               0.9 $\pm$    0.1 &                   3.11 $\pm$        0.39 &                         4 &        $\leq$  0.09  &        $\leq$  0.09  &        $\leq$  0.12  &        $\leq$  0.57  &                         7 \\       
UGCA20       &          - & - & - & -  &        - & - & - & - & - \\       
UM133        &          - & - & - & -  &        - & - & - & - & - \\       
UM311        &          - & - & - & -  &        - & - & - & - & - \\       
UM448        &                     19.3 $\pm$         0.6 &                   22.2 $\pm$         1.0 &                   19.4 $\pm$         1.0 &                         2 &               0.15 $\pm$    0.04$^b$ &           0.8 $\pm$    0.1 &                   4.01 $\pm$        0.47 &                   4.30 $\pm$        0.75 &                         7 \\       
UM461        &                     2.03 $\pm$        0.21 &                   1.49 $\pm$        0.16 &                   1.27 $\pm$        0.14 &                         4 &        $\leq$  0.09  &        $\leq$  0.09  &        $\leq$  0.12 $^b$ &    $\leq$  0.57  &                         7 \\       
VIIZw403     &                     4.31 $\pm$        0.44 &                   3.89 $\pm$        0.40 &                   2.97 $\pm$        0.31 &                         4 &               0.07 $\pm$    0.01$^b$ &           0.06 $\pm$    0.01 &               0.39 $\pm$    0.05 &               0.9 $\pm$    0.2$^b$ &                     7 \\       
\hline
\end{tabular}
}
\end{center}
{\scriptsize                                                                                        
{\bf Notes : } \\                                                                                   
\noindent                                                                                           
$^a$ The aperture used does not cover the total emission from the galaxy and the 2MASS magnitudes reported in the database are not consistent with the rest of the IR photometry. Thus we do not consider them for the modelling. \\                                                                        
$^b$ This IRAS flux density is not consistent with the rest of the IR photometry. Thus we do not consider it for the modelling. \\                                                                                                                                                                          
{\bf References: }  {\bf 2MASS} - (1) 2MASS Extended Objects Final Release 2003 ; (2) \cite{Jarrett2003} ; (3) NASA/IPAC ISA Point Source Catalog, available at http://irsa.ipac.caltech.edu(4) \cite{Engelbracht2008} ; (5) \cite{Dale2009}. {\bf IRAS} - (6) NASA/IPAC ISA IRAS Faint Source Catalog (v2.0) available at: http://irsa.ipac.caltech.edu ; (7) \cite{Engelbracht2008} ;  (8) NASA/IPAC ISA IRAS Point Source Catalog (v2.1) available at: http://irsa.ipac.caltech.edu ; (9) \cite{Sanders2003} ;   (10) \cite{Rice1988}
}                                                                                                   
\end{table}
\end{landscape}

%% file: Dustparam_DGS_KF_fin.tex
\begin{landscape}                                                                                   
\begin{table}[h!tbp]                                                                                
\begin{center}                                                                                      
\caption{DGS and KINGFISH dust parameters.}                                                         
\label{t:Dustparam}                                                                                 
{\footnotesize                                                                                      
\begin{tabular}{l ccccccc | cccc}                                                                  
\hline                                                                                              
\hline                                                                                              
Name &                 Log[\mdust] [Gr] &            Log[\mdust] [Ac] &                  Log[\ltir] &                        \fpah &                      \uav\ &                       \sigU &             T$_{{\rm dust}}$ &                      \umin &                         \DU &                    $\alpha$ &                T$_{MIR,BB}$ \\
   &                         Log[\msun] &                  Log[\msun] &                  Log[\lsun] &              [\fpah$_\odot$] &                [U$_\odot$] &                 [U$_\odot$] &                         [K] &                 [U$_\odot$] &                 [U$_\odot$] &                              &                        [K] \\
 &                                   (1) &                         (2) &                         (3) &                         (4) &                         (5) &                         (6) &                         (7) &                         (8) &                         (9) &                      (10)   &                      (11)  \\
\hline                                                                                              
{\bf DGS} & & & & & & \\
\hline                                                                                              
\end{tabular}                                                                                       
}                                                                                                   
\end{center}                                                                                        
\end{table}                                                                                         
\end{landscape}                                                                                     
\begin{landscape}                                                                                   
\addtocounter{table}{-1}                                                                            
\begin{table}[h!tbp]                                                                                
\begin{center}                                                                                      
\caption{(continued) DGS and KINGFISH dust parameters.}                                             
{\footnotesize                                                                                      
 \begin{tabular}{l ccccccc | cccc}                                                                  
\hline                                                                                              
\hline                                                                                              
Name &                 Log[\mdust] [Gr] &            Log[\mdust] [Ac] &                  Log[\ltir] &                        \fpah &                      \uav\ &                       \sigU &             T$_{{\rm dust}}$ &                      \umin &                         \DU &                    $\alpha$ &                T$_{MIR,BB}$ \\
   &                         Log[\msun] &                  Log[\msun] &                  Log[\lsun] &              [\fpah$_\odot$] &                [U$_\odot$] &                 [U$_\odot$] &                         [K] &                 [U$_\odot$] &                 [U$_\odot$] &                              &                        [K] \\
 &                                   (1) &                         (2) &                         (3) &                         (4) &                         (5) &                         (6) &                         (7) &                         (8) &                         (9) &                      (10)   &                      (11)  \\
\hline                                                                                              
UGCA20      & $             - $& $       - $&  $      - $&   $         -$ & $       -$ &  $     - $&$-$&$ - $&$ -$ &$ - $&$ -  $    \\
\hline                                                                                              
\end{tabular}                                                                                       
}                                                                                                   
\end{center}                                                                                        
\end{table}                                                                                         
\end{landscape}                                                                                     
\begin{landscape}                                                                                   
\addtocounter{table}{-1}                                                                            
\begin{table}[h!tbp]                                                                                
\begin{center}                                                                                      
\caption{(continued) DGS and KINGFISH dust parameters.}                                             
{\footnotesize                                                                                      
 \begin{tabular}{l ccccccc | cccc}                                                                  
\hline                                                                                              
\hline                                                                                              
Name &                 Log[\mdust] [Gr] &            Log[\mdust] [Ac] &                  Log[\ltir] &                        \fpah &                      \uav\ &                       \sigU &             T$_{{\rm dust}}$ &                      \umin &                         \DU &                    $\alpha$ &                T$_{MIR,BB}$ \\
   &                         Log[\msun] &                  Log[\msun] &                  Log[\lsun] &              [\fpah$_\odot$] &                [U$_\odot$] &                 [U$_\odot$] &                         [K] &                 [U$_\odot$] &                 [U$_\odot$] &                              &                        [K] \\
 &                                   (1) &                         (2) &                         (3) &                         (4) &                         (5) &                         (6) &                         (7) &                         (8) &                         (9) &                      (10)   &                      (11)  \\
\hline                                                                                              
NGC3938     & $           7.72  _{-     0.06 }^{+     0.07      }$ & $           7.34  _{-     0.03 }^{+     0.04      }$ & $          10.20  _{-     0.01 }^{+     0.01      }$ & $           1.25  _{-     0.10 }^{+     0.10      }$ & $        1.7  _{-      0.2 }^{+      0.3                      }$ & $        2.8  _{-      0.5 }^{+      0.9                      }$ & $          20.5       $ & $          0.63       $ & $       6.19   \times10^{ 1}       $ & $       2.50 $ & $         -       $  \\
\hline                                                                                              
\end{tabular}                                                                                       
}                                                                                                   
\end{center}                                                                                        
{\footnotesize                                                                                      
\noindent                                                                                           
{\bf Notes:}                                                                                        
(1) Dust masses derived using graphite for the carbonaceous component, in \msun;                    
(2) Dust masses derived using amorphous carbons for the carbonaceous component, in \msun;           
(3) Total infrared luminosity, from the best-fit model and integrated between 1 and 1000 \mic, in \lsun;                                                                                                
(4) PAH mass fraction, normalised to the Galactic value \fpah$_\odot$ = 4.57\%;                                                                                                                         
(5) Mass-averaged starlight intensity, computed with Eq. \ref{eq:U}, normalised to the Galactic value U$_\odot$ = 2.2 $\times$ 10$^{-5}$ W.m$^{-2}$;                                                                                                                                                        
(6) Standard deviation of the starlight intensity distribution, computed with Eq. \ref{eq:sigU}, normalised to the Galactic value U$_\odot$;                                                            
(7) Average dust temperature, in Kelvins, computed by integrating Eq \ref{eq:U} over $T=T_{{\rm MW}} \times$ U$^{1/(4+\beta)}$, with $\beta$=2.0, and $T_{{\rm MW}}$=19.7K from \cite{PlanckCollaboration2014XI};                                                                                                                                                                                               
(8) Minimum value of the starlight intensity, normalised to the Galactic value U$_\odot$;                                                                                                               
(9) Difference between the maximum and minimum values of the starlight intensity distribution, normalised to the Galactic value U$_\odot$;                                                              
(10) Index of the power law describing the starlight intensity distribution;                                                                                                                            
(11) Temperature of the additional MIR modBB, in Kelvins. \\                                                                                                                                            
                                                                                                    
\noindent                                                                                           
$^a$ Dust mass and temperature obtained from a modBB fit (see Section \ref{sssec:addfeat}). \\                                                                                                          
\noindent                                                                                           
$^b$ \ltir\ obtained from \cite{Galametz2013b} calibrations as no SED fit was possible for this galaxy. \\                                                                                              
\noindent                                                                                           
$^c$ In these galaxies the fit converges towards 0: no PAH are detected and it is not possible to give an upper limit with the method described in Section \ref{ssec:fpah}. \\                          
}                                                                                                   
\end{table}                                                                                         
\end{landscape}